\documentclass[paper, 12pt, letterpaper]{JHEP} 
\usepackage{amsmath}
\usepackage{amssymb}
\usepackage{graphics}
\def\Bbb{\mathbb} \def\C{{\Bbb C}}  \def\Z{{\Bbb Z}}
     
\def\boldpsi{\mbox{\boldmath $\psi$}}

\def\ZA{\Z {\cal A}} 
\def\SigmaA{\Sigma {\cal A}} 

\def\vect{\rm vect}
\def\gr{{\rm gr}}
\def\dif{{\rm dif}}
\def\Vect{\rm Vect}
\def\Gr{{\rm Gr}}
\def\Dif{{\rm Dif}}
\def\ainfrange{\tiny \begin{array}{c}i\geq 0, j\geq 1\\1\leq i+j\leq n\end{array}}

\def\Hom{\operatorname{Hom}} 
 
    \def\tr{\operatorname{tr}}\def\Tr{\operatorname{Tr}}

\def\B{{B'}}
\def\Der{{\rm Der}}

\def\tw{{\rm tw}}

\def\add{{\rm add}}
\def\per{{\rm per}}
\def\tria{{\rm tria}}
\def\ktria{{\rm ktria}}
\def\Tria{{\rm Tria}}
\def\Mod{{\rm Mod}}
\def\GrMod{{\rm GrMod}}
\def\dGMod{{\rm dGMod}}
\def\RHom{{\rm RHom}}

\def\ainf{{$A_\infty$~}}
\def\inf{{$\infty$~}}
\def\dG{{\rm dG}}
\def\cA{{\cal A}}
\def\cT{{\cal T}}
\def\cB{{\cal B}}
\def\cC{{\cal C}}
\def\cG{{\cal G}}
\def\Ob{{\rm Ob}}
\def\dual{^{\rm v}}
\def\ker{{\rm ker}}  
\def\im{{\rm im\,}}  
 \def\tr{{\rm tr\, }} \def\dim{{\rm dim}}
 \def\Tr{{\rm Tr\, }} 
\def\Card{{\rm Card}}   \def\deg{{\rm deg}}  
  
\def\End{{\rm End}} \def\Aut{{\rm Aut}}

\def\id{{\rm id}}

\newcommand{\be}{\begin{equation}} \newcommand{\ee}{\end{equation}}
\newcommand{\bea}{\begin{eqnarray}} \newcommand{\eea}{\end{eqnarray}}
\newcommand{\beann}{\begin{eqnarray*}}
  \newcommand{\eeann}{\end{eqnarray*}}
\newcommand{\bfig}{\begin{figure}} \newcommand{\efig}{\end{figure}}
\newcommand{\nn}{\nonumber}
\newcommand{\ba}{\begin{array}}\newcommand{\ea}{\end{array}}

\newtheorem{Proposition}{Proposition}[section]

\newtheorem{Theorem}{Theorem}[section]
\newtheorem{Lemma}{Lemma}[section]
\newtheorem{Corrolary}{Corrolary}[section]

\newcommand{\bp}{\begin{Proposition}}
  \newcommand{\ep}{\end{Proposition}}
\newcommand{\bt}{\begin{Theorem}} \newcommand{\et}{\end{Theorem}}
\newcommand{\bl}{\begin{Lemma}} \newcommand{\el}{\end{Lemma}}
\newcommand{\bc}{\begin{Corrolary}} \newcommand{\ec}{\end{Corrolary}}

   \def\ep{\eps}

\author{C. I.  Lazaroiu\\}
\author{~~\\
Trinity College Dublin\\
Dublin 2\\
Ireland\\
calin@maths.tcd.ie\\
}

\title{Generating the superpotential on a D-brane category:I}

\abstract{I study \ainf enhancements of algebraic Calabi-Yau
triangulated categories admitting a (triangle) generator, showing that the Serre
pairing on such categories determines and is determined by a cyclic
pairing on an enhancement of the generator.  Using this
result, I construct a formal topological string field action inducing 
an extended D-brane superpotential for such categories.  I also  give a procedure
for lifting certain 2d boundary topological field theories to open
topological string theories generated by a single D-brane.}

\preprint{}

\begin{document}

\tableofcontents

\pagebreak

\vskip .6in

\section*{Introduction} 

A basic problem in open topological string theory is the construction of  D-brane 
superpotentials on topological D-brane categories. At first sight, this might seem
hopeless since current techniques approach 
the question one D-brane at a time ---
while any interesting example involves a D-brane category with an uncountable set of objects. 

It is apparent that one needs 
a method for `generating' this quantity starting from a finite collection of 
D-branes -- in the sense that the superpotential of any 
finite D-brane system in the category should be determined by the generating branes. 

To formulate this clearly, we must distinguish from the outset between
(oriented) open-closed topological field theory in two dimensions and
(oriented) open-closed topological conformal field theory, also known
as topological string theory.  The boundary sector of such theories can be described as follows.

For a 2d topological {\em field} theory, the boundary sector is
given\cite{CIL1, Moore} by a graded associative category $\cG$
(enriched over vector spaces) endowed with nondegenerate bilinear
pairings $\langle ~,~\rangle_{ab}:\Hom_\cG(a,b)\times \Hom_{\cal
G}(b,a)\rightarrow \C$ which are homogeneous of some common degree
$-D$ and satisfy certain compatibility conditions with respect to the
composition of morphisms. The physical interpretation of such data is
as follows. The objects of $\cT$ define the boundary sectors of
the theory in the sense of \cite{CIL1}, while the morphism spaces
$\Hom_\cG(a,b)$ identify with the spaces of boundary and boundary
condition changing topological observables. The morphism compositions
describe the associative product of boundary observables, while the
bilinear pairings correspond to the boundary topological metrics and
thus come from the two-point functions on the disk. When the graded
category $\cG$ admits a shift functor which preserves the 
pairings up to sign, then one can also describe this
data through the associative category $\cT$ obtained from ${\cal
G}$ by keeping only degree zero morphisms. In this equivalent
language, the bilinear pairings give nondegenerate maps $(
~,~)_{ab}:\Hom_\cT(a,b)\times \Hom_\cT(b,a[D])\rightarrow
\C$ compatible with morphism compositions, which define a nondegenerate
`invariant' pairing on $\cT$. The original graded
category can be recovered from $\cT$ as the category ${\cal
G}=\cT^\bullet$ which has the same objects as $\cT$,
morphism spaces $\Hom_{\cT^\bullet}(a,b)=\oplus_{n\in
\Z}{\Hom_\cT(a,b[n])}$ and morphism compositions given by:
\be
\label{starcomp}
g*f=g[m]\circ f~~\forall f\in \Hom_\cT(a,b[m])~,~\forall g\in \Hom_\cT(b,c[n]). 
\ee

For a topological {\em string} theory, the boundary sector is
described \cite{HLL,Costello} by a minimal, cyclic and strictly unital
\ainf category $(\cB, \langle~,~\rangle)$, whose bilinear pairings $\langle
~,~\rangle_{ab}:\Hom_\cB(a,b)\times \Hom_\cB(b,a)\rightarrow \C$ are
nondegenerate and homogeneous of common degree $-D$.  The `suspended
forward \ainf compositions' give degree one linear maps
$\rho_{a_0\ldots a_n}:\Hom_{\cB}(a_0,a_1)[1]\otimes \ldots \otimes
\Hom_{\cB}(a_{n-1},a_n)[1]\rightarrow \Hom_{\cB}(a_0, a_n)[1]$ such that
the quantities $\langle u_0, \rho_{a_0\ldots a_n}(u_1\otimes \ldots \otimes
u_n)\rangle_{a_n a_0}$ (where $u_j\in \Hom_{\cB}(a_{j-1},a_j)$ with
$a_{-1}:=a_n$) are graded cyclically symmetric and coincide up to sign
with the integrated boundary $n$-point functions on the disk. The
\ainf, cyclicity and unitality constraints for $\rho_{a_0\ldots a_n}$
were derived in \cite{HLL, Costello} from the axioms of open-closed
topological conformal field theory. In such models, the theory
possesses a (worldsheet) boundary BRST charge such that the morphism
spaces $\Hom_{\cB}(a,b)$ arise as the BRST cohomology on the space of
boundary operators for strings stretching from $a$ to $b$.  The units
of the \ainf category are the unit boundary observables (=BRST cohomology classes of boundary operators) 
in the various boundary sectors. The bilinear pairings are induced from the BPZ form by
passage to BRST cohomology. One has\footnote{Notice that the boundary
topological metric $(~,~)$ is denoted by $\omega$ in reference
\cite{HLL}.} :
\be
\nn
\langle u_0, \rho_{a_0\ldots a_n}(u_1\otimes \ldots \otimes u_n)\rangle_{a_n a_0}={\cal F}_{u_0\ldots u_n}~~, 
\ee
with
\be
\label{amplitude}
{\cal F}_{u_0\ldots u_n}=(-1)^{{\tilde u}_1+\ldots +{\tilde u}_{n-1}}{\Big \langle}~{\cal O}_{u_0}{\cal O}_{u_1} {\cal P} 
\int {\cal O}^{(1)}_{u_2}\ldots \int {\cal O}^{(1)}_{u_{n-1}}{\cal O}_{u_n}{\Big \rangle}~~,
\ee
where the big brackets in the right hand side denote the expectation
value in the worldsheet theory on the disk. Here ${\cal O}_{u_j}$ are
the boundary (condition changing) observables associated with $u_j\in
\Hom_{\cB}(a_{j-1},a_j)$, and ${\cal O}_{u_j}^{(1)}$ are their
boundary topological descendants, which are inserted in the
clockwise order on the boundary of the disk. The portion of the
disk's boundary lying between the insertion of ${\cal O}_{u_j}$ and
${\cal O}_{u_{j+1}}$ carries the boundary label $a_j$ and corresponds
to the boundary condition associated with that D-brane. The integrals
stand for path ordered integration (hence the symbol ${\cal P}$) over
the positions of insertions of boundary descendants.  The naive
amplitude (\ref{amplitude}) receives divergent contributions when two
or more boundary insertions approach each other. These can be
regularized either as in \cite{HLL} (a version of cutoff
regularization) or geometrically by considering the moduli space of
stable punctured disks. The second regularization 
corresponds to the modular functor approach of
\cite{Costello}. Both methods lead to the same constraints on
amplitudes, which are encoded by the nondegenerate cyclic, minimal and
strictly unital \ainf category $\cB$.

Given an open topological string theory, one defines a tree-level
extended potential $W_e$ as follows.  Consider the (typically
infinite-dimensional) graded vector space ${\cal
H}_\cB:=\oplus_{a,b\in \Ob\cA}{ \Hom_{\cA}(a,b)}$. This carries a
cyclic minimal \ainf structure with products $\rho_n$ and pairing $\langle
~\rangle$ obtained from $\rho_{a_0\ldots a_n}$ and $\langle~\rangle_{ab}$ by `summing
over sectors'\cite{nc}. Picking a Grassmann algebra $G$, one considers
the right $G$-module $({\cal H}_\cB)_e:={\cal H}_\cB\otimes G$ and the natural
extensions $\langle~\rangle_e$ and $\rho_n^e$ of the pairing and \ainf
products of ${\cal H}_\cB$ to $({\cal H}_\cB)_e$. The tree-level extended
potential is the $G$-valued function $W_e:({\cal H}_\cB)_e^{\rm odd} \rightarrow G$
defined through the formal expression:
\be
\label{W}
W_e(\boldpsi)=\sum_{n\geq 2}{\frac{1}{n+1}\langle \boldpsi, \rho_n^e(\boldpsi^{\otimes n}) \rangle_e}~~
\ee
where $\boldpsi\in ({\cal H}_\cB)_e^{\rm odd}$. When $D=3$, then one can interpret the restriction of 
$W_e$ to the subspace ${\cal H}_{\cB}^1\otimes (\C \id_G)\approx {\cal H}_{\cB}^1$
as the superpotential of an $N=1$ supersymmetric field theory in four 
dimensions obtained from an abstract `compactification' of the 
string theory associated with the untwisted conformal field theory on which we base our topological model. 

Given an open topological string theory, one recovers a 2d topological field theory
by keeping only the binary \ainf products $r_{abc}$ and forgetting all higher \ainf compositions. 
Indeed, the \ainf constraints imply that the `desuspensions' 
$m_{abc}:\Hom_{\cB}(a,b)\otimes 
\Hom_{\cB}(b,c)
\rightarrow \Hom_{\cB}(a,c)$ of $r_{abc}$ give (degree zero)
associative compositions $*$ on $\cB$ via the formula 
$m_{abc}(f,g)=(-1)^{\deg f\deg g} g* f$. When endowed only with these compositions, $\cB$ 
becomes a graded associative category with `invariant' nondegenerate bilinear pairings, which 
can be identified with the category $\cG$ of a 2d 
topological field theory. We say that $(\cB, \langle~,~\rangle) $ {\em prolongs} $(\cG, \langle~,~\rangle)$.
A shift functor for the \ainf category $\cB$ induces a shift functor of the graded category $\cG$, 
and we require that the two shift functors agree. 
Since $W_e$ is determined by such a prolongation, the problem of `describing the superpotential on $\cT$' 
can be strengthened as follows:

\paragraph{Problem}{\em Given a 2d open topological field theory whose 
boundary sector is described by $(\cT, (~,~))$, find an open 
topological string theory whose boundary sector $(\cB, \langle ~,~\rangle)$ prolongs 
$(\cT^\bullet, \langle~,~\rangle)$.}

\

In the present paper, I give one solution of this problem under the assumption that 
$\cT$ is an algebraic triangulated category which is 
triangle generated by one object (in which case the pairings on $\cT$ are Serre pairings), 
and show that in this simple situation a superpotential is determined by a single D-brane. 
The construction I discuss is as follows. Assume that $\cT$ 
is triangulated and algebraic (i.e. equivalent with the stable category of a Frobenius category, 
see \cite{Keller_dG, Keller_dGcat}). Also assume given $g\in \Ob\cT$ 
such that the smallest triangulated subcategory of $\cT$ containing $g$ (and its shifts)
coincides with $\cT$. Then I construct an `off shell model' (cyclic \ainf enhancement) 
of $(\cT^\bullet, \langle~,~\rangle)$, i.e. a non-minimal but strictly unital 
nondegenerate cyclic \ainf category with shifts $(\cA, \langle ~,~\rangle_{\cA})$ 
such that $H(\cA)\approx \cT^\bullet$ and such that the nondegenerate 
pairings of $\cA$ induce 
the pairings of $\cT^\bullet$ (up to an uninteresting equivalence) when passing to cohomology. 
Via the results of \cite{Costello}, 
the data $(\cA, \langle~,~\rangle_{\cA})$ defines a topological D-brane system, which gives a 
formal extended string field action $S_e$. The extended tree-level potential of this system gives 
an extended `superpotential' $W_e$ on $\cT^\bullet$, which also carries a cyclic \ainf prolongation.  
Both of these are obtained by constructing  a strictly unital 
and shift-invariant cyclic minimal model of $(\cA, \langle~,~\rangle_\cA)$ associated with 
an appropriate choice of gauge for $S_e$. 

In detail, the argument proceeds as follows.  Since $\cT$ is algebraic and
triangle generated by $g$, the results of \cite{Keller_dG, Hasegawa}
imply that one can find a minimal and strictly unital \ainf algebra
$A_{\rm min}$ such that $H(A_{\rm min})= \Hom_{\cT^\bullet}(g,g)$
and $\cT^\bullet=H(\cA)$, where $\cA:=\tw(A_{\rm min})$ is the
\ainf category of twisted complexes over $A_{\rm min}$ \cite{HMS,
Fukaya_mirror, Hasegawa}. Any strictly unital and shift-invariant
minimal model of ${\cal A}$ gives a candidate prolongation of ${\cal
T}^\bullet$, but such a minimal model need not be cyclic.  To insure
cyclicity, we proceed in two steps:

(1) We show that existence of a nondegenerate pairing on $\cT$ implies that
one can choose $A_{\rm min}$ such that it carries a nondegenerate cyclic pairing. To avoid
computational morass, we do this by using a
quasi-isomorphic dG model, showing that it carries an
invariant and homologically nondegenerate bilinear pairing, then
transport this to a nondegenerate cyclic pairing on a minimal model via
an \ainf quasi-isomorphism.

(2) We show that any nondegenerate cyclic pairing on $A_{\rm min}$
induces a nondegenerate cyclic pairing on $\tw(A_{\rm min})$ via a
natural extension process. In turn, the latter descends to a Serre
pairing on $H(\tw(A_{\rm min}))\approx \cT^\bullet$, which induces
the original Serre pairing on $\cT$ up to an uninteresting transformation.

When endowed with the induced pairing, the cyclic \ainf category
$\cA=\tw(A_{\rm min})$ provides the cyclic off-shell model of ${\cal
T}$ promised above. The space ${\cal H}_\cA:=\oplus_{a,b\in {\cal
T}}\Hom_{\cal A}(a,b)$ carries the structure of a unital and cyclic \ainf algebra
induced from $\cA$, whose bilinear pairing is
nondegenerate.  Using this data, we construct a formal extended string
field action $S_e$ describing a topological D-brane system whose D-branes are 
the objects of $\cA$. Since
$(\cA, \langle ~,~\rangle_\cA)$ is completely determined by $A_{\rm
min}$ and its pairing, the physics of this topological D-brane system 
is determined by the latter data. Studying the
extremum conditions for $S_e$, one finds that any twisted complex in
$\cA=\tw(A_{\rm min})$ can be obtained as the result of `topological
tachyon condensation' in a system of open strings stretching between a
finite number of shifted copies of a single D-brane $a\in \Ob \cA$ such that $A_{\rm min}=\Hom_{\cA}(a,a)$. In this
sense, the single D-brane $a$ and its shifts generate
the entire \ainf D-brane category $\cA$, as well
as its pairings.  It therefore also generates all string field
amplitudes in this D-brane category.

An extended D-brane potential on $\cT^\bullet=H(\cA)$ can now be
obtained as the effective tree-level potential of this formal string
field action. The
superpotential has the form (\ref{W}), where the minimal products
$\rho$ correspond to a cyclic, strictly unital and shift-invariant
minimal model of the cyclic, strictly unital \ainf category with
shifts $({\cal A},\langle~,~\rangle_\cA)$.  For example, one can
pick a `standard' gauge fixing condition, leading to a minimal model
of the type described in \cite{Merkulov, KS_old, Fukaya_mirror}. Since
$S_e$ is determined by $A_{\rm min}$ and its pairing in the sense explained
above, the extended potential is entirely determined by this
data.  In this sense, such a
superpotential is generated by the single D-brane $a$.

The paper is organized as follows. Section 1 recalls the mathematical description 
of oriented open 2d topological field theories, paying special attention to 
the shift-equivariant and triangulated cases. In Section 2, we recall some
background on \ainf categories and their homological algebra, fixing
the notation and conventions used throughout the paper.  The important point of 
this section is the choice of signs in the construction of the \ainf category 
of twisted complexes. In Section 3,
we discuss cyclic \ainf categories and give an extension procedure
which starts with a cyclic pairing on an \ainf category $\cA$ and
induces a cyclic pairing on its category of twisted complexes
$\tw(\cA)$ (this generalizes a construction originally
discussed in \cite{CIL2, CIL3, CIL4} for the dG case with $D=3$). We also give an
explicit construction of cyclic minimal models, which enriches a
result of \cite{KS_old}, addressing the issues of 
unitality and shift-equivariance for such models.
Finally, we discuss the formal string field theory
interpretation of cyclic \ainf categories and give the construction of
extended D-brane `superpotentials' starting from a formal string field action.  In
Section 4, we consider the case of cyclic differential graded algebras
$A$.  After recalling an equivalent construction of homological
algebra over $A$, we show that the category $H^0(\tw(A))$ is
Calabi-Yau iff $A$ admits a homologically nondegenerate cyclic pairing
(which in this case amounts to the more familiar notion of a
graded-symmetric invariant pairing). We also show that a similar
statement holds when $A$ is replaced by a minimal \ainf algebra
$A_{\rm min}$. Section 5 gives the construction of our formal string field
action. After recalling a generation result due to \cite{Keller_dG}
and \cite{Hasegawa}, we use the results of Section 4 to show that any
generator of an algebraic Calabi-Yau triangulated category $\cT$ can
be enhanced to a cyclic and unital minimal \ainf generator.  Using
this fact and the extension procedure of Section 3, we build the
desired string field action and show that it defines a topological
D-brane system enriching the 2d topological field theory described by
$\cT$. We also show that all D-branes in this system can be obtained
from a single D-brane and its shifts through the process of
topological tachyon condensation.  Finally, we use the construction of
Section 3 to induce a superpotential on $\cT$ as well as a
prolongation of $\cT^\bullet$.

Certain technical results used in the paper can be found in
appendices. In Appendix A, we discuss categories with shifts, duality
structures and cyclic structures. Appendix B provides a different
perspective on cyclic minimal models, which is afforded by the geometric
formalism of cyclic \ainf algebras \cite{konts_formal, Ginzburg,
Lazarev, nc}.

\paragraph{Conventions, notation and terminology.}

Throughout this paper, we work over the field $\C$ of complex
numbers\footnote{All results extend trivially to base
fields of characteristic zero.}  and consider the following tensor
(=symmetric monoidal) categories:

\begin{itemize}
\item The category $\vect$ of vector spaces over $\C$, whose morphism
spaces we denote by $\Hom_\C(V,W):=\Hom_{\vect}(V,W)$.

\item The category $\gr$ of $\Z$-graded vector spaces over $\C$. The morphism spaces are 
\be 
\nn
\Hom_{\gr}(V,W):=\{f\in \Hom_\C(V,W)|f(V^n)\subset W^{n}~~\forall n\in \Z\}~~,
\ee
where $V=\oplus_{n\in \Z}V^n$ and $W=\oplus_{n\in \Z}W^n$.

\item The category $\dif$ of (possibly unbounded) cochain complexes of
vector spaces over $\C$. Viewing complexes as pairs $(V,d_V)$ with
$V\in \Ob[\gr]$ and $d_V$ a differential of degree $+1$ on $V$, the
morphism spaces are:
\be 
\nn
\Hom_{\dif}(V,W):=\{f\in \Hom_\gr(V,W)|d_W\circ f_k =f\circ d_V  \}~~.
\ee

\end{itemize}

In the present paper, an {\em associative category} means a small associative
category enriched over \vect.  A {\em graded associative category}
means a small associative category enriched over \gr. A {\em
differential graded (dG) category} means a small associative category
enriched over \dif. An \ainf category means an \ainf category enriched over $\gr$. 
Similar enrichment conventions apply to functors between such categories. We make
systematic use of the Koszul sign rule for graded quantities, unless
explicitly mentioned otherwise. We also use the convention of writing
various equations only for homogeneous elements (in order to indicate
the signs).  We will make use of the following enriched categories: 

\begin{itemize}
\item $\Vect$ is the $\C$-category of vector spaces over $\C$. This is
the same as $\vect$ except that $\Hom_\C(V,W)$ are viewed as vector
spaces rather than as sets.

\item $\Gr$ is the graded associative category of $\Z$-graded vector
spaces over $\C$. This has morphisms spaces $\Hom_\Gr(V,W)
=\oplus_{n\in \Z} \Hom^n_\Gr(V,W)$, where: \be \nn
\Hom^n_\Gr(V,W):=\{f\in \Hom_\C(V,W)|f(V^k)\subset W^{k+n}~~\forall
k\in \Z\}~~.  \ee A graded vector space $V$ is called {\em degreewise
finite} if $\dim_\C V^n<\infty$ for all $n\in \Z$. degreewise finite
graded vector spaces form a full graded subcategory $\Gr_{\rm df}$ of
$\Gr$.

\item $\Dif$ is the dG category of (possibly unbounded) cochain
complexes of vector spaces over $\C$. This has morphism spaces:
\be 
\nn
\Hom_{\Dif}(V,W):=\Hom_\Gr(V,W)~~
\ee
with differentials $d_{V,W}\in \Hom_{\Gr}^1(\Hom_{\Dif}(V,W))$ given by: 
\be
\label{dif}
d_{V,W}(f)=d_W\circ f-(-1)^{\deg f}f\circ d_V~~.
\ee
Notice that $Z^0(\Dif)=\dif$ and the forgetful functor gives an
embedding $\Dif\subset \Gr$. Here and below, we let $Z(\ldots),
B(\ldots), H(\ldots) $ denote passage to cocycles, coboundaries and
cohomology.

\end{itemize}

The graded category $\Gr$ is endowed with a shift functor $[1]$, which
is defined through:
\be
\nn
V[1]^n:=V^{n+1}~~,
\ee
with the obvious action on morphisms. This gives an automorphism of $\Gr$ as a graded category. Setting $[n]:=[1]^n$, we
have $\Hom^n_\Gr(V,W)=\Hom_{\gr}(V,W[n])$. Let $\id_\Gr$ be the identity 
endofunctor. The {\em suspension} of $V \in \Ob\Gr$ is the map $s_V:V\rightarrow V[1]$
(the identity map of $V$ viewed as a map of degree $-1$ from $V$ to $V[1]$). 
The {\em signed suspension} 
is the map  $\sigma_V:V\rightarrow V[1]$ of degree $-1$ given by $\sigma_V(x)=(-1)^{\deg x} x$. 
The {\em signed} suspensions give a natural transformation $\sigma:\id_{\Gr}\rightarrow [1]$ of degree $-1$ 
since one has $\sigma_W f=(-1)^{\deg f} f\sigma_V$ for homogeneous $f\in \Hom_\Gr(V,W)$ (notice the sign factor which is 
required by the Koszul rule). 

Similarly, the dG category $\Dif$ has a shift functor  which acts on objects $(V,d_V)$ as 
$(V,d_V)[1]:=(V[1],d_V)$ and on morphisms in the same way as in $\Gr$. The signed suspensions  
give maps of complexes $\sigma_V:(V,d_V)\rightarrow (V,d_V)[1]$ of degree $-1$. Together, they 
define a natural transformation $\sigma:\id_\Dif\rightarrow [1]$ of degree $-1$.

The {\em dualization functor} is the contravariant endofunctor $^{\rm
v}$ of $\Gr$ defined as follows.  For any $V\in \Ob \Gr$, set $V^{\rm
v}:=\Hom_\Gr(V,\C)$, where $\C$ is viewed as a graded vector space
concentrated in degree zero.  We have $(V^{\rm v})^n=\Hom_{\gr}(V,\C[n])= 
\{\eta\in \Hom_{\C}(V,\C)| \eta(x)=0~~{\rm unless}~\deg x=-n\}.$ This gives isomorphisms $(V^{\rm v})^n\approx
\Hom_\C(V^{-n},\C)$. The functor $^{\rm v}$ acts on homogeneous morphisms
$f\in \Hom_\Gr(V,W)$ by:
\be
\label{fdual}
f^{\rm v}(\eta):=(-1)^{\deg f \deg \eta}\eta\circ f~~,
\ee
which implies the graded contravariance condition $(f\circ g)^{\rm
v}=(-1)^{\deg f\deg g} g^{\rm v}\circ f^{\rm v}$ for composable
$f,g$. The dualization functor preserves $\Gr_{\rm df}$ and squares to the
identity on this subcategory.

The dualization functor induces a contravariant dG functor $^{\rm v}:\Dif\rightarrow \Dif$ as follows. 
For any complex $(V,d_V)$, endow $V^{\rm v}$ with the differential $d_{V^{\rm v}}=-d_V\dual$, i.e.: 
\be
\nn
d_{V^{\rm v}}\eta:=(-1)^{1+\deg \eta}\eta\circ d_V~~
\ee
and let $^{\rm v}$ act on morphisms as in (\ref{fdual}). Notice the $d$-compatibility relations 
$d_{W^{\rm v}, V^{\rm v}}(f^{\rm v})=d_{V,W}(f)^{\rm v}$.  The natural
isomorphism\footnote{This follows from $B(V^{\rm v})=\im (d_{V^{\rm
v}})=(\ker d_V)^o$, $Z(V^{\rm v})=\ker (d _{V^{\rm v}})=(\im d_V)^o$
and $(\im d_V)^o/(\ker d_V)^o\approx (\ker d_V/\im d_V)^{\rm v}$,
where $S^o:=\oplus_{n\in \Z} \{\eta\in (V^{\rm
v})^{n}|~\eta|_{S^{-n}}=0\}\subset V^{\rm v}$ is the degreewise polar
of a homogeneous linear subspace $S=\oplus_{n\in \Z}S^n\subset V^{\rm
v}$.  These relations do not require finite-dimensionality. }
$H(V^{\rm v})\approx H(V)^{\rm v}$, implies that dualization preserves
the full subcategory of acyclic complexes.

An associative category $\cA$ will be called {\em Hom-finite}
if $\Hom_\cA(a,b)$ is finite-dimensional for all objects $a,b$. A
graded associative category $\cG$ is called $\Hom$-finite if its
underlying associative category is $\Hom$-finite. It is called {\em
degreewise Hom-finite} if all $\Hom_\cG(a,b)$ are degreewise
finite.

For any unital associative ring $R$, we let $_R\Mod$, $\Mod_R$ and $_R\Mod_R$ denote the categories of 
(unital) left, right and bi-modules over $R$ and $_R\GrMod, \GrMod_R$, $_R\GrMod_R$ the categories of 
(unital) graded left, right and bi-modules over $R$. 

For a unital dG algebra $A$, we let $_A\dGMod$, $\dGMod_A$ and $_A\dGMod_A$ denote the dG categories
of (strictly) unital dG left, right and bi-modules over $A$. 

For an \ainf algebra $A$, we let  $_A\Mod$, $\Mod_A$ and $_A\Mod_A$ denote the dG categories
of strictly unital \ainf left, right and bi-modules over $A$. Similar notation applies when $A$ is replaced 
with an \ainf category $\cA$. 

\

\section{Shift-equivariant open topological field theories in two dimensions}

In this Section, we discuss the mathematical description of open topological field 
theories in two dimensions, paying special attention to the case when the category of boundary 
sectors has a shift functor. A more detailed account of certain aspects can be found 
in Appendix \ref{sec:duality}, which also discusses the relation with the usual theory of Serre functors. 

\subsection{The mathematical description of open topological field theories in two dimensions}

Given an integer $D$, a $D$-{\em cyclic structure} on a unital graded
associative category $\cG$ is a family of degree zero linear maps
$\tr_a:\Hom_{\cG}(a,a)\rightarrow \C[-D]$ indexed by the objects $a,b$
of $\cG$, which satisfy the relations:
\be
\tr_a(uv)=(-1)^{\deg u~\deg v}\tr_b(vu)~,~{\rm for}~~ v\in \Hom_{\cG}(a,b)~,~\forall 
u\in\Hom_{\cG}(b,a)~~.~~~~~~~~~\\
\ee
Defining degree zero bilinear pairings
$\langle~\rangle_{ab}:\Hom_{\cG}(a,b)\times
\Hom_{\cG}(b,a)\rightarrow \C[-D]$ via $\langle u, v
\rangle_{ab}:=\tr_b(uv)$, this
corresponds to the conditions: \bea \langle u f,
v\rangle_{a'b}&=&\langle u, fv\rangle_{a,b}~~ \forall
f\in\Hom_{\cG}(a',a),~u\in\Hom_{\cG}(a,b),~v\in\Hom_{\cG}(b,a')~~~~~~~~\\
\langle u, v\rangle_{a,b}&=&(-1)^{\deg u~\deg v}\langle v,
u\rangle_{b,a}~~, ~~\forall u\in\Hom_{\cG}(a,b)~~\forall v
\in\Hom_{\cG}(b,a)~~.~ \eea A graded category $\cG$ endowed with a
D-cyclic structure will be called a $D$-cyclic graded category.  The
cyclic structure is called {\em non-degenerate} if $\cG$ is degreewise
Hom-finite and all pairings $\langle u, v \rangle_{ab}$ are
nondegenerate as bilinear forms. The following is a trivial extension
of a result proved in \cite{CIL1}:

\

{\em An oriented open topological field theory in two-dimensions (=the
boundary sector of an oriented open-closed 2d topological field
theory) is described by a unital graded associative category $\cG$
endowed with a nondegenerate cyclic structure.}

\

As in \cite{CIL1}, this follows from the modular functor approach,
except that we allow for a $\Z$-grading on the space of worldsheet
fields. This is possible provided that the worldsheet model admits an
unbroken $U(1)$ symmetry. The objects of $\cG$ are interpreted 
as boundary sector labels, while the composition of morphisms in $\cG$ gives the 
boundary products.

\subsection{Two-dimensional open topological field theories with shifts}

In this paper, we are interested in the case when $\cG$ admits a shift
functor, which we denote by $[1]$. By definition, this is an automorphism of $\cG$
together with isomorphisms of graded vector spaces
$\Hom_{\cG}(a,b[1])\approx \Hom_{\cG}(a,b)[1]$, which are natural in
$a$ and $b$.  A $D$-cyclic structure on $(\cG,[1])$ is called
shift-equivariant if the following conditions are satisfied:
\be
\label{SE}
\tr_{a[1]}(u[1])=(-1)^{D+1}\tr_a(u)\Leftrightarrow \langle u[1],v[1] \rangle_{a[1]b[1]}=(-1)^{D+1}\langle u,v \rangle_{ab}~~.
\ee
In this case, we also say that the associated open 2d topological field theory is shift-equivariant.

Let $\cT={\cal G}^0$ be the {\em null restriction} of $\cG$, i.e. the subcategory of $\cG$ obtained by keeping only morphisms of degree zero. This admits 
an automorphism (again denoted by $[1]$) obtained by restricting the shift functor of $\cG$. 
Then $\cG$ can be reconstructed as the {\em graded completion} $\cG=\cT^\bullet$ of the unital associative category 
$\cT$. This is the category having the same objects as $\cT$ and morphism 
spaces $\Hom_{\cT^\bullet}(a,b)=\oplus_{n\in \Z}\Hom_{\cT}(a,b[n])$, with compositions defined as in (\ref{starcomp}). 
In fact, graded completion and null restriction give 
inverse equivalences between the categories of small associative categories with shifts and small graded associative 
categories with shifts (see Appendix \ref{sec:duality}).

Let us define a $D$-{\em cyclic structure} on a unital (not graded) associative category with shifts 
$(\cT,[1])$ to be a family of linear maps $tr_a:\Hom_{\cT}(a,a[D])\rightarrow \C$ 
indexed by the objects of $\cT$, which satisfy the relations:
\be
tr_a(u\circ v)=tr_b(v[D] \circ u)~,~{\rm for}~~v\in \Hom_{\cT}(a,b)~,~u\in\Hom_{\cT}(b,a[D])~~.~~~~~~~~~\\
\ee
Defining pairings $(~,~)_{ab}:\Hom_{\cT}(a,b)\times \Hom_{\cT}(b,a[D])\rightarrow \C$ via 
$( u, v )_{ab}:=tr_b(u[D]\circ v)$, this corresponds to the conditions: 
\bea
(u\circ f, v)_{a'b}&=&(u, f[D]\circ v)_{a,b}~~
\forall f\in\Hom_{\cT}(a',a),~u\in\Hom_{\cT}(a,b),~v\in\Hom_{\cT}(b,a'[D])~~~~~~~~~~~~\\
(u, v)_{a,b}&=&(v, u[D])_{b,a[D]}~~, 
~~\forall u\in\Hom_{\cT}(a,b)~~\forall v \in\Hom_{\cT}(b,a[D])~~~~~~~~~~~~~.
\eea
We say that the D-cyclic structure is shift-equivariant if the following relations are satisfied:
\be
tr_{a[1]}(u[1])=(-1)^{D+1}tr_a(u)\Leftrightarrow (u[1],v[1])_{a[1]b[1]}=(-1)^{D+1}(u,v)_{ab}~~.
\ee
We say that it is nondegenerate if $\cT$ is Hom finite and all pairings $(~)_{ab}$ are nondegenerate 
as bilinear forms.  One has the following correspondence (see Appendix \ref{sec:duality}): 

\

{\em Let $(\cG,[1])$ be a unital 
graded category with shifts and $(\cT,[1])$ the corresponding unital associative category with shifts (thus 
$\cT=\cG^0$ and $\cG=\cT^\bullet$). The following data are equivalent:

(a) A shift-equivariant $D$-cyclic structure on $(\cG,[1])$.

(b) A shift-equivariant $D$-cyclic structure on $(\cT,[1])$.

Moreover, one is nondegenerate iff the other is. In this case, a shift-equivariant topological field theory in two 
dimensions can be described by either datum. }

\

It is often convenient to work with the {\em twisted shift functor} $[[1]]$ of $\cG$, an automorphisms of $\cG$ which acts on 
objects through $a[[1]]:=a[1]$ and on homogeneous morphisms $f$ through $f[[1]]=(-1)^{\deg f} f[1]$. 
In terms of this, the shift-equivariance conditions (\ref{SE}) take the form:
\be
\label{SE_graded}
\tr_{a[[1]]}(u[[1]])=-\tr_a(u)\Leftrightarrow \langle u[1],v[1] \rangle_{a[[1]]b[[1]]}=-\langle u,v \rangle_{ab}~~.
\ee
The isomorphisms $\Hom_{\cG}(a,b[1])\approx \Hom_{\cG}(a,b)[1]$ become isomorphisms  
$\Hom_{\cG}(a,b[[1]])\approx \Hom_{\cG}(a,b)[1]$ which are natural up to missing Koszul signs (see Appendix \ref{sec:duality}).

Since their restrictions to $\cT=\cG^0$ coincide, $[1]$ and $[[1]]$ can be 
viewed as different extensions of the shift functor of $\cT$ to $\cG$. It is clear that the shift functor of $\cG$ 
can be recovered from $[[1]]$ by a further twist, i.e. we have $[[ [[1]] ]]=[1]$. 

Two shift-equivariant $D$-cyclic structures $\tr$ and $\tr'$ on $\cG$ are called {\em equivalent} if there 
exists an automorphism $f$ of the identity functor $\id_{\cG}$ of $\cG$ such that $f_{a[1]}=f_{a}[1]$ and
$\tr'_a(u)=\tr_a(u f_a)$ for all $a\in \Ob\cG$ and all $u\in \Hom_{\cG}(a,a)$. 
Equivalently, $tr'_a(u)=tr_a(u\circ f_a)$ for all $a$ and all $u\in \Hom_{\cT}(a,a[D])$ (see Appendix \ref{sec:duality}). 
The Yoneda lemma implies that any two nondegenerate $D$-cyclic structures on $\cT$ are equivalent in this sense. 
It follows that a shift-equivariant 2d topological field theory whose boundary sectors and products are specified 
by $\cT$ is determined up to such a transformation.

\subsection{Basic extension operations}

Define a {\em shift-equivariant $D$-cyclic category} to be a triplet $(\cT,[1],tr)$ where $\cT$ is a unital associative 
category, $[1]$ is a shift functor on $\cT$ and $tr$ is a shift-equivariant $D$-cyclic structure on $(\cT,[1])$. 
One has two basic unary operations on such objects, namely additive completion and idempotent completion. Both of 
them preserve the nondegeneracy condition on cyclic structures and thus induce operations on shift-equivariant two-dimensional 
open topological field theories. 

Recall that the additive completion of $\cT$ is the smallest additive category $\cT^\add$ 
containing $\cT$ as a full subcategory. Its objects are finite direct sums 
$A=\oplus_{i=1}^{n_A}{a_i}$ of objects $a_i$ of $\cT$, while its morphism
spaces are given by $\Hom_{\cT^{\add}}(A,A')=\oplus_{i,j}\Hom_{\cT}(a_i, a'_j)$, where 
$A'=\oplus_{j=1}^{n_{A'}}{a'_j}\in \Ob\cT^{\add}$. When $\cT$ admits a shift functor $[1]$, then $\cT^\add$ admits 
the shift functor $[1]^\add$ given by $A=\oplus_i{a_i}\rightarrow A[1]^\add:=\oplus_i a_i[1]$ on objects 
and by $u=\oplus_{i,j} u_{ij}\rightarrow u[1]^\add:=\oplus_{i,j}u_{ij}[1]$ on morphisms 
$u\in \Hom_{{\cT}^\add}(A,A')$. $\cT$ embeds in the obvious manner as a full subcategory of $\cT^\add$.
It is clear that $\cT^{\add}$ is Hom finite iff $\cT$ is. When $(\cT,[1],tr)$ is a shift-equivariant $D$-cyclic category, then $(\cT^\add, [1]^\add)$ 
admits a shift-equivariant $D$-cyclic structure $tr^\add$ given by: 
\be
\nn
tr^\add_A(u)=\sum_{i}tr_{a_i}(u_{ii})~~\forall u=\oplus_{i,j} u_{ij}\in \Hom_{\cT^\add}(A,A[D])~~,~~A=\oplus_i a_i~~.
\ee
We say that $(\cT^\add,[1]^\add,tr^\add)$ is the {\em additive completion} of $(\cT,[1],tr)$. It is easy to see 
that the former is nondegenerate iff the later is. 

Recall that the idempotent completion of the category $\cT$ is the category $\cT^\pi$ defined as follows.  
Its objects are pairs $(a,e)$ with $a\in \Ob\cA$ and $e\in \Hom_{\cT}(a,a)$ such that $e^2=e$. Its morphism 
spaces are $\Hom_{\cT^\pi}((a,e),(b,e')):=e'\circ \Hom_{\cT}(a,b)\circ e$. When $\cT$ has a shift functor $[1]$, 
then $\cT^\pi$ has the shift functor $[1]^\pi$ which acts on objects by $(a,e)[1]^\pi:=(a[1],e[1])$ and 
on morphisms $u\in \Hom_{\cT^\pi}((a,e),(b,e'))$ by $u[1]^\pi:=u[1]$. Notice that $\Hom_{\cT}((a,e),(b,e'))$ is a subspace 
of $\Hom_{\cT}(a,b)$. It is clear that $\cT^\pi$ is Hom finite iff $\cT$ is. $\cT$ embeds in the obvious manner 
as a full subcategory of $\cT^\pi$.  When $(\cT,[1],tr)$ is a shift-equivariant 
$D$-cyclic category, then $(\cT^\pi,[1]^\pi)$ 
admits a shift-equivariant $D$-cyclic structure $tr^\pi$ defined by restricting $tr$:
\be
\nn
tr^\pi_{(a,e)}(u):=\tr_a(u)~~\forall u\in \Hom_{\cT^\pi}((a,e),(a,e)[D]))\subset \Hom_{\cT}(a,a[D])~~.
\ee
We say that $(\cT^\pi,[1]^\pi,tr^\pi)$ is the idempotent completion of $(\cT,[1],tr)$. It is easy to check that the 
former is nondegenerate iff the later is. More details about idempotent completion can be found in Appendix \ref{sec:duality}.

\subsection{The triangulated case}
\label{sec:gens}

In this paper, we are interested in the case when the category $\cT$ is triangulated. As argued in 
\cite{CIL2, Diac}, this condition must be imposed if our 2d topological field theory 
is to admit a lift to a `dynamically closed' topological string theory, a.k.a a 2d topological 
conformal field theory. In this case, we let $[1]$ be the shift functor of $\cT$ as a triangulated category. 
The functor $[D]$ becomes exact when endowed with the isomorphism of functors 
$[D]\circ [1]\stackrel{\approx}{\rightarrow} [1]\circ [D]$ which acts trivially on objects but acts on 
morphisms through multiplication by $(-1)^{D}$. 

Assuming $\cT$ to be triangulated, a shift-equivariant and nondegenerate $D$-cyclic structure on $(\cT,[1])$ corresponds 
to a Serre duality structure whose Serre functor equals $[D]$; 
the pairings of the $D$-cyclic structure are the usual Serre pairings of $\cT$.
We say that $\cT$ is a Calabi-Yau category of dimension $D$ (or $D$-Calabi-Yau category) if it admits a non-degenerate 
shift-equivariant $D$ -cyclic structure; in this case, all such $D$-cyclic structures are equivalent.   

The triangulated structure of $\cT$ allows one to introduce various generation properties, which --- when present ---
allow for an explicit characterization of $\cT$.  We recall these below for later use. 

\paragraph{Generators of triangulated categories} 

Let $\cT$ be a triangulated category and ${\cal U}$  a set of
objects of $\cT$. We let $\add({\cal U})$ be the full subcategory of
$\cT$ whose objects are finite direct sums of shifts of objects
lying in ${\cal U}$.  The smallest strictly full triangulated
subcategory\footnote{In particular, this is assumed closed under shifts and
thus contains $a[n]$ for all $a\in {\cal U}$ and $n\in \Z$.} of $\cT$ containing ${\cal U}$
will be denoted $\tria_\cT({\cal U})$.  It consists of successive extensions
of objects of $\add({\cal U})$. Explicitly, the objects of $\tria_\cT({\cal U})$ 
are those objects of $\cT$ which admit a finite 
filtration whose associated graded belongs to $\add({\cal U})$ (the graded is defined 
by taking triangles on each morphism of the filtration, and is unique up to 
non-canonical isomorphism).  When $\tria_\cT({\cal U})=\cT$, we say
that ${\cal U}$ {\em triangle generates} $\cT$.

The smallest {\em thick}\footnote{i.e. closed under taking direct
summands (epaisse).}  and strictly full triangulated category of
$\cT$ containing ${\cal U}$ will be denoted $\ktria_\cT({\cal U})$; it consists
of direct summands of objects of $\tria_\cT({\cal U})$.  When $\cT$
is idempotent complete, we have a natural isomorphism $\ktria_{\cal
T}({\cal U})\approx \tria_\cT({\cal U})^\pi$.  When $\ktria_\cT({\cal U})={\cal
T}$, we say that ${\cal U}$ is a {\em Karoubian generating set} for ${\cal
T}$.

If $\cT$ has arbitrary coproducts, we let 
${\overline \add}({\cal U})$ be the full subcategory of $\cT$ 
whose objects are arbitrary direct sums of shifts of objects lying in ${\cal U}$.
In this case, we define $\Tria_\cT({\cal U})$ to be
the smallest strictly full triangulated subcategory of $\cT$
containing ${\cal U}$ and closed under arbitrary coproducts. It
consists of successive extensions of objects of 
${\overline \add}({\cal U})$, i.e. of those 
objects of $\cT$ admitting a finite filtration whose graded belongs to 
${\overline \add}({\cal U})$.  We say that ${\cal U}$ {\em
compactly generates} $\cT$ if $\Tria_\cT({\cal U})=\cT$ and
moreover each object $a$ of ${\cal U}$ is compact (a.k.a small) in ${\cal
T}$, i.e. $\Hom_\cT(a,\cdot)$ commutes with all coproducts on
$\cT$.  When $\cT$ is clear from the context, we write
$\Tria({\cal U})$ instead of $\Tria_\cT({\cal U})$ etc. When ${\cal U}$ consists of a
single object $a$, we write $\Tria(a)$ instead of $\Tria(\{a\})$
etc.

\paragraph{Relation with cyclic structures}

Let $(\cT,[1],tr)$ be a shift-equivariant $D$-cyclic triangulated
category, and ${\cal U}\subset \Ob\cT$ a non-void set of objects as
above.  Setting $\Z{\cal U}:=\{a[n]|a\in {\cal U}~,~n\in \Z\}$, we let
$\cA$ be the full subcategory of $\cT$ on the set of objects $\Z{\cal
U}$. When endowed with the induced shift functor and traces, this
becomes a $D$-cyclic category $(\cA,[1]^{\cA},\tr^{\cA})$. Let us
assume that ${\cal U}$ triangle generates $\cT$ or that ${\cal
T}$ is idempotent complete and ${\cal U}$ Karoubi generates ${\cal
T}$.  Then one has the following non-degeneracy criterion, which
shows that nondegeneracy of $\tr$ is equivalent with nondegeneracy of
$\tr^{\cA}$.

\paragraph{Proposition(Appendix \ref{sec:duality})}{ Assume that ${\cal U}$ triangle generates $\cT$ or that $\cT$ 
is idempotent complete and ${\cal U}$ Karoubi generates $\cT$.
Then a shift-equivariant $D$-cyclic structure $tr$ on $(\cT,[1])$
is non-degenerate iff the bilinear forms $(~, ~)_{ab}:\Hom_{\cal
T}(a,b)\times \Hom_\cT(b,a[D])\rightarrow \C$ are non-degenerate
for all $a,b\in \Z{\cal U}$.}

\section{Background on $A_\infty$ categories} 

This section recalls the basics of \ainf categories, fixing notations
and sign conventions.  The reader can consult \cite{Keller_intro,
Keller_dGcat, Keller_functor} for reviews and \cite{Hasegawa,
Fukaya_mirror, Seidel, KS} for in-depth discussion. We use `forward
suspended compositions' in order to simplify sign factors.  The sign
conventions in the definition of shift functors and of the shift
completion are somewhat non-standard, being motivated by the
application to enhanced triangulated categories.  These are chosen
consistently with those of Appendix \ref{sec:duality}, which the
reader might wish to consult while reading this section.

\subsection{Basics} 

Recall that a small \ainf category $\cA$ is specified by a set of
objects $\Ob \cA$ and by graded vector spaces $\Hom_\cA(a,b)$ for any
$a,b\in \Ob \cA$, together with linear maps $\mu_{a_n\ldots
a_0}:\Hom_\cA(a_{n-1}, a_n)\otimes \ldots \otimes \Hom_\cA (a_1,a_2)
\otimes \Hom_\cA (a_0, a_1) \rightarrow \Hom_\cA (a_0,a_n)$ of degree
$2-n$ subject to \ainf constraints (see eqs. (\ref{c_ainf})
below). Denoting the degree of homogeneous elements $x\in
\Hom_{\cA}(a,b)$ by $|x|$, the homogeneity constraints on the \ainf
products take the form:
\be
\nn
|\mu_{a_n \ldots a_0}(x_n\otimes \ldots \otimes x_1)|=|x_1|+\ldots +|x_n|+2-n~~.
\ee
In particular, $\mu_{ba}:\Hom_{\cA}(a,b)\rightarrow \Hom_{\cA}(a,b)$
have degree one, while $\mu_{cba}:\Hom_{\cA}(b,c)\otimes
\Hom_{\cA}(a,b)\rightarrow\Hom_{\cA}(a,c)$ have degree zero.

For any objects $a,b$ of $\cA$, let $s_{ab}:\Hom_{\cal
A}(a,b)\rightarrow \Hom_\cA(a,b)[1]$ be the suspension operator of the
graded vector space $\Hom_\cA(a,b)$. Denoting the degree of elements
$x\in \Hom_{\cA}(a,b)[1]$ by ${\tilde x}=|x|-1$, we have $s(x)=x$ and
$s$ is a map of degree $-1$. Notice that $\Hom_{\cA}(a,b)[1]$ is the
same vector space as $\Hom_{\cA}(a,b)$, except that we use the `tilde
grading' instead of the grading given by $|~|$.  To simplify notation,
we will often use $s_{ab}$ instead of $s_{ab}[n]$ to denote the
induced map $s_{ab}[n]:\Hom_{\cA}(a,b)[n]\rightarrow
\Hom_{\cA}(a,b)[n+1]$.  Accordingly, we write 
$s^n_{ab}:\Hom_{\cA}(a,b)\rightarrow \Hom_{\cA}(a,b)[n]$ for the
iteration $s_{ab}[n-1]\circ \ldots s_{ab}[1]\circ s_{ab}$ with $n$ a
positive integer and set $s_{ab}^0:=\id_{\Hom_\cA(a,b)}$ and
$s_{ab}^n:= s_{ab}^{-1}[n+1]\circ \ldots s_{ab}^{-1}[-1]\circ
s_{ab}^{-1}$ for $n$ a negative integer.

The \ainf constraints can be written in a few equivalent forms. In
order to obtain the maximum simplification of sign factors, it is
convenient to work not with the traditional compositions $\mu$, but
rather with equivalent maps defined as follows. First, introduce
`forward compositions' $m_{a_0\ldots a_n}:\Hom_\cA(a_0, a_1)\otimes
\Hom_\cA (a_1,a_2)\ldots \otimes \Hom_\cA (a_{n-1},a_n) \rightarrow
\Hom_\cA (a_0,a_n)$ via
\footnote{The relation between $\mu$ and $m$ is similar to but not quite
the same as passing to the opposite \ainf category, since we do not
reverse the sense of arrows in $\cA$. } :
\be
\nn
m_{a_0\ldots a_n}(x_1\otimes \ldots \otimes x_n):=(-1)^{\sum_{1\leq 1<j\leq n}{|x_i||x_j|}}
\mu_{a_n\ldots a_0}(x_n\otimes \ldots \otimes x_1)~~.
\ee
Next, introduce `suspended forward compositions' $r_{a_0\ldots
a_n}:=s_{a_0a_n}\circ m_{a_0\ldots a_n}\circ (s_{a_0a_1}^{-1}\otimes
\ldots \otimes s_{a_{n-1}a_n}^{-1}) :\Hom_\cA(a_0, a_1)[1]\otimes
\Hom_\cA (a_1, a_2)[1] \otimes\ldots \otimes \Hom_\cA (a_{n-1},
a_n)[1] \rightarrow \Hom_\cA (a_0,a_n)[1]$, which have degree $+1$. Of
course, we can also view these as maps $r_n:\Hom_\cA(a_0, a_1)\otimes
\Hom_\cA (a_1, a_2) \otimes\ldots \otimes \Hom_\cA (a_{n-1}, a_n)
\rightarrow \Hom_\cA (a_0,a_n)$, of degree $+1$ with respect to the
`tilde grading'. To keep notation manageable, we will often tacitly
change between these points of view.  For the first few compositions,
we find:
\bea
&& m_{ab}(x)=r_{ab}(x)~~,~~m_{abc}(x_1\otimes x_2)=(-1)^{{\tilde x}_1}r_{abc}(x_1\otimes x_2)~~\nn\\
&& m_{abcd}(x_1 \otimes x_2 \otimes x_3)=(-1)^{{\tilde x}_2}r_{abcd}(x_1 \otimes x_2 \otimes x_3)~~~~~\nn
\eea
and
\bea
&& \mu_{ab}(x)=r_{ba}(x)~~,~~\mu_{abc}(x_1\otimes x_2)=(-1)^{|x_1||x_2|}m_{cba}(x_2\otimes x_1)=
(-1)^{{\tilde x}_1{\tilde x}_2+{\tilde x}_1+1}r_{cba}(x_2\otimes x_1)~~\nn\\
&& \mu_{abcd}(x_1 \otimes x_2 \otimes x_3)=(-1)^{{\tilde x}_2+\sum_{i<j}{({\tilde x}_i+1)({\tilde x}_j+1)}}
r_{dcba}(x_3 \otimes x_2 \otimes x_1)~~~~~.\nn
\eea

In terms of the suspended forward compositions, the \ainf constraints take the relatively simple form:
\bea
\label{c_ainf}
\!\!\!\!\!\!\!\!\!\!\!\!\!\!\sum_{\ainfrange} (-1)^{{\tilde x}_1+\ldots +{\tilde x}_i} 
r_{a_0\ldots a_i,a_{i+j}\ldots a_n} 
(x_1\otimes \ldots \otimes x_i \otimes r_{a_i\ldots a_{i+j}}(x_{i+1}\otimes \ldots &&\otimes x_{i+j})\otimes x_{i+j+1}\otimes 
\ldots \otimes x_n)\nn\\
&& =0~~\forall n\geq 1~~~~~~~~~
\eea
where $x_j\in \Hom_{\cA}(a_{j-1},a_j)[1]$ is any sequence of
`forward-composable' morphisms.  Using the Koszul rule, these can also
be written as\footnote{Such a simple formula does not seem to exist
for the traditional `backward' compositions.}:
\be
\label{ainf_nice}
\!\!\!\!\!\!\!\!\!\!\sum_{\ainfrange}r_{a_0\ldots a_i, a_{i+j}\ldots a_n}\circ
(\id_{a_0a_1}\otimes \ldots \otimes \id_{a_{i-1}a_i}\otimes
r_{a_i\ldots a_{i+j}}\otimes \id_{a_{i+j}a_{i+j+1}}\otimes \ldots \otimes 
\id_{a_{n-1}a_n})=0~~\forall n\geq 1~~,
\ee
where $\id_{ab}:\Hom_\cA(a,b)[1]\rightarrow \Hom_\cA(a,b)[1]$ is the
identity endomorphism of the vector space $\Hom_\cA(a,b)[1]$ (which of
course can be identified with the identity endomorphism of
$\Hom_{\cA}(a,b)$ by applying the shift functor $[1]$ of the category
of graded vector spaces to the latter).

The first three constraints imply that $m_{ab}$ square to zero and act
as derivations of $m_{abc}$, which in turn are associative up to
homotopy. This also amounts to the conditions that $\mu_{ab}$ square
to zero and act as derivations of $\mu_{abc}$, which in turn are
associative up to homotopy. In this paper, we will make systematic use
of the compositions $r_{a_0\ldots a_n}$. However, we stress that the
\ainf structure is defined by the backward compositions
$\mu_{a_n\ldots a_0}$, in particular an \ainf module over $\cA$ is
understood with respect to the structure given by $\mu$; this is
important when distinguishing between left and right \ainf modules ---
a right \ainf module in this paper is the same as a right \ainf module
in the sense of \cite{Hasegawa} (even though it looks like a `left'
module when written with respect to the forward compositions $r$).

\paragraph{Observation} The suspended forward compositions considered in this paper are 
related to the suspended backward compositions 
$r^S_{a_n\ldots a_0}$ of \cite{Seidel} (denoted by $\mu_{\cA}$ in loc. cit.) via: 
\be
\nn
r_{a_0\ldots a_n}(x_1\otimes \ldots \otimes x_n)=r^S_{a_n\ldots a_0}(x_n\otimes \ldots \otimes x_0)~~,
\ee
without any sign prefactors. However the \ainf constraints for $r^S$ are not as nice as (\ref{ainf_nice}).

\paragraph{Observation} It might seem more natural to define the suspended \ainf products by using the 
signed suspensions $\sigma_{ab}$ of the spaces $\Hom_{\cA}(a,b)$ (see the introduction). However, this 
introduces unwanted sign factors in other formulas. This is why we define $r$ as above.

\paragraph{The cohomology category.}

The {\em cohomology category} $H(\cA)$ is the (possibly non-unital)
graded associative category having the same objects as $\cA$, morphism
spaces given by $\Hom_{H(\cA)}(a,b):=
H_{\mu_{ab}}(\Hom_{\cA}(a,b)):=\ker(\mu_{ab})/\im(\mu_{ab})$ and morphism compositions $\Hom_{H(\cA)}(b,c)\otimes
\Hom_{H(\cA)}(a,b)\rightarrow \Hom_{H(\cA)}(a,c)$ induced by $\mu_{cba}$:
\be
\nn
[x]*[y]:=[\mu_{cba}(x\otimes y)]=(-1)^{|x||y|}[m_{abc}(y\otimes x)]~~\forall x\in Z_{\mu_{bc}}(\Hom_{\cA}(b,c))~,
~\forall y\in  Z_{\mu_{ab}}(\Hom_{\cA}(a,b))~~.
\ee 
We let $H^0(\cA)$ be the associative subcategory obtained from $H(\cA)$ by 
considering only morphisms of degree zero. 

\paragraph{Unitality and finiteness conditions.}

The \ainf category $\cA$ is called {\em strictly unital} if every
object $a$ admits a degree zero endomorphism $u_a\in \Hom^0_{\cA}(a,a)$ such that the
following relations are satisfied:
\begin{eqnarray}
\label{cat_unitality}
r_{a_0\ldots a_{j-2},a_j, a_j, a_{j+1}\ldots a_n }(x_1\otimes \ldots
\otimes x_{j-1}\otimes u_{a_j}\otimes x_{j+1}\otimes \ldots \otimes x_n)&=&0~~ {\rm for~all}~~n\neq 2
~{\rm~and~all}~ j~~\nn\\
r_{a,a,b}(u_a\otimes x)=-x~~,~~r_{a,b,b}(x \otimes u_b)&=&(-1)^{{\tilde x}} x~~,
\end{eqnarray}
where $x_j\in \Hom_{\cA}(a_{j-1},a_j)$ etc. It is called {\em
homologically unital} if every object $a$ admits a degree zero
$r_{aa}$-closed endomorphism $u_a$ which induces an identity morphism
in the graded associative category $H(\cA)$. It is easy to check that
the units $u_a$ of a strictly unital \ainf category are uniquely
determined, as are the cohomology classes $[u_a]$ in the homologically
unital case.

An \ainf category $\cA$ is called {\em degreewise Hom-finite} if
$\dim_\C\Hom^n_{\cA}(a,b)<\infty$ for all $a,b\in \Ob\cA$ and all
$n\in \Z$.  It is {\em compact} if $H(\cA)$ is degreewise Hom-finite,
i.e. $\dim_\C H^n(\Hom_{\cA}(a,b))<\infty$ for all $a,b\in \Ob\cA$ and
all $n\in \Z$.

\paragraph{\ainf functors.}

Given two \ainf categories $\cA$, $\cB$, an \ainf functor
$F:\cA\rightarrow \cB$ is given by a map $F:\Ob \cA\rightarrow \Ob
\cB$ together with linear maps $F_{a_0\ldots
a_n}:\Hom_{\cA}(a_0,a_1)\otimes \ldots \otimes \Hom_{\cA}(a_{n-1},a_n)
\rightarrow \Hom_\cB(F(a_0),F(a_n))$ homogeneous of degree $1-n$ (here $n\geq 1$)
such that the suspended maps $F_{a_0\ldots a_n}^s:= s^{\cal
B}_{F(a_0)F(a_n)}\circ F_{a_0\ldots a_n}\circ ((s_{a_0a_1}^{\cal
A})^{-1}\otimes \ldots \otimes (s_{a_{n-1}a_n}^{\cal A})^{-1})
:\Hom_{\cA}(a_0,a_1)[1]\otimes \ldots \otimes
\Hom_{\cA}(a_{n-1},a_n)[1] \rightarrow \Hom_\cB(F(a_0),F(a_n))[1]$ ---
which are homogeneous of degree $0$ --- satisfy the conditions:
\bea
\label{func}
\!\!\!\!\!\!\!\!\!\! & & \sum_{p=1}^n
\sum_{0< i_1<i_2<\ldots <i_{p-1}< n} r_{F(a_0)F(a_{i_1}) 
F(a_{i_2})\ldots F(a_{i_{p-1}}) F(a_n)}
^\cB\circ (F^s_{a_0\ldots a_{i_1}}\otimes
F^s_{a_{i_1}\ldots a_{i_2}}\otimes \ldots \otimes F^s_{a_{i_{p-1}}\ldots a_n})\nn\\
\!\!\!\!\!\!\!\!\!\! &=&\sum_{0\leq i<j\leq n} F^s_{a_0\ldots a_i, a_j\ldots a_n}
\circ (\id^{\cA}_{a_0a_1}\otimes \ldots \otimes \id^{\cA}_{a_{i-1}a_i}\otimes 
r^{\cA}_{a_i\ldots a_j}\otimes \id^{\cA}_{a_{j}a_{j+1}}\otimes \ldots \otimes 
\id^{\cA}_{a_{n-1}a_n})~,~\forall n\geq 1~~ ~~~~~
\eea
Together with $F_{ab}$, the map on objects induces a (possibly
non-unital) functor $H(F):H(\cA)\rightarrow H(\cB)$ of graded
associative categories. $F$ is called a {\em quasi-isomorphism} if
$H(F)$ is an isomorphism.  It is called {\em strict} if $F_{a_0\ldots
a_n}=0$ unless $n=1$.  In this case, equations (\ref{func}) reduce to:
\be
\nn
r_{F(a_0)F(a_1) \ldots F(a_n)}^\cB\circ (F^s_{a_0a_1}\otimes
F^s_{a_1a_2}\otimes \ldots \otimes F^s_{a_{n-1}a_n}) =F^s_{a_0 a_n}\circ  
r^{\cA}_{a_0\ldots a_n}~,~\forall n\geq 1~~ ~~~~~.\nn
\ee
We will often not indicate the object subscripts on the maps $F_{ab}$.

An {\em \ainf
endomorphism} of $\cA$ is an \ainf functor $F:\cA\rightarrow \cA$. An
\ainf endomorphism is an {\em automorphism} if the map on objects is
bijective and $F_{ab}$ are bijective for all $a,b$.  
As in the case
of \ainf algebras, one has a notion of strictly unital \ainf functor
between strictly unital \ainf categories, as well as a notion of \ainf
equivalence of such categories, which amounts to an \ainf functor for
which $H(F)$ is an equivalence between the graded cohomology
categories. Finally, one has a notion of \ainf natural transformations etc. 
Instead of reviewing these here, we refer the reader to
\cite{Hasegawa, Fukaya_mirror, Seidel} for details.

\paragraph{Twisted shift functors.}

A {\em twisted shift functor} on $\cA$ is a strict automorphism $[[1]]$ of
$\cA$ together with isomorphisms of complexes 
$\Hom_\cA(a,b[[1]])\stackrel{\rho_{ab}}{\rightarrow} \Hom_{\cal
A}(a,b)[1]$ for all $a,b\in \Ob \cA$ which are natural up to signs in $a$ and $b$.
The last condition means the following. Endowing $\Hom_{\cA}(a,b)$ with the 
differential $\mu_{ba}$, we can view $\Hom_{\cA}(\cdot, \cdot)$ as 
an \ainf bifunctor $\Hom_{\cA}:\cA^{\rm opp}\times \cA\rightarrow \Dif$.
Then we require that the maps $\gamma_{ab}:=s_{ab}^{-1}\circ \rho_{ab}:\Hom_\cA(a,b[[1]])\rightarrow \Hom_{\cA}(a,b)$ 
give a morphism $\gamma:\Hom_{\cA}\circ(\id_{\cA}\times [[1]])\rightarrow \Hom_\cA$ of degree $+1$
in the associative category of \ainf bifunctors $\cA^{\rm op} \times \cA\rightarrow \Dif$ (whose 
morphisms are the strict natural transformations of degree zero, see \cite[paragraph (1d)]{Seidel}). 
We have $\Hom_{\cal A}(a[[m]],b[[n]])\approx \Hom_\cA(a,b)[n-m]$ for all $a,b\in \Ob
\cA$ and $m,n\in \Z$. When $\cA$ is strictly unital, the twisted shift
functor automatically preserves all units, i.e. $u_a[[1]]=u_{a[[1]]}$ for
all $a\in \Ob\cA$; in the homologically unital case, only the
cohomology classes must agree, i.e.  $[u_a[[1]]]=[u_{a[[1]]}]$.  

The strict automorphism conditions take the form: 
\be
\nn
r_{a_0[[1]] a_1[[1]] \ldots a_n[[1]]}\circ ([[1]]^s\otimes \ldots \otimes [[1]]^s) =
[[1]]^s\circ  r_{a_0\ldots a_n}~,~\forall n\geq 1~~ ~~~~~,\nn
\ee
where $[[1]]^s=s_{a[[1]]b[[1]]} \circ [[1]]\circ s_{ab}^{-1}:
\Hom_{\cA}(a,b)[1]\rightarrow \Hom_{\cA}(a[[1]],b[[1]])[1]$ are the suspended maps on morphisms as in the 
previous paragraph.  Notice that these conditions are equivalent with: 
\be
\nn
m_{a_0[[1]] a_1[[1]] \ldots a_n[[1]]}\circ ([[1]]\otimes \ldots \otimes [[1]]) =
[[1]]\circ  m_{a_0\ldots a_n}~,~\forall n\geq 1~~ ~~~~~.\nn
\ee
Passing to cohomology, we find that $[[1]]$ induces a twisted shift functor (see Appendix \ref{sec:duality})
$[[1]]^H$ of the graded associative category $H(\cA)$; this acts on objects in the same way as $[[1]]$.  $[[1]]^H$ is an
automorphism of $H(\cA)$ endowed with isomorphisms of graded vector spaces
$\Hom_{H({\cal A})}(a,b[[1]]^H)\approx \Hom_{H(\cA)}(a,b)[1]$ which are natural up to missing Koszul signs. 
These isomorphisms of graded vector spaces are induced by the isomorphisms of complexes $\Hom_\cA(a,b[[1]])\approx \Hom_{\cal A}(a,b)[1]$.

The functor $[[1]]^H$ restricts to a shift functor for the
ungraded subcategory $H^0(\cA)\subset H(\cA)$ .  Following the
notations of Appendix \ref{sec:duality}, we denote this restriction by
$[1]$ (of course, this again acts on objects in the same way as
$[[1]]$).  The relation $\Hom_{H^0(\cA)}(a,b[[n]])\approx
H^{n}(\Hom_{\cal A}(a,b)) $ implies that the graded completion
$H^0(\cA)^\bullet $ of the associative category with shifts
$(H^0(\cA),[1])$ is isomorphic with $H(\cA)$, and that $[[1]]^H$ is the twisted 
shift functor of $H^0(\cA)^\bullet $ in the sense of Appendix \ref{sec:duality}. 
To simplify notation, we will write $[[1]]$ instead of
$[[1]]^H$ --- which of them is meant should be clear from the context.

\paragraph{Minimal \ainf categories and minimal models.}

An \ainf category is called {\em minimal} if all unary compositions $r_{ab}$ vanish. Given an \ainf category 
$\cA$, a {\em minimal model} of  $\cA$ is a minimal \ainf category $\cB$ which is quasi-isomorphic with $\cA$. 
Any \ainf category admits a minimal model \cite{Hasegawa, Fukaya_mirror}.

\subsection{Sector decomposition.}
\label{sec:sectors}

\paragraph{The total Hom space.}

When working with \ainf categories, many formulas can be simplified by using the following trick \cite{nc}. 
Consider the commutative associative algebra $R:=R_{\cA}$ on generators $(\epsilon_a)_{a\in \Ob\cA}$ 
and with relations $\epsilon_a\epsilon_b=\delta_{ab}\epsilon_a$ where $\delta_{ab}$ is the Kronecker symbol. 
Notice that $R$ is unital iff $\cA$ has a finite number of objects (in which case $\sum_{a\in \Ob\cA}{\epsilon_a}$ is the unit). 
Since $\epsilon_a$ are commuting idempotents, we have a decomposition $R\approx \oplus_{a\in \Ob\cA}{\C}$ as an associative algebra. 
Consider the vector space: 
\be
\label{calH} 
{\cal H}:={\cal H}_{\cA}:=\oplus_{a,b\in \Ob\cA}\Hom_{\cA}(a,b)~~, 
\ee
with the grading:
\be
\nn
{\cal H}^n:=\oplus_{a,b\in \Ob\cA}\Hom^n_{\cA}(a,b)~~.
\ee
We let $P_{ab}:{\cal H}\rightarrow \Hom_{\cA}(a,b)$ be the projectors
of ${\cal H}$ on the subspaces $\Hom_{\cA}(a,b)$ defined by this
decomposition.  As in \cite{nc}, the `binary decomposition'
(\ref{calH}) gives a graded $R$-bimodule structure on ${\cal H}$,
obtained by defining the outer left and right multiplications with
$\epsilon_a$ respectively $\epsilon_b$ to be given by the projectors
$_aP, P_b$ of ${\cal H}$ on the subspaces $_a{\cal H}:=\oplus_{b\in
\Ob\cA}\Hom_{\cA}(a,b)$ respectively ${\cal H}_b:=\oplus_{a\in
\Ob\cA}\Hom_{\cA}(a,b)$:
\be
\nn
\epsilon_ax:=_a\!\!P(x)~~,~~x\epsilon_b:=P_b(x)~~\forall x\in {\cal H}~~.
\ee

\paragraph{Total \ainf products.}

Let us define {\em total products} $r_n:{\cal H}[1]^{\otimes n}\rightarrow {\cal H}[1]$ via: 
\be
\label{rdecomp}
r_n(x^{(1)}\otimes \ldots \otimes x^{(n)}):=
\oplus_{a_0,a_n}\sum_{a_1\ldots a_{n-1}}r_{a_0\ldots a_n}(x^{(1)}_{a_0a_1}\otimes \ldots \otimes 
x^{(n)}_{a_{n-1}a_n})
\ee
where $x^{(j)}=\oplus_{a, b\in \Ob\cA}x^{(j)}_{ab}\in {\cal H}[1]$ with
$x^{(j)}_{ab}\in \Hom_{\cA}(a,b)[1]$. The sum in the right hand side
has a finite number of nonzero terms since $u^{(j)}$ have finite
support in $\Ob\cA\times \Ob\cA$. It is clear from (\ref{rdecomp})
that we can view $r_n$ as elements of $\Hom^1_{_R\Mod_R}({\cal
H}[1]^{\otimes_R n}, {\cal H}[1])$. Moreover, they obey the \ainf
relations\footnote{Thus $({\cal H}, (r_n)_{n\geq 1})$ is an \ainf
algebra over $\C$. However it is not quite an \ainf algebra over the
commutative ring $R$ since the left and right module structures on
${\cal H}$ need not agree.}:
\be
\label{ainf_alg}
\sum_{\ainfrange} (-1)^{{\tilde x}_1+\ldots +{\tilde x}_i}
r_{n-j+1} (x_1\otimes \ldots \otimes x_i\otimes r_j(x_{i+1}\otimes \ldots \otimes x_{i+j})\otimes x_{i+j+1}\otimes \ldots \otimes x_n)=0~~.
\ee
The categorical \ainf compositions $r_{a_0\ldots a_n}$ can be recovered from $r_n$, since $R$-multilinearity 
implies the decomposition (\ref{rdecomp}), while (\ref{ainf_alg}) imply (\ref{c_ainf}).

\paragraph{Description of \ainf functors.}

Similarly, an \ainf functor $F:\cA\rightarrow \cB$ induces maps 
$F_n^s\in\Hom_{_R\GrMod_R}({\cal H}_{\cA}[1]^{\otimes_R n}, {\cal H}_\cB[1])$ defined through: 
\be
\nn
F_n^s(x^{(1)}\otimes \ldots \otimes x^{(n)}):=
\oplus_{a_0,a_n}\sum_{a_1\ldots a_{n-1}}F^s_{a_0\ldots a_n}(x^{(1)}_{a_0a_1}\otimes \ldots \otimes x^{(n)}_{a_{n-1}a_n})~~,
\ee
such that conditions (\ref{func}) amount to the constraints: 
\bea
\label{morphism}
& &\sum_{p=1}^n \sum_{0< i_1<i_2<\ldots <i_{p-1}< n} r_p^\cB\circ (F^s_{i_1}\otimes F^s_{i_2-i_1}\otimes \ldots \otimes F^s_{n-i_{p-1}})\nn\\
&=&\sum_{0\leq i<j\leq n} F^s_{n+i-j+1}
\circ (\id_{{\cal H}_{\cA}}^{\otimes i}\otimes r^{\cA}_j\otimes \id_{{\cal H}_{\cA}}^{\otimes n-j})~,~\forall n\geq 1~~.~~~~~
\eea
Once again, the \ainf functor $F$ can be recovered form the maps $F^s_n$. 

\paragraph{Description of twisted shift functors} 

Let us assume that $\cA$ has twisted shifts. Then the bijection on objects $[[1]]:\Ob \cA \rightarrow \Ob \cA$ induces an 
algebra automorphism $[[1]]^R:R\rightarrow R$ given by: 
\be
\nn
\lambda=\sum_{a\in \Ob\cA}{\lambda_a \epsilon_a}\rightarrow \lambda[[1]]^R:=\sum_{a\in \Ob\cA}{\lambda_a \epsilon_{a[[1]]}}~~,
\ee
i.e. $(\lambda[[1]]^R)_a=\lambda_{a[[-1]]}$. Here the coefficients $\lambda_a\in \C$ vanish except for a finite number of objects of $\cA$.
The automorphism $[[1]]^R$ is unital when $\cA$ has a finite number of objects. 

The maps on morphisms $[[1]]:\Hom_{\cA}(a,b)\rightarrow \Hom_{\cA}(a[[1]],b[[1]])$ induce a graded vector space 
automorphism $[[1]]^{\cal H}$ of ${\cal H}$ defined through:
\be
\nn
x=\oplus_{a,b}{x_{ab}}\rightarrow x[[1]]^{\cal H}:=\oplus_{a,b}{x_{ab}[[1]]}~~,
\ee
i.e. $(x[[1]]^{\cal H})_{ab}=x_{a[[-1]]b[[-1]]}[[1]]$. This is compatible with $[[1]]^R$ in the following sense:
\be
\nn
(\lambda x \mu)[[1]]^{\cal H}=\lambda[[1]]^R x[[1]]^{\cal H} \mu[[1]]^R~~\forall \lambda, \mu \in R~~\forall x\in {\cal H}~~.
\ee
The suspended twisted shift functor $[[1]]^s$ corresponds to the suspended graded vector space automorphism:
\be
\nn
([[1]]^{\cal H})^s:=s\circ [[1]]\circ s^{-1}~~,
\ee
where $s:{\cal H}\rightarrow {\cal H}[1]$ is the suspension map of ${\cal H}$. 
For simplicity, we will denote $[[1]]^R$ and $[[1]]^{\cal H}$ simply by $[[1]]$, and the suspended map $([[1]]^{\cal H})^s$ simply by 
$[[1]]^s$. 

It is not hard to check that the strict \ainf automorphism conditions for the twisted shift functor are equivalent with the relations:
\be
\label{shift_total}
r_n\circ ([[1]]^s \otimes \ldots \otimes [[1]]^s)=[[1]]^s\circ r_n~~\forall n\geq 1~~.
\ee

\paragraph{Observation}{When $\cA$ has a finite number of objects, 
strict unitality amounts to the existence of a {\em central} degree zero element $u$ of the $R$-bimodule ${\cal H}$ 
such that the following relations are satisfied:
\begin{eqnarray}
\label{alg_unitality}
r_n(x_1\otimes \ldots \otimes x_{j-1} \otimes u \otimes x_{j+1}\otimes \ldots \otimes x_n)&=&0~~
{\rm for~all}~~~~ n\neq 2 ~{\rm~and~all}~ j~~\nn\\
r_2(u \otimes x)=-x~~,~~r_2(x \otimes u)&=&(-1)^{{\tilde x}} x~~,
\end{eqnarray}
In this case, the categorical units are recovered as $u_a:=P_{aa}(u)$, thus $u=\oplus_{a\in \Ob\cA}{u_{a}}$.}

\subsection{The \ainf categories $\ZA$, $\SigmaA$ and $\tw(\cA)$}

\paragraph{Twisted shift completion.}

The {\em twisted shift completion} $\Z \cA$ of $\cA$ is an $A_\infty$ category
whose objects are pairs $(a,n)$ where $a\in \Ob \cA$ and $n\in \Z$,
with $\Hom_{\Z \cA}((a,m),(b,n)):=\Hom_\cA(a,b)[n-m]$.  Define a
bijective map $[[1]]$ on $\Ob \cA$ via $(a,n)[[1]]:=(a,n+1)$ and
denote its $n$-th iteration by $[[n]]$, where $[[0]]=\id_{\Ob\ZA}$
(for $n<0$, set $[[n]]:=[[-n]]^{-1}$).  Identifying $a$ with $(a,0)$
allows us to write $(a,n) \equiv a[[n]]$. For $A,B\in \Ob\ZA$, define
a bijection $[[1]]_{AB}:\Hom_{\ZA}(A,B)\rightarrow
\Hom_{\ZA}(A[[1]],B[[1]])=\Hom_{\ZA}(A,B)$ via
$f[[1]]_{AB}=(-1)^{|f|}f$ for homogeneous $f$.  We let $[[n]]_{AB}$ be
the $n$-th iteration for all $n\in \Z$ (with the obvious notational
conventions).  Thus $[[n]]_{AB}:\Hom_{\ZA}(A,B)\rightarrow
\Hom_{\ZA}(A[[n]], B[[n]])=\Hom_{\ZA}(A,B)$ acts as
$f[[n]]_{AB}=(-1)^{n|f|}f$ for homogeneous $f$. We will usually not
write the subscripts $A$ and $B$. Similarly, we define
$[[n]]^s_{AB}:=s_{A[[n]]B[[n]]}\circ [[n]]_{AB}\circ s_{AB}^{-1}:
\Hom_{\ZA}(A,B)[1]\rightarrow \Hom_{\ZA}(A[[n]],
B[[n]])[1]=\Hom_{\ZA}(A,B)[1]$, which acts as
$f[[n]]_{AB}^s=(-1)^{n(1+{\tilde f})}f$ for homogeneous $f$.
We will often not write the object subscripts of $[[n]]_{AB}$ and $[[n]]_{AB}^s$. 
With this convention, we have $[[m]]\circ [[n]]=[[m+n]]$ and  $[[m]]^s\circ [[n]]^s=[[m+n]]^s$
on both objects and morphisms. 

For all $a,b\in \Ob\cA$, we have
the bijection $[[-m]]^s:\Hom_{\ZA}(a[[m]],b[[n]])[1]\rightarrow
\Hom_{\ZA}(a,b[[n-m]])[1]=\Hom_{\cA}(a,b)[1][n-m]=\Hom_{\cA}(a,b)[n-m+1]$. 
Hence $s_{ab}^{m-n}[1]\circ [[-m]]^s$ (which we write simply as $s_{ab}^{m-n}\circ [[-m]]^s$) gives a bijection from
$\Hom_{\ZA}(a[[m]],b[[n]])[1]$ to $\Hom_{\cA}(a,b)[1]$, whose inverse is
$[[m]]^s\circ s_{ab}^{n-m}[1]$ (written simply as $[[m]]^s\circ s_{ab}^{n-m}$). 
Using these maps, we define suspended forward compositions
\bea
&&r^{\Z A}_{a_0[[k_0]]\ldots a_n[[k_n]]}: \Hom_{\ZA}(a_0[[k_0]],a_1[[k_1]])[1]\otimes \ldots 
\otimes \Hom_{\ZA}(a_{n-1}[[k_{n-1}]],a_n[[k_n]])[1]\rightarrow \nn\\
&&\rightarrow \Hom_{\ZA}(a_0[[k_0]],a_n[[k_n]])[1]~~\nn
\eea
via the expressions: 
\bea
&&r^{\Z A}_{a_0[[k_0]]\ldots a_n[[k_n]]}:=\nn\\
&&= (-1)^{k_n-k_0}[[k_0]]^s\circ s^{k_n-k_0}_{a_0a_n}\circ
r_{a_0\ldots a_n}
\circ (s_{a_0a_1}^{k_0-k_1}\otimes \ldots \otimes 
s_{a_{n-1}a_n}^{k_{n-1}-k_n})\circ ([[-k_0]]^s\otimes \ldots \otimes [[-k_{n-1}]]^s)~~\nn
\eea
These are easily seen to satisfy the \ainf constraints, thus making
$\ZA$ into an \ainf category.  Moreover, $[[1]]$ becomes a twisted shift
functor for $\ZA$, so $(\Z\cA, [[1]])$ is an \ainf category with
(twisted) shifts. If $\cA$ is strongly unital with identity morphisms $u_a$,
then $\Z \cA$ is strongly unital with identity morphisms
$u_{a[[n]]}:=u_a[[n]]$ (this follows from an easy computation).
$\Z\cA$ contains $\cA$ as the full \ainf subcategory on the objects
$a[[0]]$ ($a\in \Ob\cA$). When $\cA$ is degreewise Hom-finite respectively 
compact, then $\ZA$ has the same property.

\paragraph{Additive completion of the shift completion.}

The {\em additive completion} $\Sigma\cA$ of $\ZA$  
is the smallest additive \ainf category containing $\ZA$. Its objects are
finite direct sums of objects of $\ZA$, with morphisms defined accordingly. Thus any object 
$A$ of $\Sigma\cA$ decomposes as $A=\oplus_{i=1}^{n_A}{a_i[[n_i]]}$ for some $a_i\in \Ob\cA$
and some $n_i$ in $\Z$. Given $B=\oplus_{j=1}^{n_B}{b_j[[m_j]]}\in \Ob\Sigma\cA$, 
we have $\Hom_{\Sigma\cA}(A,B)=\oplus_{i,j}\Hom_{\ZA}(a_i[[n_i]],b_j[[m_j]])=
\oplus_{i,j}\Hom_{\cA}(a_i,b_j)[m_j-n_i]$. The compositions  $r^{\ZA}$ extend to \ainf compositions 
on ${\Sigma \cA}$ in the obvious manner, making $\Sigma\cA$ into an \ainf category. Explicitly, we have: 
\be
\nn
r_{A^{(0)}\ldots A^{(n)}}^{\Sigma\cA}(x^{(1)}\otimes \ldots \otimes 
x^{(n)})=\oplus_{i_0,i_n}\sum_{i_1\ldots i_{n-1}}r_{a^{(0)}_{i_0}[[m^{(0)}_{i_0}]]\ldots 
a^{(n)}_{i_n}[[m^{(n)}_{i_n}]]}^{\ZA}(x^{(1)}_{i_0i_1}\otimes 
x^{(2)}_{i_1i_2}\otimes \ldots 
\otimes x^{(n)}_{i_{n-1}i_n})
\ee
where $x^{(k)}=\oplus_{i,j}{x^{(k)}_{i j}}
\in \Hom_{\Sigma\cA}(A^{(k-1)},A^{(k)})$ with $A^{(k)}=\oplus_{i}{a^{(k)}_{i}
[[m^{(k)}_{i}]]}$  and 
$x^{(k)}_{ij}\in \Hom_{\ZA}(a^{(k-1)}_{i}[[m^{(k-1)}_i]], a_{j}^{(k)}[[m^{(k)}_j]])$.
The twisted shift functor of $\ZA$ extends to a strict automorphism 
of $\SigmaA$ given by:
\be
\nn
A=\oplus_{i=1}^n A_i\rightarrow A[[1]]:=\oplus_{i=1}^n{A_i[[1]]}~~
\ee
where $A_i\in \Ob \ZA$; one takes the obvious action on morphisms. 
$[[1]]$ is a twisted shift functor, so $(\SigmaA,[[1]])$ is an \ainf category with (twisted) shifts, which 
is strictly unital when $\cA$ is. The units are given by $u_{\oplus_i a_i[[n_i]]}=\oplus_i{u_{a_i}[[n_i]]}$, 
where $u_a$ are the units of $\cA$. $\SigmaA$ contains $\ZA$ as a full \ainf subcategory in the obvious manner. 
When $\cA$ is degreewise Hom-finite respectively 
compact, then $\SigmaA$ has the same property.

\paragraph{Bounded twisted complexes.}

A (strictly one-sided) {\em bounded twisted complex} $q$ 
over $\cA$ is a finite collection of morphisms $q_{ij}\in \Hom^1_{\ZA}(A_i,A_j)=
\Hom_{\ZA}(A_i,A_j)[1]^0$ 
of the shift-completed category $\ZA$ with $1\leq i, j\leq l_q$ and $q_{ij}=0$ unless $i<j$, 
which are required to obey the {\em generalized Maurer-Cartan equations}:
\be
\label{MCinf}
\sum_{n=1}^{j-i} \sum_{i_1\ldots i_{n-1}}
r^{\ZA}_{A_i A_{i_1}\ldots A_{i_{n-1}}A_j} (q_{ii_1}\otimes q_{i_1  i_2} \otimes \ldots \otimes 
q_{i_{n-2} i_{n-1}} \otimes q_{i_{n-1} j} )=0~~\forall ~1\leq i<j\leq n~~,
\ee
where the term for $n=1$ is defined to be $r^{\ZA}_{A_i A_j}(q_{ij})$. 
Notice that we denote the twisted complex by $q$, with the understanding that the objects $A_1\ldots A_{l_q}$ of $\SigmaA$ 
are implicitly given as the domains/codomains of the morphisms $q_{ij}$. This is done to simplify notation. 
The positive integer $l_q$ is called the {\em length} of $q$. For later convenience, we set
$A_q:=\oplus_{i=1}^{l_q} A_i\in \Ob \Sigma\cA$. 
The morphisms $q_{ij}$ can be combined into 
a single (endo)morphism ${\hat q}:=\oplus_{i,j}{q_{ij}}\in \Hom^1_{\Sigma\cA}(A_q,A_q)$ of $\SigmaA$. 
Recall that $A_i:=a_i[[n_i]]$ for some $a_i\in \Ob\cA$ and some $n_i\in \Z$.

Bounded twisted complexes form an \ainf category $\tw(\cA)$ if one sets 
$\Hom_{\tw(\cA)}(q,q'):=\Hom_{\Sigma\cA}(A_q,A_{q'})=\oplus_{i,j}\Hom_{\ZA}(a_i[[n_i]],a'_j[[n'_j]])$
and defines \ainf products as follows:
\be
\label{rtw}
r^{\tw(\cA)}_{q_0\ldots q_n}(x_1\otimes \ldots \otimes x_n):=
\sum_{t_0\ldots t_n\geq 0}r^{\Sigma\cA}_{(A_0)_{t_0+1}\ldots (A_n)_{t_{n}+1}}
( {\hat q}_0^{\otimes t_0} \otimes x_1 \otimes {\hat q}_1^{\otimes t_1} \otimes x_2\otimes 
\ldots \otimes x_n \otimes {\hat q}_n^{\otimes t_n})~~.
\ee
In the expression above, we take $x_i\in \Hom_{\tw(\cA)}(q_{i-1},q_i)=
\Hom_{\Sigma \cA}(A_{q_{i-1}},A_{q_i})$, with ${\hat q}_i\in \Hom_{\Sigma\cA}(A_{q_i},A_{q_i})$ defined as above.
The notation $(A)_k$ stands for the sequence $A,A,\ldots A$ consisting of $k$ copies of $A$. 
The \ainf relations for (\ref{rtw}) follow from the generalized Maurer-Cartan equations
(\ref{MCinf}).

The twisted shift functor $[[1]]$ of $\SigmaA$ extends to $\tw(A)$ as follows. Given
a twisted complex $q$ with $q_{ij}:A_i\rightarrow A_j$, we let $q[[1]]$
be the twisted complex $q'$ given by $A'_i:=A_i[[1]]$ and
$q'_{ij}=q_{ij}[[1]]$. Shift-invariance of $r^{\ZA}$ implies that
$q[[1]]$ satisfies the generalized Maurer-Cartan equations.  We let
$[[1]]$ act on morphisms in $\tw(A)$ in the same way as in
$\SigmaA$. Using definition (\ref{rtw}) and shift-invariance of $\SigmaA$, 
we find that $(\tw(\cA),[[1]])$ is an \ainf category with (twisted) shifts, which is strictly
unital if $\cA$ is. The units are $u_q=u_{A_q}$ where $u_{A}$ are the units of $\SigmaA$ (this again 
follows by an easy computation). Notice that $\tw(\cA)$ contains $\Sigma\cA$ as the full subcategory
on those twisted complexes $q$ for which all $q_{ij}$ vanish (such a
twisted complex is called {\em degenerate} and identifies with the
object $A_q$ of $\SigmaA$). When $\cA$ is degreewise Hom-finite then 
$\tw(\cA)$ has the same property. 
When $\cA$ is compact, the same is true\footnote{This follows from the spectral sequence of 
\cite[Section 3, paragraph (3l)]{Seidel}, which computes $H(\tw(\cA))$ starting from $H(\SigmaA)$ by using the obvious
finite filtration possessed by each twisted complex.} of $\tw(\cA)$. 

\subsection{The triangulated categories $D(\cA)$, $\tria (\cA)$ and $\per(\cA)$}

\paragraph{The derived category $D(\cA)$.}

Let $\cA$ be a strictly unital \ainf category and $\Mod_\cA$ be the dG
category\footnote{The morphisms in $Z^0(\Mod_\cA)$ are strictly unital \ainf morphisms of
\ainf right modules over $\cA$.} of strictly unital right \ainf
modules over $\cA$ \cite{Hasegawa}, i.e. contravariant unital \ainf functors from $\cA$ to $\Dif$. 
We can define the derived category of $\cA$ via $D(\cA):=H^0(\Mod_\cA)$ (this is one possible description
of $D(\cA)$, see Remark 5.2.0.2 ref. \cite{Hasegawa}, which however uses different notations).  For any object
$a\in \Ob\cA$, let ${\hat a}\in \Mod_\cA$ denote its image through the
(first component of the) Yoneda \ainf functor $y:\cA\rightarrow \Mod_\cA$ constructed in
\cite{Hasegawa, Fukaya_mirror} (${\hat a}$ is the \ainf functor $\Hom_{\cA}(\cdot, a)$). 
We let ${\tilde \cA}$ denote the full dG subcategory of $\Mod_\cA$ determined by the set of objects ${\cal
U}:=\{{\hat a}| a\in \Ob\cA\}$ and ${\hat \cA}=H^0({\tilde \cA})$ 
the full associative subcategory of $D(\cA)$ determined by ${\cal U}$. The Yoneda functor induces
isomorphisms $H(\Hom_{\Mod_\cA}({\hat a}, {\hat b})) \approx
H(\Hom_{\cA}(a,b))$ for all $a,b\in \Ob\cA$, which imply 
${\hat \cA}=H^0({\tilde \cA})\approx H^0(\cA)$. It is shown in \cite{Hasegawa} that $D(\cA)$ is a triangulated
category with infinite coproducts, compactly generated by ${\cal U}$ ---
in particular we have $D(\cA)=\Tria_{D(\cA)}({\cal U})$ in the notation of Appendix \ref{sec:gens}. 

\paragraph{The categories $\tria(\cA)$ and $\per(\cA)$.}

One defines $\tria(\cA)=\tria_{D(\cA)}({\cal U})$ to be the smallest strictly full
triangulated subcategory containing ${\cal U}$, and
$\per(\cA)=\ktria_{D(\cA)}({\cal U})$ to be the smallest strictly full
triangulated and idempotent-complete subcategory containing the same
set of objects (see Appendix \ref{sec:gens} for notation).  
The category $\per(\cA)$ is called the {\em perfect
derived category} of $\cA$.  It follows from \cite{Hasegawa} that
$\per(\cA)$ coincides with the full category of all compact objects of
$D(\cA)$.  We have obvious inclusions $\tria(\cA)\subset
\per(\cA)\subset D(\cA)$.

As explained in \cite{Hasegawa,Fukaya_mirror}, the Yoneda \ainf
functor factorizes as $y=y''\circ y'$, where $y':\cA\rightarrow
\tw(\cA)$ is the obvious embedding and $y'':\tw(\cA)\rightarrow
\Mod_\cA$ induces an equivalence $H^0(\tw(\cA))\approx
\tria(\cA)$. The latter gives an explicit description of
$\tria(\cA)$ through twisted complexes, presenting it as an
\ainf-enhanced triangulated category\footnote{This notion is
essentially equivalent with that considered in \cite{BK}.}. 
Since $\tw(\cA)$ is an \ainf category with shifts, 
we also have $H(\tw(\cA))=H^0(\tw(A))^\bullet=\tria(\cA)^\bullet$.

\paragraph{The case of \ainf algebras.}

When $\cA$ has a single object $a$, then $\cA$ can be identified with the
\ainf algebra $A:=\Hom_{\cA}(a,a)$. We have suspended products
$r_n:A[1]^{\otimes n}\rightarrow A[1]$ ($n\geq 1$) subject to
conditions (\ref{ainf_alg}), as well as desuspended products 
$m_n:A^{\otimes n}\rightarrow A$ given by $r_n=s\circ m_n\circ (s^{-1})^{\otimes n}$, where 
$s:A\rightarrow A[1]$ is the suspension operator. As per our conventions, the  classical \ainf structure 
of $A$ is defined by the products $\mu_n$, so  $m_n$ define a classical \ainf algebra structure on 
the opposite \ainf algebra $A^{\rm op}$.  Strict unitality amounts to existence of
an element $u\in A^0$ satisfying (\ref{alg_unitality}).  An \ainf
functor between two \ainf categories with one object corresponds to an
\ainf morphism of \ainf algebras, given by maps $F_n:A^{\otimes
n}\rightarrow B$ ($n\geq 1$) satisfying (\ref{morphism}).  The
cohomology category $H(\cA)$ reduces to the $\mu_1$-cohomology $H(A)$, which
is a graded associative algebra. In this case, $\cA$ is degreewise
Hom-finite iff $A$ is degreewise finite as a graded vector space, i.e. 
iff $A^n$ is finite-dimensional for all $n$; we then say that 
$A$ is degreewise finite. $\cA$ is compact iff the graded vector space 
$H(A)$ is degreewise finite i.e. iff $H^n(A)$ is finite-dimensional for all $n$.
In this case, we say that $A$ is compact. 

We use the notation $D(A):=D(\cA)$, $\tria(A):=\tria(\cA)$,
$\tw(A):=\tw(\cA)$ etc.  The category $D(A)$ is compactly generated by 
the single object ${\hat a}$, which in this case we denote by ${\hat A}$; 
this object of $\Mod_\cA=\Mod_{(A,\mu)}=_{(A^{\rm op},m)}\Mod$  can be identified with $(A,\mu)$ viewed as a right 
\ainf module over itself.  Then $\tria(A)$, $\per(A)$ are the strictly full 
triangulated subcategories of $D(A)$
generated by ${\hat A}$ (the second in the Karoubi sense). 

A basic result \cite{Hasegawa} states that any strictly unital \ainf
morphism $\varphi:A\rightarrow B$ between \ainf algebras $A,B$ induces
a `restriction'  exact functor $D(B)\rightarrow D(A)$
mapping ${\hat B}$ into ${\hat A}$ and thus $\tria(B)$ into $\tria(A)$ and $\per(B)$ into $\per(A)$.
When $\phi$ is an \ainf quasi-isomorphism, this functor is an exact 
equivalence $D(B)\stackrel{\approx}{\rightarrow} D(A)$, whose
restrictions give equivalences $\tria(B)
\stackrel{\approx}{\rightarrow} \tria(A)$ and $\per(B)
\stackrel{\approx}{\rightarrow} \per(A)$.  Hence the triangulated
categories $\tria(A)$, $\per(A)$ and $D(A)$ are determined up to exact
equivalence by the \ainf quasi-isomorphism class of $A$. This allows
one to replace $A$, for example, by a minimal or antiminimal (dG)
model \cite{Hasegawa}.

\section{Cyclic \ainf categories}
\label{sec:cyc_ainf}

In this section we discuss cyclic \ainf categories, then explain how a
cyclic pairing extends from an \ainf category to its category of
twisted complexes.  We also discuss a class of cyclic minimal models
of \ainf categories obtained by adapting a procedure due to
\cite{KS_old} (see also \cite{Merkulov}). After addressing the issues
of unitality and existence of shifts for such minimal models, we give a
string field theory interpretation of this construction.

\subsection{Basics}
\label{sec:basics_cyc}

Let $D$ be an integer. A $D$-{\em cyclic} structure on an \ainf category $\cA$ 
consists of morphisms of graded vector spaces 
$\langle ~\rangle_{ab}:\Hom_\cA(a,b)\otimes \Hom_\cA(b,a)\rightarrow \C[-D]$ for all $a,b\in \Ob\cA$
which satisfy the graded symmetry condition 
$\langle u\otimes v\rangle_{ab}=(-1)^{|u||v|}\langle v\otimes u\rangle_{ba}$ and the graded 
cyclicity relations: 
\be
\label{rho_cyc}
\langle x_0\otimes r_{a_0\ldots a_n}(x_1\otimes \ldots \otimes x_n) \rangle_{a_n,a_0}=
(-1)^{{\tilde x}_0({\tilde x}_1+\ldots +{\tilde x}_n)}
\langle x_1\otimes r_{a_1 \ldots a_n a_0}(x_2\otimes \ldots \otimes x_n \otimes x_0) \rangle_{a_0,a_1} ~~.
\ee
The doublet $(\cA, \langle ~\rangle)$  is called a {\em $D$-cyclic 
\ainf category}. We will often view $\langle~\rangle$ as bilinear forms 
$\langle~,~\rangle:\Hom_\cA(a,b)\times \Hom_\cA(b,a)\rightarrow \C$. Homogeneity of the pairings 
amounts to the `selection rule':
\be
\nn
\langle x,y\rangle=0~~{\rm unless}~|x|+|y|=D~~.
\ee

For reader's convenience, we list the first two cyclicity relations: 
{\footnotesize \be
\nn
\langle x_0 \otimes r_{a_0a_1}(x_1)\rangle_{a_1a_0}=(-1)^{{\tilde x}_0} \langle r_{a_1a_0}(x_0) \otimes x_1\rangle_{a_1a_0}~~,~~
\langle x_0 \otimes r_{a_0a_1a_2}(x_1 \otimes x_2)\rangle_{a_2a_0}=
(-1)^{{\tilde x}_0+{\tilde x}_1}\langle r_{a_2a_0a_1}(x_0\otimes x_1)\otimes x_2\rangle_{a_2a_1}
\ee}
and their translation in terms of the products $m_1,m_2$:
{\footnotesize \be
\nn
\langle m_{a_1a_0}(x_0)\otimes x_1\rangle_{a_1a_0}+(-1)^{|x_0|} \langle x_0 \otimes m_{a_0a_1}(x_1)\rangle_{a_1a_0}=0~~,~~
\langle m_{a_2a_0a_1}(x_0 \otimes x_1)\otimes x_2\rangle_{a_2a_1}=\langle x_0 \otimes  m_{a_0a_1a_2}(x_1 \otimes x_2)\rangle_{a_2a_0}~~,
\ee}
and in terms of $\mu_1,\mu_2$:
{\footnotesize \be
\label{mu_cyc}
\langle \mu_{a_0a_1}(x_0)\otimes x_1\rangle_{a_1a_0}+(-1)^{|x_0|} \langle x_0 \otimes \mu_{a_1a_0}(x_1)\rangle_{a_1a_0}=0~~,~~
\langle \mu_{a_2a_1a_0}(x_2 \otimes x_1)\otimes x_0\rangle_{a_0a_2}=
\langle x_2 \otimes  \mu_{a_1 a_0 a_2}(x_1 \otimes x_0)\rangle_{a_1a_2}~~.
\ee}
Here $x_i\in \Hom_{\cA}(a_{i-1~{\rm mod}~2},a_{i})$ for the first equation of each pair and 
$x_i\in \Hom_{\cA}(a_{i-1~{\rm mod}~3},a_{i})$ for the second equation of each pair. 

\paragraph{Cyclic structure induced on cohomology.}

The first equations in (\ref{mu_cyc}) imply that $\langle
~\rangle_{ab}$ descend to well-defined pairings on the cohomology
category $H(\cA)$, which we denoted by $\langle ~\rangle^H_{ab}
$. These are given by
\be
\nn
\langle [x]\otimes [y] \rangle^H_{ab}=\langle x\otimes y \rangle_{ab}~~\forall x\in Z(\Hom_{\cA}(a,b))~~
\forall x\in Z(\Hom_{\cA}(b,a))~~
\ee
and define a $D$-cyclic structure on the graded category $H(\cA)$. Indeed, the associated bilinear
forms $\langle~,~\rangle^H$ are obviously graded-symmetric (with respect to the
grading $|~|$) and satisfy $\langle u* v, w\rangle_{ac}^H=\langle u,
v* w\rangle_{bc}^H$ for all $u\in \Hom_{H(\cA)}(b,c), ~v\in
\Hom_{H(\cA)}(a,b)$ and $w\in \Hom_{H(\cA)}(c,a)$ (as implied by the
second equation in (\ref{mu_cyc})).

\paragraph{Nondegeneracy conditions.}

A $D$-cyclic structure on $\cA$ will be called {\em strictly
nondegenerate} if $\cA$ is degreewise Hom-finite and the bilinear forms $\langle
~,~\rangle_{ab}$ are non-degenerate.  It is called
{\em homologically non-degenerate} if $\cA$ is compact and the
bilinear pairings induced on cohomology are nondegenerate;
equivalently, the $D$-cyclic structure induced on the graded
associative category $H(\cA)$ is nondegenerate.  We say that $\cA$ is $D$-Calabi-Yau if it
admits at least one homologically nondegenerate D-cyclic structure.

A cyclic structure on $\cA$ is nondegenerate iff the pairings
$\langle~\rangle_{ab}$ induce isomorphisms of graded vector spaces
$\Hom_{\cA}(a,b)[D]\rightarrow \Hom_{\cA}(b,a)^{\rm v}$. Notice that
the latter condition implies\footnote{Indeed, an ungraded
vector space $V$ can be isomorphic with its linear dual $V'=\Hom_\C(V,\C)$ iff it is finite-dimensional. 
This follows from the inequality $\dim_\C V\leq \dim_\C V'$ for the transfinite dimension 
(cardinal of any basis), which is strict unless $\dim_\C V$ is finite.} degreewise Hom-finiteness of $\cA$.

The cyclic structure is homologically nondegenerate iff the bilinear
forms induce isomorphisms of graded vector spaces
$\Hom_{H(\cA)}(a,b)[D]\rightarrow \Hom_{H(\cA)}(b,a)^{\rm v}$.  Notice
that the latter condition implies compactness of $\cA$.

\paragraph{Compatibility with shifts.}

Let us assume that $\cA$ has a twisted shift functor $[[1]]$. 
A $D$-cyclic structure on $\cA$ is called {\em shift-equivariant }
if the following condition holds (cf. relation (\ref{SE_graded})): 
\be
\label{se}
\langle x[[1]]\otimes y[[1]]\rangle_{a[[1]]b[[1]]}=-\langle x\otimes y 
\rangle_{ab}~~\forall a,b\in \Ob\cA~, ~~~\forall 
x\in \Hom_{\cA}(a,b)~,~\forall y\in \Hom_{\cA}(b,a)~~,
\ee
i.e. $\langle~\rangle_{a[[1]]b[[1]]}\circ ([[1]]\otimes [[1]])=\langle~\rangle_{ab}$. 
In this case, we also say that $(\cA,[[1]],\langle~\rangle)$ is a {\em cyclic \ainf category with 
shifts.} Given such a pairing on $\cA$, the 
cyclic pairing induced on $H(\cA)$ is shift-equivariant.  

\paragraph{The strictly unital case.}

Let us assume that $\cA$ is strictly unital. 
Then all information carried by the cyclic pairings is encoded by the linear maps
$\tr_a:\Hom_{\cA}(a,a)\rightarrow \C[-D]$ of degree zero defined through: 
\be
\nn
\tr_a(x):= \langle u_a\otimes x\rangle_{aa}~~\forall x\in \Hom_{\cA}(a,a)~~.
\ee
Indeed, the second cyclicity condition in (\ref{rho_cyc}) gives $\langle~\rangle_{ab}=\tr_a\circ m_{aba}$, i.e.: 
\be
\label{e1}
\langle x \otimes y\rangle_{ab}=\tr_a (m_{aba}(x\otimes y))~~\forall x\in \Hom_{\cA}(a,b)~~\forall 
y\in \Hom_{\cA}(b,a)~~,
\ee
while graded symmetry of the pairings  reduces to: 
\be
\label{e2}
\tr_a(m_{aba}(x\otimes y))=(-1)^{|x||y|}\tr_b(m_{bab}(y\otimes x))~~
\forall x\in \Hom_{\cA}(a,b)~~\forall y\in \Hom_{\cA}(b,a)~~.
\ee
The remaining cyclicity conditions (\ref{rho_cyc}) become: 
\be
\label{e3}
\tr_a \circ m_{a a_1\ldots a_{n-1} a}=0~~\forall n\neq 2~~.
\ee
Conversely, equations (\ref{e1}), (\ref{e2}) and (\ref{e3}) imply 
(\ref{rho_cyc}) upon using the \ainf constraints. When $\cA$ admits a twisted shift functor $[[1]]$, the shift-equivariance 
condition (\ref{se}) becomes:
\be
\nn
\tr_{a[[1]]}{x[[1]]}=-\tr_a(x)~~\forall x\in \Hom_{\cA}(a,a)~~\forall a\in \Ob\cA~~.
\ee
In terms of the original \ainf compositions $\mu$, equations (\ref{e1}) take the form: 
\be
\label{e1_mu}
\langle x\otimes y\rangle_{ab}=(-1)^{|x||y|}\tr_a (\mu_{aba}(y\otimes x))~~{\rm for}~~ x\in \Hom_{\cA}(a,b)~~{\rm and}~~y\in \Hom_{\cA}(b,a)~~,
\ee
while relations (\ref{e2}) and (\ref{e3}) read:
\be
\label{e2_mu}
\tr_a(\mu_{aba}(y\otimes x))=(-1)^{|x||y|}\tr_b(\mu_{bab}(x\otimes y))~~,~~\tr_a \circ \mu_{a a_{n-1}\ldots a_1 a}=0~~\forall n\neq 2~~.
\ee
The last relation in (\ref{e2_mu}) implies $\tr_a\circ \mu_{aa}=0$, which 
shows that $\tr_a$ descend to well-defined functionals on $H(\cA)$, which we denote by 
$\tr_a^H$: 
\be
\nn
\tr_a^H([x]):=\tr_a(x)~~\forall x\in Z(\Hom_{\cA}(a,a))~~.
\ee
These satisfy:
\be
\nn
\tr_a^H(x*y)=(-1)^{|x||y|}\tr^H_b(y*x)~\forall x\in \Hom_{H(\cA)}(b,a)~~\forall 
y\in \Hom_{H(\cA)}(a,b)~~
\ee
and:
\be
\nn
\langle x,  y \rangle^H_{ab}=(-1)^{|x||y|}\tr_a^H(y*x)=\tr_b^H(x*y)~~
\forall x\in \Hom_{\cA}(a,b)~~\forall y\in \Hom_{\cA}(b,a)~~.
\ee
Therefore, they correspond to the traces defined by $\langle~\rangle^H$. 

\paragraph{Suspended pairings}

It is sometimes convenient to use {\em suspended pairings} 
$\omega_{ab}:\Hom_{\cA}(a,b)[1]\otimes \Hom_{\cA}(b,a)[1]\rightarrow \C[2-D]$ defined through:
\be
\label{omega}
\omega_{ab}:=\langle~\rangle_{ab}\circ (s_{ab}^{-1}\otimes s_{ba}^{-1})~~,
\ee
i.e. $\omega_{ab}(x\otimes y)=(-1)^{\tilde x}\langle x\otimes y\rangle_{ab}$. These are graded antisymmetric: 
\be
\nn
\omega_{ab}(x\otimes y)=(-1)^{{\tilde x}{\tilde y}+1}\omega_{ba}(y\otimes x)
\ee
and satisfy the modified cyclicity conditions: 
\be
\label{omega_cyc}
\omega_{a_na_0}( x_0\otimes r_{a_0\ldots a_n}(x_1\otimes \ldots \otimes x_n))=
(-1)^{{\tilde x}_0+{\tilde x}_1+{\tilde x}_0({\tilde x}_1+\ldots +{\tilde x}_n)}
\omega_{a_0a_1}(x_1\otimes r_{a_1\ldots a_n a_0}(x_2\otimes \ldots \otimes x_n \otimes x_0)) ~~.
\ee
When $\cA$ is strictly unital, $\omega_{ab}$ correspond to the {\em suspended traces}: 
\be
\nn
\Tr_a:=\tr_a\circ s_{aa}^{-1}:\Hom_{\cA}(a,a)[1]\rightarrow \C[1-D]
\ee 
to which they are related through the equations: 
\be
\nn
\omega_{ab}=\Tr_a\circ r_{aba}~~.
\ee
We also have $\Tr_a(x)=-\omega_{ab}(u_a\otimes x)$.  The modified cyclicity conditions (\ref{omega_cyc}) amount to: 
\be
\nn
\Tr_a\circ r_{aa_1\ldots a_{n-1}a}=0~~\forall n\neq 2
\ee
together with: 
\be
\nn
\Tr_a(r_{aba}(x\otimes y))=(-1)^{{\tilde x}{\tilde y}+1}\Tr_b(r_{bab}(y\otimes x))~~.
\ee
When $\cA$ admits a twisted shift functor, the shift-equivariance condition (\ref{se}) becomes $\omega_{a[[1]]b[[1]]}\circ ([[1]]^s\otimes [[1]]^s)=-\omega_{ab}$, i.e.: 
\be
\label{se_omega}
\omega_{a[[1]]b[[1]]}(x[[1]]^s\otimes y[[1]]^s)=-\omega_{ab}(x\otimes y)~~,
\ee
which in the strictly unital case amounts to:
\be
\nn
\Tr_{a[[1]]}\circ [[1]]^s=-\Tr_a~~.
\ee

\paragraph{Sector decomposition.}

In terms of the sector decomposition considered in Section
\ref{sec:sectors}, a $D$-cyclic pairing on $\cA$ amounts to a 
morphism of graded $R$-bimodules $\langle ~\rangle\in \Hom_{_R\GrMod_R}({\cal
H}\otimes_R {\cal H} , R[-D])$ where $R$ is endowed with the obvious
bimodule structure over itself. This map takes the form $\langle
x\otimes_R y\rangle=\oplus_{a,b\in \Ob \cA}\langle x_{ab}\otimes y_{ba}\rangle_{ab}e_a$ for all $x,y\in {\cal H}$. 
The cyclicity conditions (\ref{rho_cyc}) reduce to:
\be
\label{alg_cyc}
\langle x_0\otimes_R r_n(x_1\otimes_R\ldots \otimes_R x_n) \rangle=
(-1)^{{\tilde x}_0({\tilde x}_1+\ldots +{\tilde x}_n)}
\langle x_1\otimes_R r_n(x_2\otimes_R \ldots \otimes_R x_n\otimes_R x_0) \rangle ~~.
\ee
These can also be written as follows. Since $R$ is commutative, one 
has natural inclusions\footnote{Given $f\in \Hom_{_R\GrMod_R}({\cal H}^{\otimes_R n} , R[-D])$ we have 
$f(x)=\sum_{a,b}{f(\epsilon_a x\epsilon_b)}=
\sum_{a,b}{\epsilon_a f(x)\epsilon_b}=\sum_a {\epsilon_a f(x) \epsilon_a}=
f(\sum_a \epsilon_a x \epsilon_a)$, where we used commutativity of $R$ and the identities 
$\epsilon_a\epsilon_b=\delta_{ab}\epsilon_a$. Thus $f(x)$ is determined by its restriction to $[{\cal H}^{\otimes_R n}]^R$. This means 
that $f$ can be viewed as a degree zero $\C$-linear map from $ [{\cal H}^{\otimes_R n}]^R$ to $R[-D]$.}
$\Hom_{_R\GrMod_R}({\cal H}^{\otimes_R n} , R[-D])\subset \Hom_{\gr}([{\cal
H}^{\otimes_R n}]^R , R[-D])$, where $[{\cal H}^{\otimes_R n}]^R $ is the center of the 
$R$-bimodule ${\cal H}^{\otimes_R n}$. In particular,  $\langle ~\rangle$ can be viewed as a  morphism of graded vector spaces
$\langle~\rangle\in \Hom_{\gr}([{\cal H}\otimes_R {\cal H}]^R, R[-D])$ which commutes with the action of $\epsilon_a$. 
Consider the maps ${\cal P}_{n+1}\in \Hom_{R}({\cal H}^{\otimes_R (n+1)}, R[-D])$ defined 
through:
\be
\nn
{\cal P}_{n+1}:=\langle~\rangle\circ (\id_{\cal H}\otimes_R r_n)~~,
\ee
which we view as elements of $\Hom_{\gr}([{\cal H}^{\otimes_R (n+1)}]^R, R[-D])$ as explained above.
Let $\Pi_{n+1}:[{\cal H}^{\otimes_R (n+1)}]^R\rightarrow [{\cal H}^{\otimes_R (n+1)}]^R$ 
be the $\C$-linear automorphism given by\footnote{This is easily seen to be well-defined upon considering the sector 
decomposition of $x_0\ldots x_n$.} :
\be
\label{cycperm}
\Pi_{n+1}(x_0\otimes_R x_1\otimes_R \ldots  \otimes_R x_n)=(-1)^{{\tilde x}_0({\tilde x}_1+\ldots +{\tilde x}_n)}
x_1\otimes_R \ldots \otimes_R x_n\otimes_R x_0~~,
\ee
Then equations (\ref{alg_cyc}) amount to:
\be
\label{mod_cyc}
{\cal P}_{n+1}\circ \Pi_{n+1}={\cal P}_{n+1}~~\forall n\geq 1~~.
\ee
When $\cA$ admits a twisted shift functor, the shift-equivariance condition (\ref{se}) for the bilinear pairings reduces to:
\be
\label{se_total}
\langle~\rangle\circ ([[1]]\otimes [[1]])=-\langle~\rangle~~,
\ee
where $[[1]]:{\cal H}\rightarrow {\cal H}$ is the total shift operator defined in the previous section. When $\cA$ is 
also unital, we have total traces $\tr:{\cal H}^R\rightarrow R[-D]$ given by $\tr(x)=\sum_{a}{x_{aa}}$, and the 
shift-equivariance condition takes the form: 
\be
\nn
\tr\circ [[1]]=-\tr~~.
\ee

\subsection{Extension of cyclic pairings} 
\label{sec:extension}

Let $\cA$ be an \ainf category endowed with a $D$-cyclic structure
$\langle ~\rangle$.  In this subsection, we show that
$\langle~\rangle$ extends naturally to a shift-equivariant pairing on
$\tw(\cA)$, thereby inducing a shift-equivariant pairing on the
triangulated category $\tria(\cA)$.  When the pairing of $\cA$ is
nondegenerate or homologically nondegenerate, the same is true of the
pairing induced on $\tw(\cA)$. In each of these cases, the
triangulated category $\tria(\cA)=H^0(\tw(\cA))$ is $D$-Calabi-Yau.

\paragraph{Extension to $\ZA$.} 

Consider the linear maps $\omega_{a[[m]],b[[n]]}:\Hom_{\ZA}(a[[m]],b[[n]])[1]\otimes \Hom_{\ZA}
(b[[n]], a[[m]])[1] \rightarrow \C[2-D]$ given by: 
\be
\label{ZA_omega}
\omega_{a[[m]],b[[n]]}^{\ZA}:= (-1)^m\omega_{ab}\circ  (s_{ab}^{m-n}
\otimes s_{ba}^{n-m})\circ ([[-m]]^s\otimes [[-n]]^s)~~
\ee
and the desuspended maps $\langle~\rangle_{a[[m]],b[[n]]}^{\ZA}:\Hom_{\ZA}(a[[m]],b[[n]])\otimes \Hom_{\ZA}
(b[[n]], a[[m]]) \rightarrow \C[-D]$ defined through $\omega_{a[[m]],b[[n]]}^{\ZA}=\langle~\rangle^{\ZA}_{a[[m]],b[[n]]}
\circ (s_{a[[m]]b[[n]]}^{-1}\otimes s_{b[[n]]a[[m]]}^{-1})$, i.e.:  
\be
\label{ZA_pairing}
\langle~\rangle_{a[[m]],b[[n]]}^{\ZA}=(-1)^m\langle~\rangle_{ab} \circ (s_{ab}^{m-n}
\otimes s_{ba}^{n-m})\circ ([[-m]]\otimes [[-n]])~~.
\ee
An easy computation shows  that $\langle ~\rangle^{\ZA}$ is a cyclic pairing on the \ainf category 
$\ZA$, of the same degree as the original pairing on $\cA$. It is also clear 
from (\ref{ZA_pairing}) that $\langle~\rangle^{\ZA}$ is shift-equivariant.
Hence $(\ZA,[[1]],\langle~\rangle^{\ZA})$ is a cyclic \ainf category with twisted shifts. 

When $\cA$ (and thus $\ZA$) is strictly unital, we define traces $\tr^{\ZA}$ and $\tr$ associated with $\langle~\rangle^{\ZA}$ and $\langle~\rangle$ 
and suspended traces $\Tr^{\ZA}$ and $\Tr$ associated with $\omega^{\ZA}$ and $\omega$ as in the previous subsection:
\bea
\nn
\omega_{ab}=\Tr_{a}\circ r_{aba}~~~&,&~~\omega^{\ZA}_{a[[m]]b[[n]]}=\Tr^{\ZA}_{a[[m]]}\circ r_{a[[m]]b[[n]]a[[m]]}^{\ZA}\nn\\
\langle~\rangle_{ab}=\tr_{a}\circ m_{aba}~~&,&~~\langle~\rangle^{\ZA}_{a[[m]]b[[n]]}=\Tr^{\ZA}_{a[[m]]}\circ m_{a[[m]]b[[n]]a[[m]]}^{\ZA}\nn
\eea
and we have $\Tr_a=\tr_a\circ s_a^{-1}$ and $\Tr_{a[[m]]}^{\ZA}=\tr_{a[[m]]}\circ s_{a[[m]]}^{-1}$. 
Equations (\ref{ZA_omega}) and (\ref{ZA_pairing}) amount to: 
\be
\label{ZA_tr}
\Tr_{a[[m]]}=(-1)^m\Tr_a \circ [[-m]]^s\Leftrightarrow \tr_{a[[m]]}=(-1)^m\tr_a \circ [[-m]]
\ee
i.e.
\be
\label{ZA_traces}
\Tr_{a[[m]]}(x)=(-1)^{m{\tilde x}} \Tr_a(x)~~{\rm and}~~\tr_{a[[m]]}(x)=(-1)^{m(|x|+1)} \tr_a(x)~~{\rm for}~~ x\in \Hom_{\cA}(a[[m]],a[[m]])~~.
\ee

\paragraph{Extension to $\SigmaA$.} 

The pairing $\langle ~\rangle^{\ZA}$ extends by additivity to a
pairing $\langle ~\rangle^{\Sigma\cA}$ on $\SigmaA$: 
\be 
\nn 
\langle
f,g\rangle^{\Sigma\cA}=\sum_{i,j} \langle f_{ij}\otimes g_{ji}\rangle^{\ZA}_{A_i,B_j}~~ \ee where $f=\oplus_{ij}{f_{ij}}\in
\Hom_{\Sigma\cA}(A,B)$, $g=\oplus_{ij}{g_{ji}}\in
\Hom_{\Sigma\cA}(B,A)$ with $A=\oplus_{i}{A_i}, B=\oplus_{j}{B_j} \in
\Ob\Sigma\cA$ and $f_{ij}\in \Hom_{\ZA}(A_i, B_j)$, $g_{ji}\in \Hom_{\ZA}(B_j, A_i)$ (here $A_i,B_j\in \Ob\ZA$).  This pairing on $\SigmaA$
obeys the cyclicity conditions with respect to $r^{\SigmaA}$ and is
obviously shift-equivariant. Thus
$(\SigmaA,[[1]],\langle~\rangle^{\SigmaA})$ is a cyclic \ainf category
with twisted shifts.

\paragraph{Extension to $\tw(\cA)$.}

Recalling that the morphism spaces of $\tw(\cA)$ coincide with those
of $\Sigma\cA$, one checks by direct computation that the products
$r^{\tw(\cA)}$ are cyclic with respect to $\langle
~\rangle^{\Sigma\cA}$. Defining $\langle ~\rangle^{\tw(A)}:= \langle
~\rangle^{\Sigma\cA}$, we conclude that
$(\tw(\cA),[[1]],\langle~\rangle^{\tw(A)})$ is a cyclic \ainf category
with twisted shifts. Recall that $\tw(\cA)$ is degreewise Hom-finite, respectively
compact, iff $\cA$ has the same property.  Tracing through the steps
above, it is clear $\langle~\rangle^{\tw(A)}$ is strictly
nondegenerate iff the original pairing on $\cA$ is strictly
nondegenerate. As we will see below, a similar statement holds for homological 
nondegeneracy. 

\paragraph{The cyclic structure induced on $\tria(\cA)$.}

Passing to the cohomology category, the pairing on $\tw(\cA)$ induces
a shift-equivariant $D$-cyclic structure $\langle~\rangle^H$ on $H(\tw(\cA))$. Since
$H(\tw(\cA))=H^0(\tw(\cA))^\bullet$, this corresponds to a
shift-equivariant $D$-cyclic structure on the triangulated category
$H^0(\tw(\cA))=\tria(\cA)$.  The results of Appendix
\ref{sec:tria_duality} show that the latter is non-degenerate iff
$\langle~\rangle^H_{ab}:H(\Hom_{\cA}(a,b))\otimes
H(\Hom_{\cA}(b,a))\rightarrow \C$ are nondegenerate for all $a,b\in
\Ob{\cA}$, which amounts to homological nondegeneracy of the cyclic
pairing of $\cA$. Thus:

\paragraph{Proposition} A $D$-cyclic structure on $\cA$  induces 
shift-equivariant $D$-cyclic structures on $\tw(\cA)$ and
$\tria(\cA)=H^0(\tw(\cA))$.  Moreover:

(1) $\tw(A)$ is degreewise Hom-finite iff $\cA$ is and the cyclic
structure induced on $\tw(\cA)$ is strictly nondegenerate iff the
original cyclic structure on $\cA$ is strictly nondegenerate
  
(2) $\tw(A)$ is compact iff $\cA$ is and the cyclic structure induced on
$\tw(\cA)$ is homologically nondegenerate (in particular, $\tria(\cA)$
is $D$-Calabi-Yau) iff the original cyclic structure on $\cA$ is
homologically nondegenerate

\subsection{Minimal models induced by a cohomological splitting}
\label{sec:minmodel}

Fixing an \ainf category $\cA$, let $R$, ${\cal H}:={\cal H}_\cA$ and
$r_n$ be defined as in Section \ref{sec:sectors}.  Recall that a 
minimal model of $\cA$ is a minimal \ainf category $\cB$ which
is quasi-isomorphic with ${\cA}$. The work of \cite{Merkulov, KS_old,
Fukaya_mirror} provides an explicit construction of a particular class
of minimal models, which we recall below.  Adapting this
will allow us to build a special class of cyclic minimal models for a
cyclic \ainf category. In this section, we view $r$ as defined on 
the space ${\cal H}$ endowed with the tilde grading.

\paragraph{Retracts.} 

Define a {\em strict homotopy retraction} of ${\cA}$ to be a homotopy
retraction of the $R$-complex $({\cal H}, r_1$) (notice that $r_1=m_1$), i.e. a pair $(P, G)$
with $P\in \Hom_{_R\GrMod_R}({\cal H}, {\cal H})$ and $G\in
\Hom_{_R\GrMod_R}({\cal H}, {\cal H}[-1])$ such that:

\

\noindent (1) $P^2=P$

\noindent (2) $\id_{\cal H}-P=r_1\circ G+G\circ r_1$. 

\

In category-theoretic language, this corresponds to morphisms of graded vector spaces 
$P_{ab}:\Hom_{\cA}(a,b)\rightarrow \Hom_{\cA}(a,b)$ and
$G_{ab}:\Hom_{\cA}(a,b)\rightarrow \Hom_{\cA}(a,b)[-1]$ such that
$(P_{ab})^2=P_{ab}$ and $\id_{ab}-P_{ab}=r_{ab}\circ
G_{ab}+G_{ab}\circ r_{ab}$ (recall that
$\id_{ab}:=\id_{\Hom_{\cA}(a,b)}$).  The submodule $B:=\im P\subset
{\cal H}$ corresponds to the subspaces $B_{ab}=\im P_{ab}\subset
\Hom_{\cA}(a,b)$ (we have $B=\oplus_{a,b\in \Ob \cA}B_{ab}$).

We let $i:B\rightarrow {\cal H}$ be the inclusion and $p:{\cal
H}\rightarrow B$ be the corestriction of $P$ to $B$ (thus $P=i\circ
p$). These correspond to the inclusions $i_{ab}:B_{ab}\rightarrow
\Hom_{\cA}(a,b)$ and surjections $p_{ab}:\Hom_{\cA}(a,b)\rightarrow
B_{ab}$. Using the identity $r_1\circ r_1=0$, condition (2) above
implies $r_1\circ P=P\circ r_1$, which in turn shows that 
$r_1(B)\subset B$. 

For every $n\geq 2$, consider the set $\cT_n$ of all oriented and connected
{\em planar}\footnote{Orienting the plane clockwise, this means that
all edges meeting any given vertex of $T$ are cyclically ordered in
the clockwise direction on the plane. Such an ordering is called a
ribbon structure in \cite{Fukaya_mirror}, where a planar graph is
called a ribbon graph.}  trees $T$ such that:

\

\noindent (I) $T$ has exactly $n+1$ vertices of valency one (called {\em external vertices}), all other 
vertices having valency at least $3$ (these are called {\em internal vertices}). The edges meeting an external 
vertex are called {\em external edges}, the other being called {\em internal edges}.  
An external edge is called {\em incoming} if it leaves the corresponding external vertex, and 
{\em outgoing} if it enters the corresponding external vertex. 

\noindent (II) Exactly one external edge is outgoing (being called the {\em root} of $T$ ); the other $n$
external edges are incoming (being called the {\em leaves} of $T$).

\noindent (III) For each internal vertex of $T$, exactly one of the edges it meets leaves that vertex; the 
others enter it. 

\

We let $E(T)$ respectively $E_i(T)$, $E_e(T)$ be the sets of all edges, respectively all internal
and external edges of $T$.  We also let $e_i(T):=\Card E_i(T)$ be the number of internal edges. For each $T\in \cT_n$ we
define a morphism of graded $R$-bimodules $\rho_T\in
\Hom_{_R\Mod_R}(B^{\otimes_R n} , B)$ as follows:

\

\noindent (a) Associate the inclusion $i$ with every leaf of $T$

\noindent (b) Associate the surjection $p$ with the root of $T$

\noindent (c) Associate the product $r_k$ with each internal vertex of
$T$ of valency $k+1$

\noindent (d) Associate $G$ with each internal edge of $T$.

\

\begin{figure}[hbtp]
\begin{center}
\scalebox{0.7}{\input{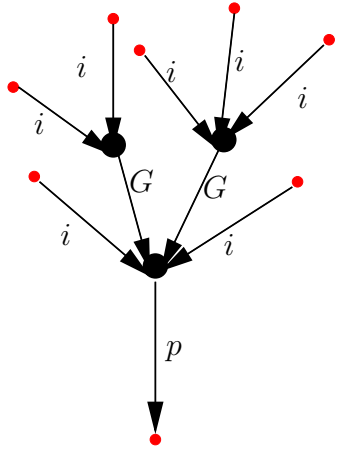}}
\end{center}
     \caption{An oriented tree $T\in \cT_7$ with $e_i(T)=2$.}
  \label{minproducts}
\end{figure}

Following the tree from its root toward its leaves, consider the
composition of the operators associated to each edge and vertex, using
tensor products and insertions of the identity map $\id_B$ wherever needed (always arranged in
clockwise order in the plane).  Finally, multiply the result by
the sign factor $(-1)^{e_i(T)}$.  For the example shown in figure
\ref{minproducts}, this gives the product:
\be
\nn
\rho_{T}=+p\circ r_4\circ (i\otimes G\otimes G\otimes i)\circ (\id_B\otimes r_3\otimes r_2\otimes \id_B)\circ 
(\id_B \otimes i\otimes i\otimes i\otimes i\otimes i\otimes \id_B):B^{\otimes 7}\rightarrow B~~.
\ee

\noindent We now define $\rho_1:=r_1|_B=p\circ r_1\circ i$ (recall that $r_1(B)\subset B$)
and $\rho_n:=\sum_{T\in \cT_n}{\rho_T}\in
\Hom_{_R\Mod_R}(B^{\otimes_R n}, B)$ for all $n\geq 2$. Notice that $\rho_2=p\circ r_2\circ (i^{\otimes 2})=p\circ
r_2|_{B\otimes B}$. Expanding into sectors, one can write the compositions $\rho_{a_0\ldots a_n}$ as sums over decorated 
trees, i.e. trees $T\in \cT_n$ together with labels chosen coherently for the two sides of each edge. One can visualize 
this by considering the ribbon associated with $T$, and placing labels in the obvious manner. This is entirely trivial and 
we leave it as an exercise for the reader. One has the following result:

\paragraph{Theorem\cite{KS_old, Fukaya_mirror}}{ The products $(\rho_n)_{n\geq 1}$ satisfy the (forward) \ainf relations. Hence they define an 
\ainf category $\cB$ having the same objects as $\cA$ and
morphism spaces $\Hom_\cB(a,b):=B_{ab}$. }

\

\noindent The \ainf category $\cB$ will be called the {\em retract} of $\cA$ along $(P,G)$. Its \ainf compositions 
$\rho_{a_0\ldots a_n}:\Hom_\cB(a_0,a_1)\otimes \ldots \otimes \Hom_\cB(a_{n-1},a_n)\rightarrow \Hom_\cB(a_0,a_n)$ 
are obtained by decomposing:
\be
\nn
\rho_n=\oplus_{a_0,a_n}\sum_{a_1\ldots a_{n-1}}\rho_{a_0\ldots a_n}~~,
\ee
which is possible by $R$-multilinearity. In the notation of Section \ref{sec:sectors}, we
have $B={\cal H}_\cB=\oplus_{a,b\in \Ob \cB}\Hom_{\cal
B}(a,b)$.

\paragraph{Observation} Notice the relations $\rho_T=p\circ \lambda_T$ and $\rho_n=p\circ \lambda_n$  
where $\lambda_n=\sum_{T\in \cT_n}{\lambda_T}$, with
$\lambda_T\in \Hom_{_R\Mod_R}(B^{\otimes_R n}, {\cal H})$ ($n\geq 2$)
defined exactly as $\rho_T$ except that we insert $\id_{\cal H}$
instead of the map $p$ along the root of the tree $T\in {\cal
T}_n$. Consider the maps $\iota_n\in \Hom_{_R\Mod_R}(B^{\otimes_R n},
{\cal H})$ given by $\iota_1:=i$ and $\iota_n=G\circ \lambda_n$ for
$n\geq 2$.  It was shown in \cite{Fukaya_mirror} that
$\iota=(i_n)_{n\geq 1}$ gives an \ainf quasi-isomorphism between
$(B,\rho)$ and $({\cal H},r)$. Of course, this corresponds to an
\ainf quasi-isomorphism between the \ainf categories $\cB$ and $\cA$.
In particular, $i$ induces an isomorphism of graded $R$-bimodules $i_*:H({\cal H})\rightarrow
H(B)$, i.e. an isomorphism between the graded associative categories
$H(\cB)$ and $H(\cA)$.

\

A strict homotopy retraction $(P,G)$ of $\cA$ is called a {\em
cohomological splitting} if $r_1|_B=0$, i.e. $\rho_1=0$. In this case,
$i_*$ and $p_*$ give inverse isomorphisms between $B$ and
$H_{r_1}({\cal H})$. Moreover, $(B,\rho)$ is a minimal model of
$({\cal H},r)$ and the category $\cB$ is a minimal model of
$\cA$.

\

\noindent {\em Proof}.  We give the proof of the proposition for completeness\footnote{This is essentially the proof given 
in \cite{KS_old}, except that we give a clear accounting of the signs.}. Let 
$(r)_1^n:=r_1\circ r_n+\sum_{i=0}^{n-1}r_n\circ (\id_{\cal H}^{\otimes i}\otimes r_1\otimes \id_{\cal H}^{\otimes (n-i-1)})$, 
with a similar notation $(\rho)_1^n$ for the products $\rho$. The \ainf relations (\ref{ainf_alg})  are equivalent
with $r_1^2=0$ together with the equations: 
\be
\label{req}
(r)_1^n=-\sum_{i\geq 0}^{n-2}\sum_{j=2}^{n-i} r_{n-j+1}\circ (\id_{\cal H}^{\otimes i}\otimes r_j\otimes 
\id_{\cal H}^{\otimes (n-j-i)})~~\forall n\geq 2~~.
\ee
Since $r_1$ preserves the subspace $B$, we have  $\rho_1^2=r_1^2|_B^B=0$, so it suffices to prove that $\rho_n$  
satisfy: 
\be
\label{rhoeq}
(\rho)_1^n=-\sum_{i\geq 0}^{n-2}\sum_{j=2}^{n-i} \rho_{n-j+1}\circ (\id_\cB^{\otimes i}\otimes \rho_j\otimes 
\id_\cB^{\otimes (n-j-i)})~~\forall n\geq 2~~.
\ee 

Given $f\in \End_{_R\Mod_R}({\cal H})$,  a tree $T\in \cT_n$ having at least one internal edge 
and $e\in E_i(T)$, we let $\rho_{T,e}^f$ be the product obtained from $\rho_T$
upon replacing the insertion of $G$ along the internal edge $e$ with
the operator $f$. We define $\rho_n^f=\sum_{T\in \cT_n, e_i(T)\geq 1}
\sum_{e\in E_i(T)}\rho_{T,e}^f \in \Hom_{_R\Mod_R}(B^{\otimes_R n},
B)$.  We also let ${\hat \rho}_{T,e\pm}$ be the maps obtained from
$\rho_T$ upon inserting the operator $r_1$ before, respectively
after the insertion of $G$ along that edge and define ${\hat
\rho}_{T,e}:= {\hat \rho}_{T,e+} + {\hat \rho}_{T,e-}$.  Given an arbitrary tree $T\in \cT_n$ 
and $e\in E_e(T)$, we let ${\hat \rho}_{T,e}$ be the map obtained from $\rho_T$
upon inserting the operator $\rho_1=r_1|_B^B$ near the external vertex lying on the edge 
$e$. Since $r_1\circ P=P\circ r_1$ and $r_1(B)\subset B$, we have $\rho_1\circ p=p\circ r_1$ and 
$i\circ \rho_1=r_1\circ i$, so ${\hat \rho}_{T,e}$ coincides with the map obtained by inserting $r_1$ 
next to the internal vertex meeting $e$. Finally, define ${\hat \rho}_n:=\sum_{T\in \cT_n}\sum_{e\in
E(T)}{\hat \rho}_{T,e}$.  Using equation $r_1\circ G+G\circ
r_1=\id_{\cal H}-P$, we find:
\be
\label{he1}
{\hat \rho}_n=(\rho)_1^n+\rho_n^{\id_{\cal H}}-\rho_n^P~~,
\ee
where the first term comes from the $\rho_1$-insertions along external edges. 

The map ${\hat \rho}_n$ can also be computed by using equations
(\ref{req}). Indeed, consider the sum of those contributions to ${\hat
\rho}_T$ coming from insertions of $r_1$ immediately next to the
output or immediately next to one of the inputs of the product $r_k$
associated with a fixed internal vertex $v$ of $T$ of valency $k+1$. Equation
(\ref{req}) (with $n$ replaced by $k$) allows us to replace the sum of
such contributions with the sum of those contributions to the product
$\rho^{\id_{\cal H}}_n$ which arise from trees $T'$ obtained from $T$ upon
replacing the vertex $v$ with two vertices of valency $k-j+2$ and
$j+1$ connected by an internal edge (here $j$ runs from $2$ to $k$).
This edge can be chosen in 
$k -j+1$ distinct ways which correspond to the sum over $i$ in
(\ref{req}). This edge of $T'$ carries the insertion of the identity
operator $\id_{\cal H}$ required by the definition of $\rho^{\id_{\cal H}}_{T'}$. Since
$e_i(T')=e_i(T)+1$, the minus sign from (\ref{req}) produces the extra
minus sign required by the sign prefactor in the definition of
$\rho^{\id}_{T'}$. Applying this procedure to all internal vertices, it is
clear that ${\hat \rho}_n$ can be expressed as: 
\be
\label{he2}
{\hat \rho}_n=\rho_n^{\id_{\cal H}}~~.
\ee
Combining this with equation (\ref{he1}) gives
$(\rho)_1^n=\rho_n^P$. On the other hand, a moment's thought shows
that $\rho_n^P=-\sum_{i\geq 0}^{n-2}\sum_{j=2}^{n-i} \rho_{n-j+1}\circ
(\id_B^{\otimes i}\otimes \rho_j\otimes \id_B^{\otimes (n-j-i)})$, where
we used the decomposition $P=i\circ p$ and the minus sign is again due
to the prefactor used in the definition of $\rho_T$. This shows that (\ref{rhoeq}) 
are satisfied.

\paragraph{The strictly unital case.}

Now assume that $\cA$ is strictly unital.  A strict homotopy
retraction $(P,G)$ of $\cA$ is called {\em strictly unital} if it
satisfies the supplementary conditions:

\

\noindent (3) $P_{ab}\circ G_{ab}=0$ for all $a,b \in \Ob \cA$ and $G_{aa}(u_a)=0~~\forall a\in \Ob\cA$~~.

\

\noindent Since $r_{aa}(u_a)=0$, conditions (3) imply
$P_{aa}(u_a)=u_a$ i.e. $u_{a}\in B_{aa}$. They also imply $p_{ab}\circ G_{ab}=0$. We have the following:

\paragraph{Proposition} Assume that $\cA$ is strictly unital and let $(P,G)$ be a strict homotopy retraction of $\cA$. 
Then the \ainf category $\cB$ is strictly unital with the same
units as $\cA$.

\

\noindent {\em Proof}. To avoid notational morass, let us first assume
that $\cA$ has a finite number of objects.  With this assumption,
set $u:=\oplus_{a\in Ob\cA}{u_a}\in {\cal H}^R$ and notice that
conditions (3) amounts to $G(u)=0$ and $p\circ G=0$ and that we have $u\in B$. It suffices to
show that the products $\rho_n$ satisfy
(\ref{alg_unitality}). For this, let $T\in \cT_n$. Condition (3)
and unitality of $r_n$ imply that $\rho_T(x_1\otimes \ldots \otimes
u \otimes \ldots \otimes x_n)$ vanishes unless the internal vertex of $T$
which meets the root of $T$ has valency 3 and $u$ is inserted along an incoming edge flowing
directly into this vertex (otherwise $u$ is killed either by a
product $r_k$ with $k\neq 2$ or by an insertion of $G$). If $T$ satisfies
these conditions, then the unitality condition for $r_2$ implies that the insertion of $r_2$
at this vertex of $T$ reproduces whatever flows into it from
the other branch up to a sign. Since $p\circ G=0$, the result is killed by
the final insertion of $p$ at the root unless this other branch is
again reduced to an external edge. Thus $\rho_T(x_1\otimes \ldots
\otimes u\otimes \ldots \otimes x_n)$ vanishes unless $T$ has a single
internal vertex, which is of valency $2$ (i.e. unless $T$ is the unique tree in $\cT_2$ having 
exactly one internal vertex). In particular, this requires
$n=2$, which gives the unitality property $\rho_n(x_1\otimes \ldots \otimes u\otimes
\ldots \otimes x_n)=0$ for $n\neq 2$. The unitality constraint for $\rho_2$
follows from $\rho_2=p\circ r_2|_{B^{\otimes 2}}$ by using the unitality property of $r_2$ and 
the fact that $u\in B$.

If $\cA$ has an infinity of objects, a trivial adaptation of proof above goes through if
one adds object labels to all trees and maps involved, as alluded to above. We leave
this as an exercise for the reader.

\paragraph{The cyclic case.}

Now suppose that $\cA$ is endowed with a cyclic pairing $\langle
~\rangle$. A strict homotopy retraction $(P,G)$ of $\cA$ is called {\em cyclic}
if the following condition is satisfied:

\

\noindent (4) $\langle G(x) \otimes y\rangle=(-1)^{|x|}\langle x \otimes G(y)\rangle~~\forall x,y\in {\cal H}$~~. 

\

Combining (4) and (2) above and the cyclicity condition $\langle r_1(x)\otimes y\rangle=(-1)^{\tilde x}\langle x \otimes 
r_1(y)\rangle$ (see Section \ref{sec:basics_cyc}) gives:
\be
\label{P_cyc}
\langle P(x) \otimes y\rangle=\langle x \otimes P(y)\rangle~~.
\ee

\noindent The following proposition generalizes a result of \cite{Polishchuk} and \cite{CIL4}. 

\paragraph{Proposition}{Let $(P,G)$ be a cyclic strict homotopy retraction of 
a $D$-cyclic \ainf category $(\cA,\langle~\rangle)$.  Then the \ainf
products $\rho_n$ on $B$ defined above are cyclic with respect to the
restriction of $\langle~\rangle$ to $B$. 
Together with this restricted pairing, they define a $D$-cyclic \ainf structure on the category $\cB$.}

\

\noindent The restricted pairing $\langle~\rangle^B=\langle~\rangle\circ (i \otimes i)$ corresponds to 
pairings $\langle~\rangle^\cB_{ab}:\Hom_\cB(a,b)\otimes \Hom_\cB(b,a)\rightarrow \C$ as explained in Section 
\ref{sec:cyc_ainf}. The cyclic \ainf category $(\cB, \langle~\rangle^\cB)$ will be called the {\em retract} of  
$(\cA,\langle~\rangle)$ along $(P,G)$. 
A cohomological splitting $(P,G)$ is called cyclic if it is cyclic as a strict homotopy retraction. 
In this case, the cyclic \ainf category $(\cB, \langle~\rangle^\cB)$ is minimal. 

\

\noindent {\em Proof}. Writing (\ref{P_cyc}) as $\langle ~\rangle\circ (P\otimes
\id_{\cal H})=\langle~\rangle\circ(\id_{\cal H} \otimes P)$  and using $P\circ i=i$ and
the decomposition $P=i\circ p$, we find: 
\be
\label{P_reduction}
\langle~\rangle^B\circ (\id_B\otimes p)=\langle~\rangle\circ (i\otimes \id_{\cal H})~~, 
\ee
i.e. $\langle x \otimes p(y)\rangle^B=\langle x \otimes y\rangle$ for all $x\in B$ and $y\in {\cal H}$. To prove cyclicity of 
$\langle ~\rangle^B$, we 
must show that the maps ${\cal P}_{n+1}:=\langle ~\rangle^B\circ (\id_B\otimes \rho_n)\in 
\Hom_\C([B^{\otimes_R(n+1)}]^R, R)$ satisfy ${\cal P}_{n+1}\circ \Pi_{n+1}={\cal P}_{n+1}$ 
(see equations (\ref{mod_cyc})).  This follows from the following argument. 

\begin{figure}[hbtp]
\begin{center}
\scalebox{0.7}{\input{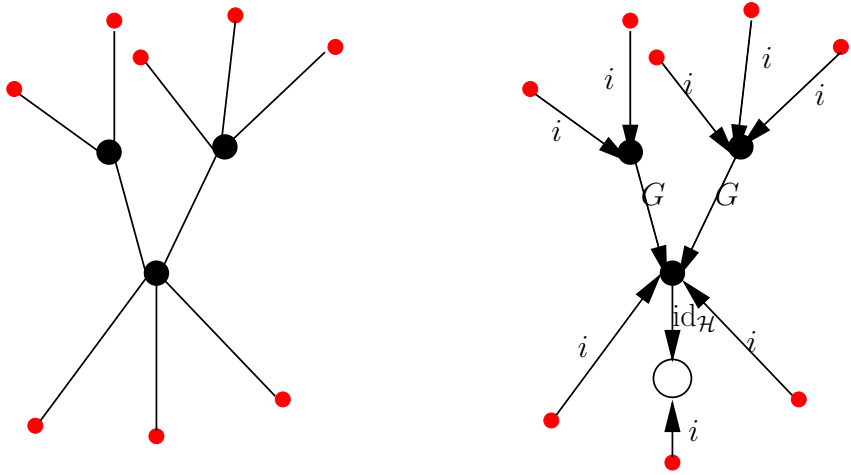}}
\end{center}
     \caption{A graph $\theta\in \Theta_8$ (left) and one of the contributions it brings to ${\cal P}_8$. 
The pairing $\langle ~\rangle$ is associated with the empty circle, which also indicates the choice of external vertex 
$e$ which determines the presentation $\theta=T'_e$ and thus the contribution shown to the right.}
  \label{amplitudes}
\end{figure}
Using the definition of $\rho_n$, we write $\rho_n=p\circ\lambda_n$.
Equation (\ref{P_reduction}) shows
that ${\cal P}_{n+1}=\langle~\rangle\circ (i\otimes \lambda_n)$. 
Thus ${\cal P}_{n+1}= \sum_{T\in \cT_n}{{\cal P}_T}$,
where ${\cal P}_T=\langle~\rangle\circ (i\otimes \lambda_T)$
have the following graphical description.

Let $\Theta_{n+1}$ be the set of all unoriented simply connected {\em
planar} graphs having $n+1$ vertices of valency one (=external
vertices) and such that all other vertices have valency at least 3
(=internal vertices).  Given a tree $T\in \cT_n$, we let $T'\in
\Theta_{n+1}$ be the unoriented graph obtained from $T$ by forgetting
the orientation of all edges. This gives a surjection ${\cal
T}_n\stackrel{\pi}{\rightarrow} \Theta_{n+1}$, $\pi(T)=T'$,  which is an $n+1$-fold cover.  Let us fix $\tau\in \Theta_{n+1}$.
Picking any external edge $e$ of $\tau$ gives a
presentation $\tau=T_e'$ where $T_e\in \cT_n$ is obtained by
orienting $e$ outwards (and viewing it as the root edge) and orienting
all other edges toward the root edge (in fact all trees in the preimage $\pi^{-1}(\tau)$ are obtained in this way).    
Now split the edge $e$ in the 
middle by inserting a new vertex (depicted as an empty circle), thus creating two edges $e', e''$, 
where $e''$ is the new external edge. We give $e'$ the orientation originally carried by $e$ and $e''$ the 
opposite orientation (see figure \ref{amplitudes}). Forgetting $e''$ for a moment, we obtain a tree in $\cT_n$ which 
is isotopic with $T_e$. To its incoming external edges and internal vertices we associate the maps $i$ and $r_k$ as before.
To its root edge $e'$ we associate the map $\id_{\cal H}$. Finally, to the edge $e''$ of the new tree we associate the map $i$.
Reading the diagram in the obvious way gives the contribution ${\cal
P}_{\tau,e}:={\cal P}_{T_e}$ to ${\cal P}_{n+1}$.  These observations allow
us to write ${\cal P}_{n+1}=\sum_{\tau \in \Theta_{n+1}}\sum_{e\in
E_e(\tau)} {\cal P}_{\tau,e}$. Equations (\ref{mod_cyc}) now follow
essentially from the fact that the last expression is invariant under cyclic permutations of the $n+1$ external edges (we leave the details of 
this last step to the reader). 

\paragraph{Shift equivariance.}

Let us assume that $\cA$ has a twisted shift functor $[[1]]$ and let $[[1]]:{\cal H}\rightarrow {\cal H}$ be the total shift operator defined in 
the previous section. A strict homotopy retraction $(P,G)$ is called {\em shift-invariant} if the following condition is 
satisfied: 

\

\noindent (5) $G\circ [[1]]=[[1]]\circ G$

\

\noindent In this case, equations (1) and shift-invariance of $m_1=r_1$ imply $P\circ [[1]]=[[1]]\circ P$.

\paragraph{Proposition} Let $\cA$ be a cyclic \ainf category with shifts whose pairing $\langle~\rangle$ is shift-equivariant, and 
let $(P,G)$ be a shift-invariant and cyclic strict homotopy retraction. Then the retract category ${\cal B}$ constructed as above has a twisted shift functor 
and its pairing is shift-equivariant. 

\

\noindent{\em Proof.} Equations $P\circ [[1]]=[[1]]\circ P$ show that $[[1]]$ preserves the subspace $B=\im P$, on which it restricts to 
a total shift functor. Thus $[[1]]_{ab}:\Hom_{\cA}(a,b)\rightarrow \Hom_{\cA}(a[[1]],b[[1]])$ map the subspace $\Hom_{\cal B}(a,b)$ of 
$\Hom_{\cA}(a,b)$ into the subspace $\Hom_{\cB}(a[[1]],b[[1]])$ of $\Hom_{\cA}(a[[1]],b[[1]])$. 
Since the total pairing $\langle~\rangle$ of $\cA$ satisfies  $\langle~\rangle\circ ([[1]]\otimes [[1]])=-\langle~\rangle$, 
it is clear that the restricted bilinear pairing $\langle~\rangle^B$ satisfies $\langle~\rangle^B\circ ([[1]]\otimes [[1]])=-\langle~\rangle^B$.
Since both $P$ and $G$ commute with $[[1]]$, and since $r_n\circ ([[1]]^s\otimes \ldots \otimes [[1]]^s)=[[1]]^s\otimes r_n$, 
it is clear from the definition of $\rho_n$ that $\rho_n\circ ([[1]]^s\otimes \ldots \otimes [[1]]^s)=[[1]]^s\otimes \rho_n$. 

\paragraph{The cyclic and strictly unital case.}

\noindent Combining everything, we find:

\paragraph{Corollary} {Any strictly unital and cyclic strict homotopy retraction $(P,G)$ 
of a cyclic and strictly unital \ainf category 
$(\cA, \langle~\rangle)$ determines a strictly unital and
cyclic \ainf category $(\cB,\langle~\rangle^\cB)$. Moreover:

\

\noindent (1) Assume that ${\cal A}$ has a twisted shift functor and its cyclic structure is shift-equivariant, and that $(P,G)$ is shift-invariant. 
Then ${\cB}$ has a twisted shift functor induced from $\cA$ and its cyclic structure is shift-equivariant.

\

\noindent (2) When $(P,G)$ is a cohomological splitting, then $\cB$ is a minimal model of $\cA$. }

\paragraph{Observation} Assume that the cyclic structure 
on $\cA$ is homologically nondegenerate and let $(P,G)$ be a cyclic
cohomological splitting of $\cA$.  Then the cyclic structure induced
on $\cB$ is strictly nondegenerate. Indeed, we have $\langle~\rangle^B=\langle~\rangle\circ (i\otimes i)$, 
which gives  $\langle~\rangle^B=\langle~\rangle^H\circ (i_*\otimes i_*)$, where 
$i_*:B\rightarrow H_{r_1}({\cal H})$ 
is the map induced by $i$ on cohomology
and $\langle~\rangle^H$ is the pairing induced by $\langle~\rangle$ on
$r_1$-cohomology.  Since $i_*$ is bijective and $\langle~\rangle^H$
is nondegenerate, we have the desired conclusion.

\subsection{Interpretation through formal open string field theory}
\label{sec:sft}

\paragraph{The formal extended action.}

In the nondegenerate cyclic case, the construction given above has a string field theory interpretation, which generalizes a result of 
\cite{CIL4}. Fixing a strictly unital,  nondegenerate $D$-cyclic \ainf category $(\cA, \langle~\rangle)$, let $R=\oplus_{a\in \Ob\cA}{\C\epsilon_a}$
and consider the graded $R$-bimodule 
${\cal H}:={\cal H}_\cA:=\oplus_{a,b\in \Ob\cA}\Hom_{\cA}(a,b)$ of Section \ref{sec:sectors}, together with its \ainf products 
$r_n$. Fixing a unital Grassmann $\C$-algebra $G$, consider the $G$-supermodule ${\cal H}_e:={\cal H} \otimes G$, and notice that it carries 
a graded $R$-bimodule structure induced in the obvious manner from that of ${\cal H}$. 
Consider the natural extensions of the 
pairing and \ainf products of ${\cal H}$ 
to maps $\langle~\rangle_e:{\cal H}_e\otimes {\cal H}_e\rightarrow G$ and $r_n^e:{\cal H}_e^{\otimes n}\rightarrow G$:
\be
\nn
\langle x \otimes \alpha, y\otimes \beta\rangle_e=(-1)^{{\deg \alpha}~|y|}\langle x, y\rangle\alpha\beta
\ee 
and: 
\be
\nn
r_n^e((x_1\otimes \alpha_1) \ldots (x_n\otimes \alpha)n)=(-1)^{\sum_{i<j}{\deg \alpha_i ~{\tilde x}_j}} r_n(x_1\ldots x_n)\alpha_1\ldots \alpha_n~~.
\ee
Following \cite{CIL5, HLL,nc}, we define a (Grassmann-valued) formal 
action $S_e:{\cal H}_e^{\rm odd}=(\Pi {\cal H}_e)^{\rm even}\rightarrow G$ by the formal sum:
\be
\label{Sfa}
S_e(\varphi):=\sum_{n\geq 1}{\frac{1}{n+1}\langle \varphi \otimes r_n^e (\varphi^{\otimes n})\rangle_e}~~,
\ee
where $\varphi\in {\cal H}_e^{\rm odd}$ is the dynamical variable. The term $\frac{1}{2}\langle \varphi\otimes r_1^e (\varphi)\rangle_e$ 
in (\ref{Sfa}) plays the role of kinetic term. To make sense of the sum in (\ref{Sfa}), one can introduce a topology on ${\cal H}_e$ or simply 
restrict to the subspace ${\cal H}_{e, {\rm tors}}^{\rm odd}:=\{\varphi\in {\cal H}^{\rm odd}_e|\exists N_\varphi\in \Z_+: 
\varphi^{\otimes n}=0~\forall n\geq N_\varphi\}$. The extremum conditions 
of (\ref{Sfa}) amount to the equations: 
\be
\label{MC_Total}
\sum_{n\geq 1}{r_n^e(\varphi^{\otimes n})}=0~~
\ee
which also read: 
\be
\label{MC_Total_expanded_ext}
\sum_{n\geq 1}{
r^e_{a a_1\ldots a_{n-1}b}(\varphi_{aa_1}\otimes \ldots \otimes \varphi_{a_{n-1}b})}=0~~\forall 
a,b\in \Ob\cA~~,
\ee
with implicit summation over $a_1\ldots a_{n-1}$. The formal sums above make sense at least for 
$\varphi\in {\cal H}_{e, \rm {tors}}^{\rm odd}$. In particular, taking $\varphi$ in (\ref{MC_Total_expanded_ext}) to have the form 
$\varphi=\phi\otimes 1_G$ with $\phi\in {\cal H}^1$ and $1_G$ the unit of $G$ gives the equations:
\be
\label{MC_Total_expanded}
\sum_{n\geq 1}{
r_{a a_1\ldots a_{n-1}b}(\phi_{aa_1}\otimes \ldots \otimes \phi_{a_{n-1}b})}=0~~\forall 
a,b\in \Ob\cA~~,
\ee
which make sense at least when $\phi$ belongs to the subspace ${\cal H}^1_{\rm tors}:=\{\phi\in {\cal H}^1|\exists N_\phi\in \Z_+: 
\phi^{\otimes n}=0~\forall n\geq N_\phi\}$. It follows that any solution of (\ref{MC_Total_expanded}) induces a solution of (\ref{MC_Total_expanded_ext}).

\paragraph{The tree-level potential induced by a cohomological splitting.}

When $r_1\neq 0$, the action (\ref{Sfa}) has to be gauge-fixed. 
Any  consistent gauge-fixing procedure determines a low energy, $G$-valued, formal potential via the
semiclassical (WKB) approximation. A particular class of gauges is
provided by strictly unital cyclic cohomological splittings $(P,G)$ of $(\cA,
\langle~\rangle)$.  Defining $P_e:=P\otimes \id_G$ and $G_e:=G\otimes \id_G$, 
we can consider the gauge condition: 
\be
\label{gauge}
(\id_{{\cal H}_e}-P_e)(\varphi)=0\Leftrightarrow \varphi\in B_e^{\rm odd}~~,  
\ee
where $B_e:=\im P_e$. Working out the Feynman rules as in \cite{CIL4, CIL6}\footnote{One can in fact
study the gauge-fixing procedure in the BV formalism, as done for the
dG case in \cite{CIL7, CIL8}.}, one finds that $G$ plays the role of
propagator in the gauge (\ref{gauge}). The tree-level Feynman diagrams
are given by graphs $\theta \in \sqcup_{n\geq 2}{\Theta_{n+1}}$ (in the
notation of Section \ref{sec:minmodel}); see the first diagram in
figure \ref{amplitudes} for an example. The higher terms in
(\ref{Sfa}) give vertices in the perturbative expansion. One finds
that the tree-level potential defined in the gauge (\ref{gauge}) takes
the following form up to an uninteresting prefactor (this generalizes
a result of \cite{CIL4}): 
\be
\label{W_gen}
W_e(\varphi):=\sum_{n\geq 2}{\frac{1}{n+1}\langle \varphi \otimes
\rho_n^e(\varphi^{\otimes n})\rangle_e^B}~~{\rm for}~~\varphi\in B_e^{\rm odd}~~.  
\ee 
Here $\rho_n$ are the unital minimal \ainf products of Section \ref{sec:minmodel} (and $\rho_n^e$ their extensions to $B_e$), while
$\langle~\rangle^B_e$ is the restriction of the pairing $\langle~\rangle_e$ to the subspace $B$. 
The former pairing can be viewed as the extension of $\langle~\rangle^B$ to $B_e$. 
By the results of the previous subsection, 
the products $\rho_n$ are cyclic with respect to $\langle~\rangle^B$, which in turn is nondegenerate. 
In categorical language,  $(B,\rho, \langle~\rangle^B)$ corresponds to a cyclic, 
minimal, strictly unital \ainf category $(\cB, \langle~\rangle^\cB)$ having the same objects as $\cA$ and whose 
cyclic structure is nondegenerate; 
as explained in the previous subsection, $\cB$ is a minimal model of $\cA$.  
Since $B$ is isomorphic with $H_{r_1}({\cal H})$, we can identify $\cB$ with $H(\cA)$ and view the
minimal \ainf structure determined by $(P,G)$ as a cyclic \ainf prolongation
of the cyclic graded associative category $(H(\cA), \langle~\rangle^H)$. By the results of the 
previous subsection, ${\cal B}$ has a twisted shift functor and its pairings are shift-equivariant provided that ${\cal A}$ has the 
same properties. 

\paragraph{Topological string field theory interpretation.}

Using the modular functor approach initiated in \cite{Moore,CIL1}, it
was proved in \cite{Costello} that any strictly unital, minimal,
nondegenerate cyclic \ainf category defines an oriented open string
theory (=oriented open topological conformal field theory); this
provides a converse to the work of \cite{HLL}.  Applying this to our
case, we find that $(\cB, \langle~\rangle^\cB)$ describes an
open topological string theory, allowing us to view (\ref{Sfa}) as a formal string {\em
field} theory description of the latter.

With this interpretation, the objects of $\cA$ (which are the same as the objects of $\cB$) 
are topological D-branes. The space $H_{r_{ab}}(\Hom_{\cA}(a,b))\approx \Hom_{\cal
B}(a,b)$ becomes the spaces of topological boundary observables for
the open string stretching from $a$ to $b$. The minimal \ainf
compositions $\rho_{a_0\ldots a_n}$ are the string products
(associated with the integrated $n+1$-point functions on the disk),
while the non-minimal \ainf compositions of $\cA$ are string {\em
field} products. The latter correspond to geometric string field
vertices constructed as in \cite{Zwiebach_oc}.

Solutions of (\ref{MC_Total_expanded_ext}) describe classical vacua of the formal action (\ref{Sfa}). 
As in \cite{CIL2, CIL3}, a Grassmann-even solution $\varphi$ can be viewed as the result of `condensing target space fields' 
in the finite D-brane system described by the
elements of the set $\{a\in \Ob{\cA}|\exists b\in \Ob\cA:\varphi_{ab}\neq 0 ~{\rm or}~\exists b\in \Ob\cA:
\varphi_{ba}\neq 0 \}$, where the nonzero components $\varphi_{ab}$
are associated with those `target space fields' which acquire a
VEV. Notice that in topological string theory there is no way to determine which elements 
$\varphi_{ab}$ correspond to massless, massive or tachyonic `target space' excitations. To do this one needs a 
stability condition, which is not visible at the level of topological string theory.  

\paragraph{The generation property.}

Considering an \ainf subcategory $\cA_0\subset {\cal A}$ endowed with the restricted D-cyclic structure, 
let us assume that $\cA=\tw(\cA_0)$.
In this case, it turns out that any object of $\cA$ (=twisted complex over $\cA_0$) can be viewed
as the result of a condensation process taking place between appropriately
shifted copies of D-branes belonging to $\cA_0$. 

Indeed, let $q\in
\Ob[\tw(\cA_0)]$ be given by morphisms $q_{ij}\in \Hom^1_{\Z
\cA_0}(a_i[[n_i]], a_j[[n_j]])$, where $i,j\in I:=\{1\ldots l_q \}$, $a_i\in
\Ob\cA_0$ and $q_{ij}=0$ unless $i<j$.  Consider the sets ${\cal
S}:=\{a_i[[n_i]]|1\leq i\leq l_q\}\subset \Ob \Z \cA_0$ and
$I_\alpha:=\{i\in I|a_i[[n_i]]=\alpha\}$ for each $\alpha\in {\cal
S}$. For every $\alpha\in {\cal S}$, we set $s_\alpha:={\rm
Card}(I_\alpha)$ and define $A_\alpha:=\alpha^{\oplus s_\alpha} \in
\Ob\Sigma \cA_0$. For every $\alpha,\beta\in {\cal S}$, let
$\phi_{A_\alpha, A_\beta}:=\oplus_{i\in I_\alpha , j\in
I_\beta}{q_{ij}}\in \Hom^1_{\Sigma \cA_0}(A_\alpha,
A_\beta)=\Hom^1_{\tw(\cA_0)}(A_\alpha, A_\beta)$, where we view
$A_\alpha$ as degenerate twisted complexes via the canonical embedding
of $\Sigma \cA_0$ into $\tw(\cA_0)$ (i.e. $A_\alpha$ are viewed as twisted
complexes with zero maps).  Then $\phi:=\oplus_{\alpha,\beta\in {\cal
S} }{\phi_{A_\alpha,A_\beta}}$ is an element of ${\cal H}_{\rm tors}^1$
and it is easy to check that equations (\ref{MC_Total_expanded}) for
$\phi$ amount to the generalized Maurer-Cartan equations (\ref{MCinf})
for $q$. Thus $q$ can be viewed as the result of condensing
$\phi_{A_\alpha,A_\beta}\otimes 1_G$, which arise from strings stretching between
$A_\alpha$ and $A_\beta$.  Since $\Hom^1_{\Sigma \cA_0}(A_\alpha,
A_\beta)=\Hom^1_{\Z \cA_0}(\alpha,\beta)^{\oplus (s_\alpha s_\beta)}$,
this can also be viewed as a condensation process
taking place between the D-branes of ${\cal S}\subset \Z\cA_0$.
Summarizing this discussion, we obtain:

\

{\em Let $\cA_0$ be a strictly unital \ainf category endowed with a
nondegenerate $D$-cyclic structure and assume that ${\cal
A}=\tw(\cA_0)$, endowed with the nondegenerate $D$-cyclic
structure induced from $\cA_0$.  Then every twisted complex $q\in
\cA$ is the result of a condensation process involving open
strings stretching between a finite number of D-branes belonging to
$\Z\cA_0$.  Hence the category $\Z\cA_0$ generates ${\cal
A}$ via condensation processes. }

\

\paragraph{Observation} For the case of $3$-cyclic dG categories, the arguments of this section are due to
\cite{CIL2,CIL3} (see also \cite{Diac}). Since any \ainf category
admits a quasi-equivalent dG model\cite{Hasegawa, Fukaya_mirror}, the
treatment given in \cite{CIL2,CIL3} is essentially equivalent with the
more general discussion above.

\subsection{The case of \ainf algebras}

Recall that a strictly unital \ainf category $\cA$ with a single object $a$
identifies with a strictly unital \ainf algebra $A$.  A $D$-cyclic
pairing on $\cA$ amounts to a bilinear and graded-symmetric form
$\langle ~,~\rangle:A\times A\rightarrow \C$ of degree $-D$ satisfying
relations (\ref{alg_cyc}).  A cyclic structure $\langle ~\rangle$ on
$A$ induces a cyclic pairing on $H(A)$, as well as a cyclic pairing on
$\tw(A)$ and therefore on $\tria(A)$. The latter is nondegenerate iff
the pairing on $A$ is homologically nondegenerate.  All constructions
described above apply with trivial simplifications.

When $(A, \langle~,~\rangle)$ is homologically nondegenerate (and thus
compact), one can show
\footnote{This amounts to checking that the restricted pairing
$\langle~\rangle^B$ can be obtained from $\langle~\rangle$ by pullback
through the \ainf quasi-isomorphism induced by $i$. We omit the proof
since we have no need for it in the present paper. }  by direct
computation that the cyclic minimal categories induced by
cohomological splittings of $A$ give explicit representatives of the
isomorphism class of cyclic minimal models considered from a different
perspective in Appendix \ref{sec:cyc_geom}.

\section{ Cyclic differential graded algebras and their minimal models}
\label{sec:dG}

A differential graded algebra $A$ corresponds to an \ainf algebra
having $\mu_n=0$ for all $n\geq 3$. Setting $d:=\mu_1$ and
$xy:=\mu_2(x,y)$ for all $x,y\in A$, the \ainf constraints show that
$\mu_2$ is an associative composition and $d$ a degree one derivation
which squares to zero. We will assume that $A$ is strictly unital,
which amounts to the existence of a unit $1$ for the associative
multiplication such that $d(1)=0$. In this section, we give an
equivalent description of cyclic pairings on $A$ and briefly recall
the basics of homological algebra over $A$, following \cite{Keller_dG,
Keller_dGcat}.  We then show that $A$ admits a nondegenerate $D$-cyclic
structure iff $\tria(A)$ is $D$-Calabi-Yau, and prove a similar result
for a minimal model of $A$.

\subsection{dG modules and bimodules over a dGA}

Recall that a unital right dG module over $A$ is a unital $\Z$-graded
right module $M$ over the unital graded associative algebra underlying
$A$, together with a differential $d_M:M\rightarrow M$ of degree $+1$
which satisfies the compatibility conditions:
\be
\nn
d_M(mx)=(d_Mm)x+(-1)^{|m|} m~d x
\ee
for homogeneous elements $x\in A$ and $m\in M$. A unital left dG
module is a (unital) left $\Z$-graded right module $M$ over the unital
graded associative algebra $A$, together with a differential
$d_M:M\rightarrow M$ of degree $+1$ which satisfies:
\be
\nn
d_M(xm)=(dx)m+(-1)^{|x|} x~d_Mm~~.
\ee
Of course, a unital right dG module over $A$ is the same as a unital
left dG module over the opposite dG algebra $A^{\rm op}$, which is
defined on the underlying set of $A$ by the differential and
multiplication:
\be
\nn
d^{\rm op}(x):=d(x)~~,~~x\cdot^{\rm op} y:=(-1)^{|x||y|}yx~~.
\ee
A unital dG bimodule is a unital $\Z$-graded bimodule $M$ over the
unital graded associative algebra underlying $A$, together with a
differential $d_M:M\rightarrow M$ of degree $+1$ which satisfies the
compatibility conditions:
\be
\nn
d_M(xm)=(dx)m+(-1)^{|x|} x~d_Mm~~,~~d_M(mx)=(d_Mm)x+(-1)^{|m|} m~d x~~.
\ee
This is the same as a right dG module over the unital dG algebra $A^{\rm op}\otimes A$,
whose differential and multiplication are defined through:
\be
\nn
d^{A^{\rm op}\otimes A} (x\otimes y):=
(d^{\rm op}x)\otimes y+(-1)^{|x|}x\otimes d y~~,~~(x_1\otimes y_1)~~
\cdot^{A^{\rm op}\otimes A}(x_2\otimes y_2):=(-1)^{|x_2||y_1|}x_1x_2\otimes y_1y_2~~;
\ee
the outer multiplications of $M$ are recovered as
$xmy=(-1)^{|x||m|}m(x\otimes y)$.

We let $\dGMod_A$, $_A\dGMod$ and $_A\dGMod_A$ denote the dG
categories of unital right, left and bi- dG modules over $A$; their
morphisms are those morphisms in $\Gr$ which are compatible with the
module structures (though not necessarily with the differentials); the
differentials on morphisms are defined as in (\ref{dif}).  

\paragraph{Dualization.}

One has dualization functors $\dGMod_A{\tiny}\leftrightarrows
_A\dGMod$ and $_A\dGMod_A \rightarrow _A\dGMod_A$ defined
as follows.  Given a unital right dG module $M$ over $A$, consider the
dual complex $(M^{\rm v}, d_{M^{\rm v}})$, endowed with the outer left
multiplication $(x\eta)(m):=(-1)^{|x|(|\eta|+|m|)} \eta(m x)$.  
This makes $M^{\rm v}$ into a unital left dG-module over
$A$. For a unital left dG module $M$, endow $M^{\rm v}$ with the outer
right multiplication $(\eta x)(m):= \eta(m x)$ and with the same
differential as above; this makes it into a unital right dG-module
over $A$.  Given an $A$-bimodule $M$, endow $M^{\rm v}$ with the outer
multiplications $(x\eta y)(m):=(-1)^{|x|(|\eta| +|y|+|m|)} \eta(y m
x)$ for all $x,y\in A$ and $m\in M$ and with the differential
$d_{M^{\rm v}}$; this make it into a unital dG-bimodule over $A$. The
functors $^{\rm v}$ act on morphisms as in equation (\ref{fdual}) of the introduction.
They square to the identity on the corresponding subcategories of
finite-dimensional dG modules.

\paragraph{Tensor product.} 

Given a unital right dG module $M$ and a unital left dG module $N$,
the usual tensor product as modules $M\otimes_A N$ becomes a complex
when endowed with the differential $d_{M\otimes_A
N}(m\otimes_A n)=(d_M m)\otimes_A n+ (-1)^{|m|}m\otimes_A (d_N
n)$. When $M$ (respectively $N$) is a unital dG $A$-bimodule, this
complex is a unital left (resp. right) $A$-module when endowed with
the outer multiplication induced from the left outer multiplication of
$M$ (resp. the right outer multiplication of $N$). It is a unital dG
$A$-bimodule when both $M$ and $N$ are dG bimodules over $A$.

\paragraph{Center of a dG bimodule.}

For any dG bimodule $M$ over $A$, we let $M^A$ denote its center as a
graded $A$-bimodule, i.e. the linear subspace of all elements of $M$
which {\em graded}-commute with all elements of $A$. Notice that $M^A$
is a subcomplex of $M$, as well as a graded central bimodule over the 
graded associative algebra underlying $A$ (together with the induced 
differential, $M^A$ is a `central dG bimodule' over $A$).

\subsection{Cyclic structures on a dGA}
\label{sec:dgcyc}

Given a pairing $\langle ~\rangle:A\otimes A\rightarrow \C[-D]$, the
cyclicity conditions of Section \ref{sec:cyc_ainf} reduce to:
\be
\label{invariance}
\langle x,y\rangle=(-1)^{|x||y|}\langle y, x \rangle~~,~~
\langle dx, y\rangle + (-1)^{|x|} \langle x, dy\rangle = 0~~,~~
\langle xy,z\rangle=\langle x,yz\rangle~~\forall x,y,z\in A~~,
\ee
where we identified the pairing with the corresponding bilinear form. 
Thus a cyclic structure 
on $A$ is the same as a homogeneous `invariant bilinear form', where invariance is understood as compatibility with 
both the differential and multiplication.  Since $A$ is unital, 
we can also describe this through the linear map $\tr:A\rightarrow\C$ defined through 
$\tr(x):=\langle 1,x\rangle=\langle x, 1\rangle$. The last condition in (\ref{invariance}) 
reduces to $\langle x, y \rangle=\tr(xy)$, while the remaining constraints state that $\tr$ is 
an invariant trace: 
\be
\label{tr_invariance}
\tr(xy)=(-1)^{|x||y|}\tr(yx)~~,~~\tr(dx)=0~~.
\ee
The cohomology $H(A):=H_d(A)$ is a unital graded associative algebra with respect to the multiplication 
induced from $A$.
If $\tr^H:H(A)\rightarrow \C$ and $\langle ~,~ \rangle^H:H(A)\times H(A)\rightarrow \C$ denote 
the maps induced on cohomology, then $\tr^H$ is a (possibly degenerate)  
invariant trace on the graded associative algebra $H(A)$, and we have $\langle u,v \rangle^H=\tr^H(uv)$.

Viewing $A$ as a dG-bimodule over itself, consider its
dual dG bimodule $A^{\rm v}$. Then  
equations (\ref{tr_invariance}) state that $\tr$ is a $d_{A^{\rm v}}$-closed central 
element of this bimodule. Generalizing this, we define:

\paragraph{Definition} Let $M$ be a dG bimodule over $A$. 
A {\em $D$-trace} on $M$ is an element $\tr\in Z^{-D}((M^{\rm v})^A)$. 
Explicitly, this is a degree zero linear map $\tr:A\rightarrow\C[-D]$  which obeys:

\

\noindent (1) $\tr(d m)=0 ~~\forall m\in M$

\noindent (2) $\tr(xm)=(-1)^{|x||m|}\tr(mx)~~\forall {\rm~homogeneous~}x\in A~~{\rm and}~~m\in M$~~.

\

\noindent With this definition, we can describe cyclic structures on $A$ as follows. 

\paragraph{Proposition} {Giving a $D$-cyclic structure on $A$ amounts to giving a trace 
$\tr\in Z^{-D}((A^{\rm v})^A)$. }

\
 
The pairing $\langle ~\rangle$ induces a morphism of graded vector spaces  
$\Phi:A[D]\rightarrow A^{\rm v}$ via the relation:
\be
\label{bil_form}
\Phi(x)(y):=\langle x, y \rangle=\tr(xy)~~.
\ee
Notice that the trace can be recovered as $\tr:=\Phi(1)$, where $1$ is
the unit of $A$.  Equations (\ref{invariance}) amount to the condition
that $\Phi$ is a morphism of dG $A$-bimodules from $A[D]$ to $A^{\rm
v}$.  We have $\tr^H=\Phi_*([1])$, where $\Phi_*:H(A)\rightarrow
H(A^{\rm v})$ is the map induced by $\Phi$ on cohomology.  It is clear
that $\langle ~\rangle$ is nondegenerate in the sense of Section \ref{sec:cyc_ainf} iff $A$ is degreewise finite and 
$\Phi$ is bijective.  It is homologically nondegenerate iff $A$ is compact and
$\Phi$ is a quasi-isomorphism. Thus:

\paragraph{Proposition} {Giving a $D$-cyclic structure on a differential 
graded algebra $A$ amounts to giving a morphism of dG bimodules
$\Phi:A[D]\rightarrow A^{\rm v}$. Moreover, the cyclic structure is
nondegenerate iff $\Phi$ is an isomorphism, and homologically
nondegenerate iff $\Phi$ is a quasi-isomorphism.}

\paragraph{Observation}{The dG bimodule $A^{\rm v}$ carries a canonical degree zero trace 
$\theta_A :A^{\rm v}\rightarrow \C$ given by evaluation at the unit of
$A$, i.e. $\theta_A (\eta)=\eta(1)$ for all $\eta\in A^{\rm v}$.
Given a dG-bimodule morphism $\Phi:A[D]\rightarrow A^{\rm v}$, the
trace $\tr_\Phi=\Phi(1)$ it induces on $A$ can be expressed as
$\tr_\Phi=\theta\circ \Phi$.}

\subsection{Homological algebra over a dGA}

In this subsection, we 
recall a few basic results about the derived category of a dGA, which will be used 
later to give a homological characterization of cyclic differential graded algebras. Let $A$ be a unital 
dGA. Since $A$ is a particular type of \ainf algebra, its homological 
algebra can be treated as in the \ainf case; this amounts to considering \ainf modules over 
$A$, \ainf morphisms of \ainf modules etc.  Because this is rather complicated, it is 
advantageous to follow the direct approach of \cite{Keller_dG}, which works instead with dG 
modules over $A$. The price one pays in the direct approach is that a quasi-isomorphism  of dG modules need not be a 
homotopy equivalence of dG modules, hence the dG derived category of $A$ does not coincide with the homotopy category 
of dG modules over $A$. 

\paragraph{The dG derived category.}
 
Recall that $\dGMod_A$ denotes the dG category of unital right dG
modules over $A$.  We let ${\cal C}_{\dG}(A):=Z^0(\dGMod_A)$ be the
Abelian category of right dG modules, and ${\cal
H}_{\dG}(A)=H^0(\dGMod_A)$ be the homotopy category taken in the dG
sense\footnote{This should not be confused with the homotopy category
taken in the \ainf sense. The latter is obtained by working with \ainf
homotopy classes of \ainf morphisms of dG modules.}; the latter is a
triangulated category.  The {\em dG derived category} $D_{\dG}(A)$ of
$A$ is the triangulated category obtained by localizing ${\cal
H}_{\dG}(A)$ with respect to quasi-isomorphisms of dG-modules
\cite{Keller_dG}.  Since unital dG modules over $A$ are particular
instances of strictly unital \ainf modules, one has a faithful 
non-full functor $\dGMod_A\rightarrow \Mod_A$. It was shown in \cite{Hasegawa} that this functor
induces an equivalence between $D_{\dG}(A)$ and the \ainf derived
category $D(A)$ of $A$. Hence we can
view $D_{\dG}(A)$ as an equivalent model of $D(A)$, and throughout
this paper we shall identify it with $D(A)$ via the equivalence
above. Due to this identification, we denote $D_{\dG}(A)$ simply by
$D(A)$.

\paragraph{(P)-resolutions and (I)-resolutions.}
  
The category ${\cal C}_{\dG}(A)$ admits two Quillen model structures
whose weak equivalences are the quasi-isomorphisms.  In the
`projective' model structure, a dG module $M$ is cofibrant iff it has
{\em property (P)} \cite{Keller_dG}, which amounts to ${\cal
H}_{\dG}(M,G)=0$ for all acyclic dG modules $G$. In the `injective'
model structure, $M$ is cofibrant iff it has {\em property (I)}
\cite{Keller_dG}, which amounts to ${\cal H}_{\dG}(G,M)=0$ for all
acyclic $G$. An explicit description of the cofibrant objects of these
two model structures can be found in \cite{Keller_dG}.  The full
subcategories ${\cal H}_p(A)$ (resp. ${\cal H}_i(A)$) of ${\cal
H}_{\dG}(A)$ formed by all dG modules having property (P) (resp. (I))
are triangulated subcategories of ${\cal H}_{\dG}$. For each right
dG-module $M$, there exist triangles in ${\cal H}_{\dG}(A)$ (unique up
to isomorphism): \be \nn pM\rightarrow M\rightarrow aM\rightarrow
(pM)[1]~~,~~ a'M\rightarrow M\rightarrow iM\rightarrow (a'M)[1] \ee
where $aM, a'M$ are acyclic, $pM$ has property (P) and $iM$ has
property (I). Then $pM\rightarrow M$ is called a (P)-resolution of M,
while $M\rightarrow iM$ is called an (I)-resolution. The exact
functors $p:{\cal H}_{\dG}(A)\rightarrow {\cal H}_p(A)$ and $i:{\cal
H}_{\dG}(A)\rightarrow {\cal H}_i(A)$ commute with arbitrary
coproducts and are right respectively left adjoint to the inclusion
${\cal H}_p(A)\subset {\cal H}_{\dG}(A)$.  The localization functor
${\cal H}_{\dG}(A)\rightarrow D(A)$ restricts to exact equivalences
${\cal H}_p(A)\stackrel{\sim}{\rightarrow} D(A)$ and ${\cal
H}_i(A)\stackrel{\sim}{\rightarrow} D(A)$, with inverses given by the
functors induced by $p$ and $i$, which we denote by the same letters.
Thus $\Hom_{D(A)}(M,N)=\Hom_{{\cal H}_{\dG}(A)}(pM,N)= \Hom_{{\cal
H}_{\dG}(A)}(M,iN)$ for all right dG modules $M,N$. This allows one to
construct a homological calculus much as one does for modules over a
unital associative algebra (see \cite{Keller_dG} for details).  Each
of the categories ${\cal C}_{\dG}(A), {\cal H}_{\dG}(A), D(A), {\cal
H}_p(A)$ and ${\cal H}_i(A)$ has infinite coproducts.

\paragraph{Description of $\tria(A)$ and $\per(A)$ through dG modules.}

Let ${\hat A}$ be $A$ viewed as a (unital) right dG module over
itself.  As in the \ainf case define $\tria(A):=\tria_{D(A)}({\hat
A})$ and $\per(A):=\ktria_{D(A)}({\hat A})$ (see Appendix
\ref{sec:gens}) to be the smallest triangulated (resp triangulated and
idempotent complete) strictly full subcategories of $D(A)$ containing
${\hat A}$; these can be identified with the categories denoted by the
same symbols but defined in the \ainf sense.  Thus ${\hat A}$ is a
compact generator of $D(A)$, in particular $D(A)=\Tria_{D(A)}(A)$ etc.
When $A$ is concentrated in degree zero, $\per(A)$ is the category of
perfect complexes ($=$ complexes quasi-isomorphic with bounded
complexes of finitely generated projective modules), while $\tria(A)$
consists of complexes quasi-isomorphic with bounded complexes of
finitely-generated free modules.

\subsection{Serre duality on $\tria(A)$ and $\per(A)$} 
\label{sec:Serre}

Fixing a unital differential graded algebra $A$, we let
$\nu:=(.)\otimes^L_A A^{\rm v}:D(A)\rightarrow D(A)$ be the left
derived functor \cite{Keller_dG} of tensorization from the right with
the $A$-bimodule $A^{\rm v}$. This is defined by $\nu(M)=M\otimes^L_A
A^{\rm v}:=\pi[(pM)\otimes_A A^{\rm v}]$, where $\pi$ is the canonical
surjection ${\cal H}_{\dG}(A)\rightarrow D(A)$. Notice that $\nu$ commutes 
with arbitrary coproducts. 

In this section, we often write ${\hat A}$ simply as $A$ in order to simplify notation. 
Consider the linear maps $\beta_n:=\nu_{{\hat A}, {\hat
A}[n]}:\Hom_{D(A)}(A,A[n])\rightarrow \Hom_{D(A)}(\nu(A),\nu(A[n]))$
defined by the functor $\nu$. Recall that $A$ is called compact if
$H^n(A)$ is finite-dimensional for all $n\in \Z$.

\paragraph{Proposition}{$A$ is compact iff each of the maps $\beta_n$ is bijective.}

\

\noindent {\em Proof}. For any $M$ in $\dGMod_A$, we have a natural isomorphism of complexes
$\Hom_{\dGMod_A}(A,M)\approx M$  (given by evaluation at $1$) and a natural isomorphism of complexes
 $\Hom_{\dGMod_A}(M,A^{\rm v})\approx M^{\rm v}$ given by  
$\phi\rightarrow \eta_\phi$, where $\eta_\phi$ is the functional  on $M$ given by $\eta_\phi(m)=\phi(m)(1)$. 
For any acyclic right dG module $G$ over $A$,  these isomorphisms show that $\Hom_{\dGMod_A}(A,G)$
and $\Hom_{\dGMod_A}(G,A^{\rm v})$ are acyclic. Thus $\Hom_{{\cal H}_{\dG}(A)}(A,G)=\Hom_{{\cal H}_{\dG}(A)}(G,A^{\rm v})=0$, 
which show that $A$ has property (P) and $A^{\rm v}$ has property (I). Of course, the same is true for 
$A[n]$ and $A^{\rm v}[n]$ given any integer $n$. 

Since $A$ has property (P), we have $pA=A$  
and $\nu(A[n])=A[n]\otimes_A A^{\rm v}\approx A^{\rm v}[n]$ for all $n\in \Z$. Hence $\beta_n$ can be viewed as linear maps
$\beta_n:\Hom_{D(A)}(A,A[n])\rightarrow \Hom_{D(A)}(A^{\rm v}, A^{\rm v}[n])$.
Moreover, we have: 

\

\noindent $\Hom_{D(A)}(A,A[n])=\Hom_{{\cal H}_{\dG}(A)}(A,A[n])= H^0(\Hom_{\dGMod_A}(A,A[n]))\approx H^0(A[n])=H^n(A)$ 

\

\noindent and: 

\

\noindent $\Hom_{D(A)}(A^{\rm v},A^{\rm v}[n])=\Hom_{{\cal H}_{\dG}(A)}(A^{\rm v},A^{\rm v}[n])=
H^0(\Hom_{\dGMod_A}(A^{\rm v},A^{\rm v}[n]))=$\\
\noindent 
$=H^0(\Hom_{\dGMod_A}(A^{\rm v}[-n],A^{\rm v}))=H^0(\Hom_{\dGMod_A}(A[n]^{\rm v},A^{\rm v}))
\approx H^0(A[n]^{\rm vv})\approx H^{n}(A)^{\rm vv}$, 

\

\noindent where the first equalities in each chain follow from $pA=A$ and $i (A^{\rm v}[n])=A^{\rm v}[n]$. 

Combining everything,  we see that $\beta_n$ identify with the linear maps
$\gamma^H_n:H^n(A)\rightarrow H^n(A^{\rm vv})\approx [H^n(A)]^{\rm vv}$ 
induced by the dG bimodule morphism  $\gamma:A\rightarrow A^{\rm vv}$,
$\gamma(a)(\eta)=(-1)^{|a||\eta|}\eta(a)$. Since $\gamma_n^H$ coincides up to sign with 
the bidualization morphism of the vector space $H^n(A)$, we know that it 
is bijective iff $H^n(A)$ is finite-dimensional for all $n\in \Z$.  
It follows that all $\beta_n$ are bijective iff $A$ is compact.

\paragraph{Observation} The following are equivalent: 

\noindent (a) $A$ is compact

\noindent (b) $\tria(A)$ is Hom-finite

\noindent (c) $\per(A)$ is Hom-finite

\

\noindent {\em Proof}. The equivalence $(b)\Leftrightarrow (c)$ follows trivially from $\per(A)=\tria(A)^\pi$. The equivalence 
$(a)\Leftrightarrow (b)$ follows from $\tria(A)=H^0(\tw(A))$ upon using the equivalence 
$A={\rm compact} \Leftrightarrow \tw(A)={\rm compact}$. One can also prove the equivalence $(a)\Leftrightarrow (b)$ directly by 
using $\Hom_{D(A)}=H^{n-m}(A)$ and the fact that 
$\Hom(.,.)$ is a  cohomological functor in the first variable and a homological functor in the second.

\paragraph{Lemma} The following conditions are 
equivalent: 

\noindent (a) $\nu:D(A)\rightarrow D(A)$ is fully faithful

\noindent (b) $A$ is compact and $A^{\rm v}$ belongs to $\per(A)$ (in particular, $\nu$ preserves $\per(A)$). 

\noindent In this case, $\nu$ restricts to an autoequivalence of $\per(A)$ iff $A^{\rm v}$ is a Karoubian generator of $\per(A)$.

\

\noindent {\em Proof}. Recall from \cite{Keller_dG} that an object of
$D(A)$ is compact iff it belongs to $\per(A)$.  Using this fact, a
result of \cite[paragraph 4.2]{ Keller_dG} states that an exact
functor $F:D(A)\rightarrow D(A)$ which commutes with arbitrary
coproducts is fully faithful iff:

\noindent ($\alpha$) $F({\hat A})$ belongs to $\per(A)$

\noindent ($\beta$) the maps $F_{{\hat A},{\hat
A}[n]}:\Hom_{D(A)}({\hat A}, {\hat A}[n])\rightarrow
\Hom_{D(A)}(F({\hat A}), F({\hat A}[n]))$ are bijective for all $n\in
\Z$.

\noindent Applying this to $F=\nu$, the previous Proposition shows
that condition $(\beta)$ is equivalent with compactness of $A$, while
condition $(\alpha)$ amounts to the requirement that $A\dual=\nu({\hat
A})$ belongs to $\per(A)$. In this case, $\nu$ preserves $\per(A)$
since ${\hat A}$ is a Karoubian generator of the latter. The last
statement of the lemma is obvious.

Recall that a {\em Serre functor} on a Hom-finite triangulated category $\cT$ is an exact autoequivalence $S$ of $\cT$ 
together with isomorphisms $\Hom_{\cT}(a,b)\approx \Hom_{\cT}(b,S(a))$. In this case, $S$ is unique up to isomorphism of functors 
\cite{BK_Serre}.

\paragraph{Proposition} The category $\per(A)$ is Hom-finite and has a Serre functor $S$ 
iff the following conditions are satisfied:

\noindent (1) $A$ is compact

\noindent (2) $A^{\rm v}$ belongs to $\per(A)$ and is a Karoubian
generator of the latter.

\noindent In this case, we have $S\approx \nu|_{\per(A)}=(.)\otimes^L_A A^{\rm v}$.

\

\noindent {\em Proof}. ($\Leftarrow$) Assume that (1) and (2) hold. Since a Serre 
functor is unique up to isomorphism, it suffices to show that $\nu|_{\per(A)}$ is a Serre functor on $\per(A)$. 
By the Lemma, assumptions (1) and (2) imply that $\nu$ restricts to an autoequivalence of $\per(A)$. 
Given $P$ in $\per(A)$ and $M$ in $D(A)$, we have natural isomorphisms:
\be
\nn
\RHom_A(P,M)^{\rm v}\approx \RHom_A(M,\RHom_A(P,A)^{\rm v})~~,~~
\RHom_A(P,A)^{\rm v}\approx P\otimes_A^L A^{\rm v}~~, 
\ee
which it suffices to check for $P=A[n]$, when they hold trivially. Combining these gives:
\be
\nn
\RHom_A(P,M)^{\rm v}
\approx \RHom_A(M,P\otimes_A^L A^{\rm v})~~\forall P \in \Ob[\per(A)]~,~\forall M\in \Ob D(A)~~.
\ee
Taking $P=P_1\in \Ob[\per(A)]$ and $M=P_2\in \Ob[\per(A)]$ and applying $H^0$ gives natural isomorphisms: 
\be
\nn
\Hom_{D(A)}(P_1,P_2)^{\rm v}\approx \Hom_{D(A)}(P_2,\nu(P_1))~~.
\ee
Since $A$ is compact, the category $\per(A)$ is Hom-finite. 
Hence dualizing the last equation gives: 
\be
\label{S_isom}
\Hom_{D(A)}(P_1,P_2)\approx \Hom_{D(A)}(P_2,\nu(P_1))^{\rm v}~~,
\ee
which shows that $\nu|_{\per(A)}$ is a Serre functor. 

($\Rightarrow$) Assume that $\per(A)$ is Hom-finite with Serre functor $S$. Hom-finiteness of $\per(A)$ implies (1) by the observation 
above. To prove (2), start by combining (\ref{S_isom}) with the Serre isomorphism:
\be
\label{Serre}
\Hom_{D(A)}(P_1,P_2)\approx \Hom_{D(A)}(P_2,S(P_1))^{\rm v}~~\forall P_1, P_2 \in \Ob[\per(A)]~~,
\ee
which gives natural isomorphisms:
\be
\label{Snu}
\Hom_{D(A)}(P_2,S(P_1))\approx \Hom_{D(A)}(P_2,\nu(P_1))~~\forall ~P_j\in \Ob[\per(A)]~~.
\ee
Applying this for $P_2:=P$ and $P_1=A$ gives isomorphisms:
\be
\label{SA}
\Hom_{D(A)}(P,S(A)){\tiny \begin{array}{c}\phi_P\\\longrightarrow\\ \approx \end{array}} 
\Hom_{D(A)}(P,A^{\rm v})~~\forall ~P\in \Ob[\per(A)]~~
\ee
which are natural in $P$. Setting $P=S(A)$ in (\ref{SA}) gives a morphism $\theta=\phi_{S(A)}(\id_{S(A)})\in \Hom_{D(A)}(S(A),A^{\rm v})$. 
Using this, we define maps 
\be
\psi_M:\Hom_{D(A)}(M,S(A))\rightarrow \Hom_{D(A)}(M,A^{\rm v})~~\forall M\in \Ob D(A)
\ee
by setting $\psi_M(u):=\theta\circ u$ for all  $u\in \Hom_{D(A)}(M,S(A))$; these are clearly natural in $M$. 
Given $u\in \Hom_{D(A)}(P,S(A))$, we have 
$\phi_P(u)=\phi_P(\id_{S(A)}\circ u)=\phi_P(\id_{S(A)})\circ u=\theta\circ u=\psi_P(u)$ by naturality of $\phi_P$. Thus 
$\psi_P=\phi_P$ for all $P\in \Ob[\per(A)]$. Consider the full subcategory ${\cal T}$ of $D(A)$ whose objects are those 
$M\in \Ob D(A)$ for which $\psi_M$ is an isomorphism. This is a triangulated subcategory by the 5-lemma. 
Moreover, it is closed under taking arbitrary coproducts. Indeed, given $M_\alpha\in \Ob\cT$, 
we have  
$\Hom_{D(A)}(\sqcup_{\alpha} M_\alpha, S(A))\approx \prod_{\alpha} \Hom_{D(A)}(M_\alpha, S(A))\stackrel{\prod_\alpha\psi_{M_\alpha}}
{\rightarrow} \prod_{\alpha}\Hom_{D(A)}(M_\alpha, A^{\rm v})$
$\approx \Hom_{D(A)}(\sqcup_{\alpha} M_\alpha, A^{\rm v})$, where the first and last 
isomorphisms follow from the definition of the categorical coproduct and the map in the middle is bijective 
because $\psi_{M_\alpha}$ are. Since $D(A)=\Tria(A)$ and $\cT$ contains $A[n]$, it follows that $\cT=D(A)$, so $\psi_M$ 
are bijective for all objects $M$ of $D(A)$. The Yoneda lemma now shows that $\theta$ is an isomorphism, so 
$A^{\rm v}$ is isomorphic with $S(A)$ in $D(A)$. This implies that $A^{\rm v}=\nu(A)$ belongs to $\per(A)$ 
because the latter is a {\em strictly} full subcategory of $D(A)$. Since $\nu$ is exact and $A^{\rm v}$ a Karoubi generator of $\per(A)$, 
we find that $\nu$ preserves $\per(A)$. We can now apply the Yoneda lemma to (\ref{Snu}). This gives an isomorphism 
of functors $\nu|_{\per(A)}\approx S$, showing that $\nu|_{\per(A)}$ is an autoequivalence of $\per(A)$. 
Thus $A^{\rm v}$ is a Karoubi generator of $\per(A)$, which completes the proof of (1).

\paragraph{Lemma} The following conditions are 
equivalent: 

\noindent (a) $\nu$ preserves $\tria(A)$ and restricts to an autoequivalence of $\tria(A)$

\noindent (b) $A$ is compact and $A^{\rm v}$ belongs to $\tria(A)$ and is a triangle generator of the latter.

\

\noindent {\em Proof}.  A result of \cite[paragraph 4.2]{ Keller_dG} shows that $\nu|_{\tria(A)}:\tria(A)\rightarrow D(A)$ is 
fully faithful iff $\beta_n$ are bijective for all $n\in \Z$, which amounts to compactness of $A$. On the other hand, $\nu$ preserves 
$\tria(A)$ iff $\nu(A)=A^{\rm v}$ belongs to $\tria(A)$. The conclusion now follows.

\

\paragraph{Proposition} The category $\tria(A)$ is compact and has a Serre functor $S$ 
iff the following conditions are satisfied: 

\noindent (1) $A$ is compact 

\noindent (2) $A^{\rm v}$ belongs to $\tria(A)$ and is a triangle generator of the latter. 

\noindent In this case, we have $S\approx (.)\otimes^L_A A^{\rm v}$.

\

\noindent {\em Proof}.  Virtually identical to that of the previous proposition.

\paragraph{Corollary} The following are equivalent:

\noindent (a) The category $\per(A)$ is $D$-Calabi-Yau

\noindent (b) The category $\tria(A)$ is $D$-Calabi-Yau

\noindent (c) $A$ is compact and admits a homologically nondegenerate
$D$-cyclic pairing.

\

\noindent {\em Proof}. $(a)\Rightarrow (b)$ Obvious.

~~~~~$(b) \Rightarrow (c)$ If $\tria(A)$ is $D$-Calabi-Yau, then the
previous proposition implies that $A$ is compact and $\nu|_{\tria(A)} \approx
[D]$.  Applying this to the generator $A$, we find $\nu(A)=A^{\rm
v}\approx A[D]$ in $D(A)$. Since $\Hom_{D(A)}(A[D],A^{\rm v})=H^0(\Hom_{_A\dGMod_A}(A[D],A^{\rm v}))$ 
(because $A[D]$ has property (P)), this means that there exists a quasi-isomorphism from $A[D]$ to $A^{\rm v}$.  
The results of Section \ref{sec:dgcyc} imply that $A$ carries
a homologically nondegenerate $D$-cyclic structure (which induces a Serre pairing on 
$\tria(A)$).  

~~~$(c)\Rightarrow (a)$ If $A$ is homologically nondegenerate
$D$-cyclic and compact, then $A[D]$ is quasi-isomorphic with $A^{\rm v}$ as 
shown in Section \ref{sec:dgcyc}. Thus $A^{\rm v}\approx A[D]$ in $D(A)$, which
implies that $A^{\rm v}$ belongs to $\per(A)$ and is a Karoubian generator of the latter. By a
previous proposition, we find that $\per(A)$ has Serre duality with
Serre functor $\nu|_{\per(A)}= (.) \otimes^L_A A^{\rm v}\approx (.)
\otimes^L_A A[D]\approx (.) \otimes_A A[D]\approx [D]$, where we used
the fact that perfect dG modules have property (P). Thus $\per(A)$ is $D$-Calabi-Yau.

\subsection{Reconstruction of Serre pairings}
\label{sec:mingens}

\

\noindent The results of Sections \ref{sec:extension} and \ref{sec:Serre} give:

\paragraph{Proposition}{Let $A$ be a unital differential graded algebra. Then the following statements are equivalent: 

\noindent (a) $A$ is compact and admits a homologically nondegenerate $D$-cyclic structure 

\noindent (b) $\tria(A)$ is Hom-finite and $D$-Calabi-Yau. 

In this case, any homologically nondegenerate $D$-cyclic structure on $A$ induces a Serre duality structure on $\tria(A)$
via the construction of Section \ref{sec:extension}.}

\

\noindent{\em Proof}.  The implication $(a)\Rightarrow (b)$ follows from the results of  
Section \ref{sec:extension}, while the inverse implication follows from the previous subsection. 
The rest follows from Section \ref{sec:extension}.

\

\noindent On the other hand, Section \ref{sec:minmodel} (see Appendix
\ref{sec:cyc_geom} for a different point of view) gives the following:

\paragraph{Proposition}{Let $(A, \langle ~\rangle)$ be a compact $D$-cyclic 
and strictly unital dGA whose cyclic pairing is homologically
nondegenerate. Then there exists a finite-dimensional unital \ainf
minimal model $A_{\rm min}$ of $A$ which admits a 
nondegenerate $D$-cyclic pairing. For example, one can pick the cyclic
minimal model induced by a strictly unital and cyclic cohomological
splitting of $A$. }

\

\noindent Combining with the previous result, we obtain:

\paragraph{Proposition}{ Let $A_{\rm min}$ be a unital and minimal \ainf algebra
such that $\tria(A_{\rm min})$ is Hom-finite and $D$-Calabi-Yau. Then $A_{\rm min}$
is finite-dimensional and there exists a unital and minimal \ainf
algebra $A'_{\rm min}$ which is \ainf isomorphic with $A_{\rm min}$
and admits a nondegenerate $D$-cyclic structure which induces a Serre
duality structure on $\tria(A'_{\rm min})\approx \tria(A_{\rm min})$ via the
construction of Section \ref{sec:extension}.}

\

\noindent{\em Proof}.  It follows from \cite[Proposition
7.5.0.2]{Hasegawa} that there exists a strictly unital differential
graded algebra $A$ such that $A_{\rm min}$ is a minimal model of $A$.
By the first proposition above, $A$ is compact and admits a
homologically nondegenerate $D$-cyclic structure. The second
proposition gives a finite-dimensional, unital minimal model $A'_{\rm
min}$ of $A$ which admits a nondegenerate $D$-cyclic structure.  This
must be \ainf isomorphic with $A_{\rm min}$ since the minimal model of
$A$ is determined up to \ainf isomorphism.  The extension procedure of
Section \ref{sec:extension} shows that the cyclic structure of
$A'_{\rm min}$ induces a shift-invariant and nondegenerate $D$-cyclic structure on
$\tw(A_{\rm min})$ and thus a Serre pairing on $\tria(A'_{\rm
min})\approx \tria(A_{\rm min})$.

\paragraph{Observation} As explained in Appendix
\ref{sec:cyc_geom}, the $D$-cyclic structure of the dG algebra $A$
considered in the proof pulls back to a nondegenerate `symmetric
\inf-inner product' on $A_{\rm min}$ via any quasi-isomorphism
$A_{\rm min}\rightarrow A$; this is a generally infinite collection
of multilinear forms on $A_{\rm min}$, the first of which is a
nondegenerate bilinear pairing. The `noncommutative Darboux theorem'
discussed in Appendix (\ref{sec:cyc_geom}) implies that one can find a
$D$-cyclic minimal model $A'_{\rm min}$ such that the pull-back of the
\inf-inner product through an \ainf isomorphism $A'_{\rm
min}\rightarrow A_{\rm min}$ reduces to a $D$-cyclic structure on
$A'_{\rm min}$, i.e. all of its higher multilinear forms vanish. This
gives a different proof of the theorem above (see Appendix
\ref{sec:cyc_geom}).

\

Recall that any two Serre pairings on $\cT:=\tria(A_{\rm min})$
are equivalent in the sense that they are related by a shift-invariant
automorphism of the identity functor of $\cT$ (see Appendix
\ref{sec:duality}). It follows that any Serre pairing on
$\tria(A_{\rm min})$ is induced from a nondegenerate cyclic structure
on $A_{\rm min}$ up to such an equivalence. The equivalence 
$\tria(A)\approx \tria(A'_{\rm min})$ in the proposition above also induces 
an equivalence of cyclic structures. We refer the reader to Appendix \ref{sec:duality} 
for details about the transport of cyclic structures through functors.

\section{Generating the superpotential}

In this section we give the construction of the \ainf prolongation promised in the introduction. 
We start by discussing a result of \cite{Keller_dG} and its generalization due to \cite{Hasegawa}.

\subsection{$A_\infty$ generators of a triangulated category}

Following \cite{Keller_dGcat}, we say that a triangulated category
$\cT$ is {\em algebraic} if it is triangle equivalent with the
stable category of a Frobenius category. Given $g\in \Ob \cT$, we
set $H_g:=\Hom_{\cT^\bullet}(g,g)=\oplus_{n\in \Z}{\Hom_{\cal
T}(g,g[n])}$, viewed as a unital graded associative algebra with the
composition induced from $\cT^\bullet$.  Let $\GrMod_{H_g}$ be the
category of graded right modules over $H_g$.  Consider the functor
${\bar F}_g:\cT\rightarrow \GrMod_{H_g}$ which sends $a\in \Ob
\cT$ to the right graded $H_g$-module ${\bar F}_g(a)=\Hom_{{\cal
T}^\bullet}(g,a)$.  One has the following result:

\paragraph{Proposition \cite{Keller_dG, Keller_dGcat}}{Assume that $\cT$ is 
algebraic and let $g\in \Ob \cT$. Then there exists a unital dG
algebra $A$ such that $H(A)$ is isomorphic to $H_g$ and an exact
functor $F:\cT\rightarrow D(A)$ mapping $g$ into ${\hat A}$ such
that $H\circ F$ is isomorphic with ${\bar F}$ (here $H:D(A)\rightarrow
\GrMod_{H(A)}$ is the functor obtained by taking total cohomology).
Moreover, $F$ induces an equivalence from $\cT$ to $\tria(A)$ iff
$g$ triangle generates $\cT$, i.e. $\tria_\cT(g)={\cal
T}$.}

\

Recall from \cite[Proposition 3.2.4.1]{Hasegawa} that a strictly
unital \ainf algebra admits a strictly unital minimal model related to
the original algebra by a strictly unital quasi-isomorphism.  Since
pullback through quasi-isomorphism induces an equivalence of derived
categories \cite[Theorem 4.1.2.4]{Hasegawa}, one finds that a strictly
unital minimal model $A_{\rm min}$ of $A$ gives triangle equivalences $D(A)\approx D(A_{\rm
min})$, $\per(A)\approx \per(A_{\rm min})$ and $\tria(A)\approx
\tria(A_{\rm min})$.  This gives the following version of the result above:

\paragraph{Proposition \cite{Hasegawa}}{Assume that $\cT$ is 
algebraic and let $g\in \Ob \cT$ such that $\tria_\cT(g)=\cT$. 
Then there exists a unital minimal
\ainf algebra $(A_{\rm min},(\mu_n)_{n\geq 2})$ such that the associative algebra 
$(A_{\rm min},\mu_2)$ is isomorphic with $H_g$ and an exact functor $F:\cT\rightarrow D(A_{\rm min})$ mapping 
$g$ into ${\hat A}_{\rm min}$, which corestricts to an equivalence from $\cT$ to $\tria(A_{\rm min})$.

\

A unital minimal \ainf algebra $A_{\rm min}$ 
as in the proposition will be called a {\em minimal \ainf generator} of $\cT$. 
Given such a generator, any minimal and unital \ainf algebra $A'_{\rm min}$  
isomorphic with $A_{\rm min}$ is again an \ainf generator. Using the results of Section \ref{sec:mingens}, we obtain the 
following `cyclic' variant of the previous proposition.

\paragraph{Proposition}{{Assume that $\cT$ is 
algebraic and Calabi-Yau of dimension $D$ and let $g\in \Ob \cT$.
If $\tria_\cT(g)=\cT$, then there exists a unital minimal
$D$-cyclic \ainf algebra $(A_{\rm min}, \langle ~\rangle)$ whose
pairing is nondegenerate such that $\tria(A_{\rm min})\approx \cT$ via a
triangle equivalence mapping ${\hat A}_{\rm min}$ to $g$. Moreover,
any Serre pairing on $\cT$ is equivalent with the Serre pairing
on $\tria(A_{\rm min})$ induced by $\langle ~\rangle$ via the procedure of
Section \ref{sec:extension}. }

\

A unital, minimal and nondegenerate cyclic \ainf algebra $A_{\rm min}$ as in the proposition will be called a 
{\em cyclic minimal \ainf generator for $\cT$.} 

\subsection{The open string field action determined by a cyclic minimal \ainf generator}

Let $\cT$ be a triangulated algebraic category which is Calabi-Yau of dimension $D$, 
and assume given $g\in \Ob \cT$ such that $\tria_\cT(g)=\cT$. Fix a cyclic and unital minimal 
\ainf generator $(A_{\rm min},\langle ~\rangle)$ associated with $g$, and view $A_{\rm min}$ as the morphism 
space $\Hom_{\cA_0}(a,a)$ for an \ainf category $\cA_0$ having a single object $a$. Then ${\hat a}$ is identified with ${\hat A}_{\rm min}$ etc. 
Since the pairing on $A_{\rm min}$ is (strictly) nondegenerate, the same is true of the shift-equivariant cyclic structure 
induced on $\tw(A_{\rm min})=\tw(\cA_0)$ via the extension procedure of Section \ref{sec:extension}. 
Thus $\cA:=\tw(A_{\rm min})=\tw(\cA_0)$ is a unital and strictly nondegenerate $D$-cyclic \ainf category  with a twisted shift functor 
(which of course need not be minimal). This allows us to define a formal topological string field action 
as in section \ref{sec:sft}. Consider the {\em total boundary state space} 
${\cal H}_\cA:=\oplus_{q,q'\in \Ob[\tw(A_{\rm min})]}\Hom_{\tw(A_{\rm min})}(q,q')$, endowed with the 
total bilinear pairing $\langle~\rangle$ and total \ainf products 
$r_n$. As in Section \ref{sec:sft}, pick a Grassmann algebra $G$ and 
define a formal extended open string field action $S_e:({\cal H}_\cA)_e^{\rm odd}\rightarrow G$ by the formal sum (\ref{Sfa}).

\paragraph{The generation property.}

From Section \ref{sec:sft}, we know that $\Hom_{\tw(A_{\rm
min})}(q,q')$ provides an `off-shell model' for the space of boundary
observables $H(\Hom_{\tw(A_{\rm min})}(q,q'))\approx \Hom_{{\cal
T}^\bullet} (q,q')$ of the open string stretching from $q$ to $q'$,
while the \ainf products $r^{\tw(A_{\rm min})}_{q_0\ldots q_n}$ are
string field products. The discussion of Section \ref{sec:sft} implies:

\

{\em Every twisted complex $q\in \cA=\tw(\cA_0)=\tw(A_{\rm min})$ is the result of a
condensation process involving open strings stretching between a
finite number of shifted copies $a[[n_k]]$ of the
D-brane described by $a$. In this sense, condensation
processes among the D-branes $a[[n]]$ generate the
entire \ainf category $\cA$.}

\

Notice that the string field action (\ref{Sfa}) is {\em entirely}
determined by the minimal cyclic \ainf generator $(A_{\rm min},
\langle~\rangle)$ of $\cT$.  Indeed, we just showed that $a$ generates our D-brane category $\cA$. On
the other hand, the cyclic pairing on $\cA$ is induced by
the pairing on $A_{\rm min}=\Hom_{\cA}(a,a)$ through the extension procedure of
Section \ref{sec:extension}. Hence the entire open string field theory
is determined by $(A_{\rm min},\langle~\rangle)$. Thus:

\

{\em Every minimal cyclic and unital \ainf generator $A_{\rm min}$ of
an algebraic Calabi-Yau triangulated category defines a topological
open string field theory governing the dynamics of a topological
D-brane system whose zeroth cohomology as an \ainf category recovers
the triangulated category $\cT$. This topological D-brane system
consists of topological D-brane composites which can be obtained as
condensates between a finite number of shifted copies of a single
topological D-brane.  }

\subsection{The induced prolongation and superpotential}

The formal string field action introduced above determines a cyclic
minimal model of $(\tw(A_{\rm min}), \langle~\rangle)$ and thus
an extended `superpotential' for $H(\tw(A_{\rm
min}))=\tria(A_{\rm min})^\bullet \approx \cT^\bullet$ via the
construction of Sections \ref{sec:minmodel} and \ref{sec:sft}. As
explained in Section \ref{sec:sft}, the shift-equivariant pairing
$\langle~\rangle^{\tria(A_{\rm min})}$ determined by the pairing
induced by $\langle~\rangle$ on $\tw(A_{\rm min})$ is equivalent with the
shift-equivariant pairing $(~)$ induced on $\cT^\bullet$ by the
original Serre pairing of $\cT$.  Since everything is determined
by the cyclic \ainf generator $(A_{\rm min},\langle~\rangle)$, we
conclude:

\

{\em Every minimal cyclic \ainf generator $A_{\rm min}$ of an
algebraic Calabi-Yau triangulated category $\cT$ defines a Serre
pairing on $\cT$ together with a cyclic \ainf prolongation of the
resulting cyclic graded associative category $\cT^\bullet$ (and
thus determines an extended `superpotential' for $\cT$). These
can be constructed explicitly via the extension procedure of Section
\ref{sec:extension} and the procedure of Sections
\ref{sec:minmodel}. }

\

This result allows one to lift the open 2d topological field theory
described by $\cT$ to a topological open string theory.

\appendix

\section{Categories with shifts, duality structures and cyclic structures}
\label{sec:duality}

In this Appendix, we discuss duality structures and cyclic structures on associative
categories with shifts.  This is a slight generalization of the usual
theory of Serre functors \cite{BK_Serre}, obtained by relaxing the
nondegeneracy condition. We are interested in the description through
pairings and traces, which affords a direct link with physics.  The
treatment of signs is inspired by \cite{Bocklandt}.  In order to keep
the discussion reasonably short, we leave the proof of most statements
to the reader -- they are straightforward diagram arguments,
though a few are somewhat lengthy. 

\subsection{Associative and graded associative categories with shifts}

\paragraph{Associative categories with shifts.}

An {\em associative category with shifts} is a pair $(\cA, [1])$ where $\cA$ is a 
(possibly non-unital) associative category and $[1]:\cA\rightarrow \cA$ is a fixed automorphism
of $\cA$, called the {\em shift functor}. Given a category with shifts, we identify 
$\Hom_{\cA}(a,b)$ and $\Hom_{\cA}(a[1],b[1])$ through the linear isomorphism induced by $[1]$. 
We let $[n]:=[1]^n$ for all $n\in \Z$, where $[1]^0=[0]=\id_\cA$ is the identity automorphism 
of $\cA$ and $[1]^{-1}=[-1]$ is the inverse of $[1]$. When $\cA$ is unital, the shift functor 
satisfies $\id_a[n]=\id_{a[n]}$ for all objects $a$ and all $n\in \Z$. 
Small associative categories with shifts form an associative category ${\rm SCat}$
whose morphisms $(\cA,[1]_\cA)\rightarrow (\cB,[1]_\cB) $ are the {\em shift-invariant functors}, 
i.e. functors $F:\cA\rightarrow \cB$ obeying $F\circ [1]_\cA = [1]_\cB\circ F$.

\paragraph{Graded associative categories with shifts.}

Let $\cG$ be a graded associative category. A {\em shift functor} (in the sense of graded 
categories) on $\cG$ is an automorphism $[1]$ of $\cG$ together with isomorphisms 
$\Hom_\cG(a,b[1])\stackrel{\rho_{ab}}{\rightarrow} \Hom_\cG(a,b)[1]$ for all $a,b\in \Ob \cG$, which are natural in $a$ and $b$. 
Equivalently, $\rho:\Hom_{\cG}\circ (\id_\cA\times [1])\rightarrow [1]_{\Gr}\circ \Hom_{\cG}$ is an isomorphism of functors, 
where $[1]_{\Gr}$ is the shift functor of the category of graded vector spaces $\Gr$. 
In this case, the pair $(\cG,[1])$ is called a 
{\em graded category with shifts}.  We let $[n]:=[1]^n$ as before. Small graded associative 
categories with shifts form an associative category ${\rm SGrCat}$ 
whose morphisms are the shift-invariant functors of graded categories, i.e. functors of 
graded categories which strictly commute with shifts. 

\paragraph{Equivalence of ${\rm SCat}$ and ${\rm SGrCat}$.}

It is easy to see that the categories ${\rm SCat}$ and ${\rm SGrCat}$ are equivalent. A pair 
of quasi-inverse equivalences is given by the functors $^0:{\rm SCat}\rightarrow {\rm SGrCat}$ and 
$^\bullet:{\rm SGrCat}\rightarrow {\rm SCat}$ constructed as follows. 
Given $(\cG,[1])\in {\rm SGrCat}$, we 
let $\cG^0$ be the {\em null restriction} of $\cG$, i.e. the category 
obtained from $\cG$ by keeping only morphisms of degree zero. Thus $\Ob\cG^0=\Ob\cG$
and 
$\Hom_{\cG^0}(a,b)=\Hom^0_\cG(a,b)$ for all objects $a,b$. Then $\cG^0$ 
is an associative category with shifts, with shift functor given by restricting the action of $[1]$ on morphisms. 
We let $^0$ act in the obvious manner on morphisms of ${\rm SGrCat}$. 
Conversely, given $(\cA,[1])$ in ${\rm SCat}$, we define its 
{\em graded completion} $\cA^{\bullet}\in {\rm SGrCat}$ as follows. We take 
$\Ob\cA^{\bullet}=\Ob\cA$ and $\Hom_{\cA^\bullet}(a,b)=\oplus_{k\in \Z}\Hom_{\cA}(a,b[k])$ for 
all objects $a,b$, with the composition of morphisms: 
\be
\label{star}
g*f=\bigoplus_{n\in \Z} \sum_{k+l=n}{g_l[k]\circ f_k}~~\forall 
f=\oplus_{k\in \Z}{f_k}\in \Hom_{\cA^\bullet}(a,b)~~\forall 
g=\oplus_{l\in \Z}{g_l}\in \Hom_{\cA^\bullet}(b,c)~~,
\ee
where $f_k\in \Hom_{\cA}(a,b[k])$ and $g_l\in \Hom_{\cA}(b,c[l])$. The
morphism spaces of $\cA^{\bullet}$ are graded with homogeneous
components $\Hom^k_{\cA^\bullet}(a,b):= \Hom_{\cA}(a,b[k])$. The
compositions (\ref{star}) have degree zero, so $\cA^\bullet$ is a
graded associative category.  The relations
$\Hom_{\cA}(a[1],b[1])\approx \Hom_{\cA}(a,b)$ imply that
$\cA^\bullet$ is a graded category with shifts.  We let $^\bullet$ act
in the obvious manner on morphisms of ${\rm SCat}$.  The category
$\cA^\bullet$ is sometimes denoted by $\cA/[1]$ (the `quotient' of
$\cA$ by the group of automorphisms generated by $[1]$).  It is clear
that $^0$ and $^\bullet$ interchange unital associative categories
with unital graded associative categories.

\paragraph{Twisting the shift functor of a graded associative category with shifts.}

Given a graded associative category with shifts $(\cG, [1])$, we
can define a new endofunctor $[[1]]$ of $\cG$ as follows. We
let $[[1]]$ act on objects via $a[[1]]:=a[1]$ and on morphisms $f\in
\Hom_\cG(a,b)$ through: \be \nn f[[1]]:=(-1)^{\deg f} f[1]~~.
\ee It is clear that $[[1]]$ is an automorphism of $\cG$, which we
call the {\em twist} of $[1]$. When $\cG$ is unital, we have
$\id_a[[1]]=\id_{a[[1]]}$ because $\deg (\id_a)=0$.

Notice that the restrictions of $[1]$ and $[[1]]$ to the subcategory
$\cG^0$ coincide. When $\cG={\cal A}^\bullet$ for some
associative category with shifts, this remark allows us to view $[1]$
and $[[1]]$ as different extensions of the shift functor of $\cA$. Also 
notice that the twist of $[[1]]$ recovers $[1]$, i.e. $[[ [[1]] ]]=[1]$. 

\paragraph{Observation} Let  $\rho:\Hom_{\cG}\circ (\id_\cG\times [1])\rightarrow [1]_{\Gr}\circ \Hom_{\cG}$
be the isomorphism of functors defined by $[1]$ and let
$\beta=\sigma^{-1}\circ \rho: \Hom_{\cG}\circ (\id_\cA\times [1])\rightarrow \Hom_{\cG}$ be 
the morphism of functors of degree $+1$
obtained by composing with the inverse of the signed suspension $\sigma:\id_\cA\rightarrow [1]_{\Gr}$. 
Functoriality of $\rho$ amounts to the conditions:
\be
\nn
\beta(uv)=\beta(u)v~~,~~\beta(u[1]v)=(-1)^{\deg u} u\beta(v)~~
\ee 
for all composable morphisms $u,v$ of $\cG$. These show that $[[1]]$ satisfies: 
\be
\nn
\beta(uv)=\beta(u)v~~,~~\beta(u[[1]]v)=u\beta(v)~~,
\ee 
which means that  $\rho:\Hom_{\cG}\circ (\id_\cG\times [[1]])\rightarrow [1]_{\Gr}\circ \Hom_{\cG}$
is a `twisted natural transformation', i.e. it satisfies the naturality conditions {\em without} Koszul signs.
Equivalently, the maps $\gamma_{ab}:=s_{ab}^{-1}\circ \rho_{ab}$ give a morphism of functors 
$\gamma:\Hom_{\cG}\circ (\id_\cA\times [[1]])\rightarrow \Hom_{\cG}$ of degree $+1$.  
An automorphism $[[1]]$ of $\cG$ endowed with 
isomorphisms of graded vector spaces 
$\Hom_{\cG}(a,b[[1]])\stackrel{\rho_{ab}}{\rightarrow} \Hom_{\cG}(a,b)[1]$ which are 
natural up to missing Koszul signs will be called a {\em twisted shift functor} of $\cG$.

\paragraph{Graded functors between unital associative categories with shifts.}

Let $(\cA,[1])$ and $(\cB,[1])$ be unital categories with shifts.  A
covariant {\em graded functor} from $(\cA,[1])$ to $(\cB,[1])$ is a
pair $(F,\eta)$ where $F:\cA\rightarrow \cB$ is a covariant functor
and $\eta:F\circ [1]\rightarrow [1]\circ F$ is an isomorphism of
functors. Notice that $\eta$
induces an isomorphism of functors $F\circ
[k]\stackrel{\eta^k}{\rightarrow } [k]\circ F$, given by the
compositions $\eta^k_{a}:=\eta_{a}[k-1]\circ \eta_{a[1]}[k-2]\circ
\ldots \circ \eta_{a[k-1]}:F(a[k])\rightarrow F(a)[k]$ for all $a\in
\Ob\cA$ (we have $\eta^1=\eta$).
Also notice that the morphisms of ${\rm SCat}$ are those graded
endofunctors for which $\eta$ is the identity. 
Given two graded functors $F:(\cA,[1]_{\cA})\rightarrow (\cB,[1]_{\cB})$ and 
$G:(\cB,[1]_{\cB})\rightarrow ({\cal C},[1]_{\cal C})$, their composition 
$G\circ F$ is a graded functor from $(\cA,[1]_{\cA})$ to $({\cal C},[1]_{\cal C})$
whose grading is given by $\eta^{G\circ F}=\eta^{G}\circ G(\eta^F)$, where $\eta^{F,G}$ 
are the gradings of $F$ and $G$. Using naturality of $\eta^{F}$ and $\eta^{G}$, this 
implies $(\eta^{G\circ F})^k=(\eta^G)^k\circ G(\eta^F)^k$ for all $k\in \Z$. 

\paragraph{Examples} A simple example is given by the shift functor $[1]$ 
of $\cA$, graded through the trivial isomorphism $\id:[1]\circ
[1]\rightarrow [1]\circ [1]$.  The resulting graded functor
$\sigma:=([1],\id)$ will be called the {\em unsigned graded shift
functor of $\cA$.}  Another useful example is the {\em signed graded
shift functor} $s=([1],\eta^{(s)})$, where $\eta^{(s)}: [1]\circ
[1]\rightarrow [1]\circ [1]$ is the isomorphism of functors given by
$\eta^{(s)}_a:=-\id_{a[2]}:a[1][1]=a[2]\rightarrow a[1][1]=a[2]$. In
this case, we have $(\eta_a^{(s)})^k=(-1)^k\id_{a[k+1]}$ for all $a\in
\Ob\cA$, which gives an isomorphism of functors $(\eta^{(s)})^k:
[1]\circ[k]\rightarrow [k]\circ [1]$.

\paragraph{Observation} Let $\cA$ and $\cB$ be two triangulated categories. 
Then an exact functor $F:\cA \rightarrow \cB$ is in particular a
graded functor from $(\cA,[1])$ to $(\cB,[1])$.  Given a triangulated
category $\cA$, the signed graded shift functor $s$ is exact, while the
unsigned graded shift functor $\sigma$ fails to be exact when endowed with the trivial
grading \cite{Bocklandt}. This application motivates our interest in
twisted shift functors, as well as the choice of sign in equation
(\ref{compat}) below.

\paragraph{Graded functors between unital graded associative categories with shifts.}

Given unital graded categories with shifts $({\cal F},[1])$ and
$(\cG,[1])$, a graded functor from $({\cal F},[1])$ to $({\cal
G},[1])$ is a pair $(F,\eta)$ where $F:\cA\rightarrow \cB$ is a
covariant functor of graded categories and $\eta:F\circ [1]\rightarrow
[1]\circ F$ is an isomorphism of functors. 

\paragraph{Observation} A grading $[1]\circ [1]\stackrel{\eta}{\rightarrow}[1]\circ [1]$ 
of the shift functor $[1]$ of $\cG$ obviously corresponds to the same grading
$[[1]]\circ [1]\stackrel{\eta}{\rightarrow}[1]\circ [[1]]$ of the twisted 
shift functor $[[1]]$. As for ungraded categories, we let $\eta^{(s)}$ be the grading 
given by $\eta^{(s)}_a:=-\id_{a[2]}:a[1][1]=a[2]\rightarrow a[1][1]=a[2]$. 

\paragraph{Graded completion of graded functors.}

Let $(\cA,[1]_{\cal A})$ and $(\cB,[1]_{\cal B})$ be unital categories
with shifts.  Given a covariant graded functor
$(F,\eta):(\cA,[1]_{\cA}) \rightarrow (\cB,[1]_{\cB})$, its {\em
graded completion} is the covariant graded functor of graded
categories $(F^\bullet,\eta^\bullet): (\cA^\bullet, [1]_{\cA^\bullet})
\rightarrow (\cB^\bullet, [1]_{\cB^\bullet})$ defined as follows. We
set $F^\bullet(a)=F(a)$ for all objects $a$ and:

\

\noindent $F^\bullet(u):=\eta^k_b\circ F(u)\in
\Hom_\cB(F(a),F(b)[k])=\Hom^k_{\cB^\bullet}(F^\bullet(a),F^\bullet(b))$

\

\noindent for all $u\in
\Hom^k_{\cA^\bullet}(a,b)=\Hom_\cA(a,b[k])$. Finally, we set
$\eta^\bullet=\eta$. A somewhat lengthy diagrammatic argument using the
definition of $\eta_a^k$ shows that $\eta^\bullet:F^\bullet\circ
[1]_{{\cal A}^\bullet}\rightarrow [1]_{{\cal B}^\bullet}\circ
F^\bullet$ {\em is} an isomorphism of functors. Here $
[1]_{\cA^\bullet}$ and $ [1]_{\cB^\bullet}$ are the shift functors of
$\cA^\bullet$ and $\cB^\bullet$ induced by the shift functors of $\cA$
and $\cB$

\paragraph{Examples} The graded completion of the unsigned graded shift functor $\sigma=([1]_{\cA},\id)$ of 
$\cA$ is the unsigned graded shift functor $\sigma^\bullet=([1]_{\cA^\bullet},\id)$ of $\cA^\bullet$. The graded
completion of the signed graded shift functor $s=([1]_{\cA},\eta^{(s)})$ is the signed graded twisted shift functor
$s^\bullet=([[1]],\eta^{(s)})$ of $\cA^\bullet$.

\paragraph{Observation}

Recall that $[1]_{\cA^\bullet}$ is defined through the identifications
$\Hom^k_{\cA^\bullet}(a,b)\stackrel{\rm def}{=}
\Hom_{\cA}(a,b[k])\stackrel{[1]}{\approx} \Hom_{\cA}(a[1],b[k][1])=
\cA(a[1],b[1][k])\stackrel{\rm
def}{=}\Hom^k_{\cA^\bullet}(a[1],b[1])$, where the isomorphism in the
middle is trivial since $[k]\circ [1]=[1]\circ [k]$.  On the other
hand, $[[1]]_{\cA^\bullet}$ results by performing the identification
in the middle through the nontrivial isomorphism
$(\eta^{(s)})^{-k}:[k]\circ [1]\approx [1]\circ [k]$. 

\paragraph{Idempotent completion.}

Given a unital associative category $\cA$ and an object $a\in \Ob {\cal
A}$, an idempotent endomorphism of $a$ is an element $e\in \Hom_{\cal
A}(a,a)$ satisfying $e^2=e$. We say that $e$ is {\em split} if there
exists an object $b$ of $\cA$ and morphisms $s\in \Hom_{\cal
A}(b,a)$, $r\in \Hom_\cA(a,b)$ such that $r\circ s={\rm id}_b$ and
$e=s\circ r$. In this case, $b$ is called a retraction of $a$; two
retractions of $a$ are easily seen to be isomorphic.  When $\cA$
is additive, this condition amounts to the existence of a direct sum
decomposition $a=\ker e\oplus \im e$ for any idempotent endomorphism
$e$. 

We say that $\cA$ is {\em idempotent complete}\footnote{One
also says that $\cA$ is {\em Karoubi closed}, {\em Karoubian} or
{\em split-closed} and also that $\cA$ has split idempotents.} if
any idempotent of $\cA$ is split. Given an associative category
$\cA$, its {\em idempotent completion}\footnote{Also called the
Karoubi closure, or split closure of $\cA$.} is the smallest
idempotent complete category $\cA^\pi$ which contains $\cA$ as a full
subcategory; it has the property that any object of $\cA^\pi$ is
a retract of an object of $\cA$. Any associative category ${\cal
A}$ admits an idempotent completion, determined up to an equivalence
which restricts to the identity on $\cA$. In particular, ${\cal
A}$ is idempotent complete iff $\cA^\pi\approx \cA$. A
canonical representative can be constructed by taking the objects of
$\cA^\pi$ to be the pairs $(a,e)$ where $a$ is an object of
$\cA$ and $e$ an idempotent endomorphism of $a$, and setting $\Hom_{{\cal
A}^\pi}((a,e),(a',e')):=e'\circ \Hom_\cA(a,a')\circ e\subset \Hom_{\cal
A}(a,a')$, with the composition of morphisms induced from ${\cal
A}$. We will always understand $\cA^\pi$ to be this canonical
representative. It is clear that $\cA^\pi$ is Hom-finite iff $\cA$ is. 

Any functor $F:\cA\rightarrow \cB$ extends
to a functor $F^\pi:\cA^\pi\rightarrow \cB^\pi$ defined through 
$F^\pi(a,e)=(F(a), F_{aa}(e))$ and $F_\pi(f)=F_{a_1a_2}(f)$ for all 
$f\in \Hom_{\cA^\pi}((a_1,e_1),(a_2,e_2))$. Given a functor $G:\cB\rightarrow \cC$, 
we have $(G\circ F)^\pi=G^\pi\circ F^\pi$. Given $F,G:\cA\rightarrow \cB$ and a 
natural transformation $\phi:F\rightarrow G$, one has a natural transformation 
$\phi^\pi:F^\pi\rightarrow G^\pi$ given by 
$\phi_{(a,e)}=G_{aa}(e)\circ \phi_a\circ F_{aa}(e)=G_{aa}(e)\circ \phi_a=
\phi_a\circ F_{aa}(e)$, where the last two equalities follow from naturality of 
$\phi$ and $e^2=e$. 

The idempotent completion of an additive category is
additive.  Less obviously \cite{BS}, the idempotent completion of a
triangulated category is canonically triangulated. It is also known
\cite{Neeman} that a triangulated category with countable coproducts
is idempotent complete.

An almost identical discussion holds for graded categories. In this case, an 
idempotent $e$ is required by definition to be a homogeneous morphism of degree zero, 
and the same condition is imposed on the maps $r,s$ for a split idempotent. The 
idempotent completion is now a graded category, which is constructed 
the same way as above.

\paragraph{Idempotent completion of (graded) categories with shifts.}

Given an associative category with shifts $(\cA,[1])$, the idempotent completion 
of the shift functor $[1]^\pi$ is a shift functor for $\cA^\pi$; thus 
$(\cA^\pi,[1]^\pi)$ is a category with shifts. Given a graded functor 
$(F,\eta):(\cA,[1]_\cA)\rightarrow (\cB,[1]_\cB)$, its idempotent 
completion $F^\pi$ is a graded functor from $(\cA,[1]_\cA)$ to $(\cB,[1]_\cB)$ when endowed with the 
grading $\eta^\pi$. Similar statements holds for graded categories 
with shifts.

\subsection{Duality structures and cyclic structures}

\paragraph{Duality structures.}

Let $\cA$ be an associative category. A {\em duality structure} on $\cA$ 
is a pair $(S,\phi)$ where $S$ is an automorphism of $\cA$ and 
$\phi$ is a family of linear maps $\phi_{ab}:
\Hom_\cA(a, b)\rightarrow \Hom_\cA(b,S(a))^{\rm v}$ which are natural in $a$ and $b$.
We say that  $(S,\phi)$ is {\em nondegenerate} if $\cA$ is Hom-finite and all $\phi_{ab}$ 
are  bijective.

\paragraph{Observation} One can generalize the notion of duality structure by allowing $S$ to be
an autoequivalence of $\cA$. This doesn't give anything essentially new since such $S$ correspond to 
automorphisms of a skeleton of $\cA$. In this paper, we are interested mostly in the case 
$S=[D]=[1]^D$ for some shift functor $[1]$; then $S$ is an automorphism. 

\

Fixing a duality structure $(S,\phi)$ on $\cA$, consider the pairings 
$( ~)_{ab}:\Hom_\cA(a,b)\otimes \Hom_\cA(b,S(a))\rightarrow \C$  
defined through $(u\otimes v)_{ab}:=\phi_{ab}(u)(v)$ (we will 
tacitly identify $( ~)_{ab}$ with the associated bilinear form). 
Then naturality of $\phi$ amounts to the conditions: 
\be
\nn
(u\circ f, v)_{a'b}=( u, S(f)\circ v)_{a,b}~~
\forall f\in\Hom_{\cA}(a',a)~,~u\in\Hom_{\cA}(a,b)~,~v\in\Hom_{\cA}(b,S(a'))~
\ee 
and
\be
\nn
(g\circ u, v)_{a,b'}=(u, v\circ g)_{a,b}~~
\forall g\in\Hom_{\cA}(b,b')~,~u\in\Hom_{\cA}(a,b)~,~v\in\Hom_{\cA}(b',S(a))~.
\ee
Hence a duality structure on $\cA$ amounts to an automorphism $S$ together 
with pairings $(~)_{ab}$ obeying the conditions above. Nondegeneracy of $(S,\phi)$ 
amounts to Hom-finiteness of $\cA$ plus nondegeneracy of these pairings as bilinear forms.

When $\cA$ is unital, a duality structure  can also be described as follows.  
Defining linear maps $tr_a:\Hom_{\cA}(a,S(a))\rightarrow \C$ via $tr_a(u):=(\id_a,u)_{a,a}$, the first condition above is equivalent with: 
\be
\label{ttr}
(u,v)_{a,b}=tr_b(S(u)\circ v)~~, 
\ee
while the second becomes: 
\be
\label{tprop}
tr_a(u\circ v)=tr_b(S(v)\circ u)~~\forall v\in \Hom_{\cA}(a,b)~,~u\in\Hom_{\cA}(b,S(a))~~.
\ee
Hence a duality structure on $\cA$ amounts to an autoequivalence $S$ together 
with linear maps $tr_a$ obeying (\ref{tprop}). The information 
carried by the traces is equivalent with that carried by $\phi_{ab}$, which can be 
recovered as $\phi_{ab}(u)(v)=tr_b(S(u)\circ v)$.

The notion of duality structure has an obvious graded analogue, which we spell out in detail for later reference. 
Thus a duality structure on a graded associative category $\cG$ is a pair $(S,\phi)$ where $S$  is an 
automorphism of $\cG$ as a graded associative category and 
$\phi_{ab}:\Hom_\cG(a,b)\rightarrow \Hom_\cG(b,S(a))^{\rm v}$ are morphisms of graded vector spaces 
which are 
natural in $a$ and $b$. Equivalently, this is specified by degree zero pairings 
$(~)_{ab}:\Hom_\cG(a,b)\otimes \Hom_\cG(b,S(a))\rightarrow \C$ satisfying the graded 
analogues of the conditions above: 
\be
\nn
( u f, v)_{a'b}=( u, S(f) v)_{a,b}~~
{\rm for}~~f\in\Hom_\cG(a',a)~,~u\in\Hom_\cG(a,b)~,~v\in\Hom_\cG(b,S(a'))~
\ee 
and
\!\!\!\!\!\!\!\!\!\!\be
\nn
( g u, v)_{a,b'}=(-1)^{\deg g(\deg u +\deg v)}( u, v g)_{a,b}~~
{\rm for}~~ g\in\Hom_\cG(b,b')~,~u\in\Hom_\cG(a,b)~,~v\in\Hom_{\cG}(b',S(a))~.
\ee
When $\cG$ is unital, this data is encoded by degree zero linear maps $tr_a:\Hom_\cG(a,S(a))\rightarrow \C$ 
subject to the graded analogue of conditions (\ref{tprop}): 
\be
\label{gtprop}
tr_a(u v)=(-1)^{\deg u ~\deg v}tr_b(S(v) u)~~{\rm for}~~ v\in \Hom_\cG(a,b)~,~u\in\Hom_\cG(b,S(a))~~.
\ee
Once again, the traces and bilinear pairings are related through (\ref{ttr}). We also have 
$tr_a(u):=(\id_a,u)_{a,a}$. We say that  $(S,\phi)$ is {\em nondegenerate} if $\cG$ is degreewise 
Hom-finite and all $\phi_{ab}$ are  bijective.

\paragraph{Idempotent completion of duality structures.} 

Consider a unital associative category with shifts $(\cA,[1])$. 
A duality structure $(S,\phi)$ on $\cA$ extends to a duality structure $(S^\pi, \phi^\pi)$ on 
the shift completion $\cA^\pi$, where  
$\phi^\pi_{(a_1,e_1),(a_2,e_2)}:\Hom_{\cA^\pi}((a_1,e_1),(a_2,e_2))=
e_2\circ \Hom_{\cA}(a_1,a_2)\circ e_1\rightarrow 
\Hom_{\cA^\pi}((a_2,e_2),S^\pi(a_1,e_1))^{\rm v}=
[S(e_1)\circ \Hom_{\cA}(a_2,S(a_1))\circ e_2]^{\rm v}$ is defined by the restriction: 
\be
\nn
\phi^\pi_{(a_1,e_1),(a_2,e_2)}(x)=\phi_{a_1,a_2}(x)|_{S(e_1)\circ \Hom_{\cA}(a_2,S(a_1))
\circ e_2}~~\forall x\in \Hom_{\cA^\pi}((a_1,e_1),(a_2,e_2))
\ee
This amounts to defining pairings and traces $tr^\pi$ on $\cA^\pi$ by restricting the pairings 
of $\cA$: 
\be
\nn
(u,v)^\pi_{(a_1,e_1), (a_2,e_2)}:=( u, v)_{ab}
\ee
for all $u\in \Hom_{\cA^\pi}((a_1,e_1),(a_2,e_2))\subset \Hom_{\cA}(a_1,a_2)$ and 
$v\in \Hom_{\cA^\pi}((a_2,e_2),S^\pi(a_1,e_1))\subset \Hom_{\cA}(a_2,S(a_1))$, i.e. 
\be
\nn
tr_{(a,e)}^\pi(u)=tr_a(u)~~\forall u\in \Hom_{\cA^\pi}((a,e),(S(a),S(e)))=S(e)\Hom_{\cA}(a,S(a))e\subset \Hom_{\cA}(a,S(a))~~.
\ee

Recall that $\cA^\pi$ is Hom-finite iff $\cA$ is. In this case, 
it is easy to see that $(S^\pi,\phi^\pi)$ is nondegenerate iff $(S,\phi)$ is. Indeed, the direct implication is 
obvious while the inverse implication follow from the relations
$tr^\pi_{(a,e)}(u exe)=tr_a(uexe)=tr_a(S(e) uex)=tr_a(ux)$, which hold for all $u\in S(e)\Hom_{\cA}(a,S(a))e$ and 
all $x\in \Hom_{\cA}(a,a)$. These show that $tr^\pi_a(uv)$ vanishes for all $v\in e\Hom_{\cA}(a,a)e$ 
iff. $tr_a(ux)$ vanishes for all $x\in \Hom_{\cA}(a,a)$, which requires $x=0$ by non-degeneracy of $tr_a$.

An almost identical discussion holds for duality structures on graded associative categories.

\paragraph{Cyclic structures on graded associative categories.}

Given an integer $D$, a {\em $D$-cyclic structure} on a graded associative category $\cG$
is a collection of morphisms of graded vector spaces\footnote{Of course, one can also work 
with the morphisms $\psi_{ab}[D]: \Hom_\cG(a,b)[D]\rightarrow \Hom_\cG(b,a)^{\rm v}$.}
$\psi_{ab}:\Hom_\cG(a,b)\rightarrow \Hom_\cG(b,a)[D]^{\rm v}$ which are natural in $a$ and $b$.
Equivalently, it is specified by a collection of degree zero pairings 
$\langle ~\rangle_{ab}:\Hom_\cG(a,b)\otimes \Hom_\cG(b,a)\rightarrow \C[-D]$ (viewed as linear 
maps of degree $-D$ from $\Hom_\cG(a,b)\otimes \Hom_\cG(b,a)$ to $\C$) which satisfy: 
\be
\label{gcy0bil}
\langle u f, v\rangle_{a'b}=\langle u, f v\rangle_{a,b}~~
\forall f\in\Hom_\cG(a',a)~,~u\in\Hom_\cG(a,b)~,~v\in\Hom_\cG(b,a')~
\ee 
as well as
\be
\label{gcy1bil}
\langle f,g\rangle_{a,b}=(-1)^{\deg f~\deg g}\langle g,f\rangle_{b,a}~,~{\rm for}~ f\in 
\Hom_\cG(a,b)~,~g\in \Hom_\cG(b,a)~~.
\ee
The homogeneity condition on the pairings amounts to the selection rule:
\be
\label{selrule}
\langle f,g\rangle =0~~{\rm unless}~~\deg f+\deg g=D~~.
\ee
Let $s_{ab}^{D}:\Hom_{\cG}(a,b)\rightarrow \Hom_{\cG}(a,b)[D]$ be the map of degree $-D$ 
induced by the suspension operator; we denote its inverse by $s_{ab}^{-D}$ through a slight abuse of notation. 
Then the relation with the maps $\psi_{ab}$ is given by $\langle u, v\rangle_{ab}=\psi_{ab}(u)(s_{ba}^D(v))$. 
When $\cG$ is unital, we can also describe this 
in terms of traces $\tr_a:\Hom_{\cG}(a,a)\rightarrow \C[-D]$ (viewed as homogeneous linear maps of degree $-D$ from 
$:\Hom_{\cG}(a,a)$ to $\C$) defined through $\tr_a(u)=\langle \id_a, u\rangle_{aa}$.
These satisfy: 
\be
\nn
\tr_a(u v)=(-1)^{\deg u ~\deg v}\tr_b(v u)~~{\rm for}~~ v\in \Hom_\cG(a,b)~,~u\in\Hom_\cG(b,a)~~.
\ee
The bilinear pairings can be recovered as $\langle u, v\rangle_{ab}=\tr_{b}(uv)=(-1)^{\deg u~\deg v}\tr_a(vu)$. 

The cyclic structure is called {\em non-degenerate} 
when $\cG$ is degreewise $\Hom$-finite and $\psi_{ab}$ are bijective for all $a,b\in \Ob\cA$; the latter 
condition amounts to nondegeneracy of the bilinear pairings $\langle~,~\rangle_{ab}$. 

\paragraph{Idempotent completion of cyclic structures on graded categories.}

Given a $D$-cyclic structure $\tr$ on a unital graded category $\cG$, we define traces $\tr^\pi$ on $\cG^\pi$ via: 
\be
\nn
\tr^\pi_{(a,e)}(u):=\tr_a(u)~~\forall u\in \Hom_{\cG^\pi}((a,e),(b,e'))=e'\Hom_\cG(a,b)e\subset \Hom_\cG(a,b)~~.
\ee
These define a $D$-cyclic structure on $\cG^\pi$, called the {\em idempotent completion} of $\tr$. 

Recall that $\cG^\pi$ is degreewise Hom-finite iff $\cG$ is. In this case, it is easy to see that $\tr^\pi$ is nondegenerate 
iff $\tr$ is. 

\paragraph{Cyclic structures on categories with shifts}

Consider a graded associative category with shifts $(\cG,[1])$, and let 
$\rho_{ab}:\Hom_{\cG}(a,b[1])\stackrel{\approx}{\rightarrow}  \Hom_{\cG}(a,b)[1]$ be the isomorphism 
determined by the shift functor of $\cG$. 
For any integer $n$, let $\rho_{ab}^n$ denote the induced isomorphism 
$\Hom_{\cG}(a,b[n])\stackrel{\approx}{\rightarrow}\Hom_{\cG}(a,b)[n]$ (when $n>0$, we have
$\rho_{ab}^n:=\rho_{ab}[n-1]\circ \ldots \circ\rho_{ab[n-2]}[1]\circ \rho_{ab[n-1]}$ etc).

Using these isomorphisms, a $D$-cyclic structure $\psi$ on ${\cal G}$ 
can be identified with a duality structure $(S,\phi)$ on $\cG$ having $S=[[D]]$ and 
$\phi_{ab}=(\rho_{ba}^D)^{\rm v}\circ \psi_{ab}:\Hom_{\cG}(a, b)\rightarrow \Hom_{\cG}(b,a[D])^{\rm v}$.
This corresponds to setting $( u,v)_{ab}=\langle u, (s_{ba}^{-D}\circ\rho_{ba}^D)(v)\rangle_{ab}$ 
for the bilinear pairings $(~,~)_{ab}:\Hom_{\cG}(a,b)\times \Hom_{\cG}(b,a[D])\rightarrow \C$ of 
$(S,\phi)$.
When $\cG$ is unital, the traces $tr_a$ of 
$(S,\phi)$ are related to those of $\psi$ via $tr_a=\tr_a\circ s_{aa}^{-D}\circ \rho^D_{aa}$. Thus:

\

{\em Given a graded associative category with shifts $(\cG,[1])$, a $D$-cyclic structure on $\cG$ amounts to 
a duality structure $(S,\phi)$ on $\cG$ having $S=[[D]]$.}

\

Restricting $\phi$ to morphisms of degree zero gives a duality structure $(S_0,\phi_0)$ on the associative 
category $\cG^0$ having $S_0=[D]$. This justifies the following:

\paragraph{Definition} A duality structure $(S,\phi)$ on an associative category with shifts 
$(\cA,[1])$ is called a {\em $D$-cyclic structure} if $S=[D]$.

\paragraph{Observation} Since $[1]^\pi$ is the shift functor of $\cA^\pi$, it is clear that the idempotent
completion of a $D$-cyclic structure on $\cA$ (viewed as a duality structure) is again a $D$-cyclic structure.

\subsection{Graded duality structures}
\label{sec:shift_invar}

\paragraph{Graded duality structures on unital associative categories with shifts.}

Let  $(\cA,[1])$ be a unital associative category with shifts.
A {\em graded duality structure} on $(\cA,[1])$ is a triple $(S,\phi,\eta)$ 
where $(S,\phi)$ 
is a duality structure on $\cA$ and $(S,\eta)$ is a 
graded functor, subject to the compatibility conditions: 
\be
( u[1],v[1])_{a[1]b[1]} =-( u ,\eta_a[-1]\circ v )_{ab}
~~{\rm for} ~u\in \Hom_{\cA}(a,b)~,~ v\in \Hom_{\cA}(b,S(a[1])[-1])~~~~~~~~~~~~~
\ee
for all $a,b\in \Ob\cA$. Naturality of $\eta$ implies that these conditions are equivalent with: 
\be
\label{compat}
tr_{a[1]}(u[1])=-tr_a(\eta_a[-1]\circ u)~~\forall a\in \Ob\cA~,~\forall u\in 
\Hom_{\cA}(a,S(a[1])[-1])~~.~~~~~~~~~
\ee  

\paragraph{Graded duality structures on unital graded associative categories with shifts.}

A graded duality structure on a unital graded associative category
with shifts $(\cG,[1])$ is a triple $(S,\phi,\eta)$ where
$(S,\phi)$ is a duality structure on $\cA$ and $(S,\eta)$ is a graded
functor, subject to the same compatibility conditions as above. In
this case, the maps $\phi_{ab}$, bilinear pairings
$(~,~)_{ab}$ and traces $tr_a$ are homogeneous of degree
zero.

\paragraph{Idempotent completion of graded duality structures.} 

Let  $(\cA,[1])$ be a unital associative category with shifts. The idempotent completion of 
a duality structure $(S,\phi,\eta)$ on $(\cA,[1])$ is the triple $(S^\pi,\phi^\pi,\eta^\pi)$, 
which is easily seen to be a duality structure on $(\cA^\pi,[1]^\pi)$. A similar definition 
works for graded associative categories.

\paragraph{Graded completion of graded duality structures.} 

Consider a unital associative category with shifts $(\cA,[1])$ endowed 
with a graded duality structure $(S,\phi,\eta)$. The {\em graded
completion} of $(S,\phi,\eta)$ is the graded duality structure
$(S^\bullet,\phi^\bullet,\eta^\bullet)$, where
$(S^\bullet,\eta^\bullet)$ is the graded completion of the graded
functor $(S,\eta)$,  while
$\phi_{ab}^\bullet:\Hom_{\cA^\bullet}(a,b)\rightarrow
\Hom_{\cA^\bullet}(b,S^\bullet(a))^{\rm v}$  are defined through the
compositions:
\be
\nn
\!\!\!\!\!\!\!\Hom^k_{\cA^\bullet}(a,b)=\Hom_{\cA}(a,b[k])\stackrel{\phi_{ab}}{\rightarrow} 
\Hom_{\cA}(b[k],S(a))^{\rm v}\stackrel{[k]^{\rm v}}{\rightarrow}
\Hom_{\cA}(b,S(a)[-k])^{\rm v}= ([\Hom_{\cA^\bullet}(b,S^\bullet(a))]^{\rm v})^{k}~~,
\ee
i.e.
\be
\nn
\phi_{ab}^\bullet(u,v)=\phi_{a,b[k]}(u)(v[k])~~{\rm for}~~ u\in \Hom_{\cA}(a,b[k])~~{\rm and}~~ v\in \Hom_{\cA}(b,S(a)[-k])~~.  
\ee
Hence the pairings of $(S^\bullet,\phi^\bullet,\eta^\bullet)$ are the
homogeneous bilinear maps of degree zero
$(~,~)^\bullet_{ab}:\Hom_{\cA^\bullet}(a,b)\times
\Hom_{\cA^\bullet}(b,S^\bullet(a))\rightarrow \C$ given by:
\be
\nn
( u, v)^\bullet_{ab}=(u, v[k])_{ab[k]}~~{\rm for} ~~u\in \Hom^k_{\cA^\bullet}(a,b)~ {\rm and} ~v\in
\Hom^{-k}_{\cA^\bullet}(b,S(a))~~,
\ee
while the traces are the homogeneous degree zero linear maps
$tr^\bullet_a:\Hom_{\cA^\bullet}(a,S^\bullet(a)) \rightarrow \C$
obtained from $tr_a$ through extension by zero. It is clear that $(S^\bullet,\phi^\bullet,\eta^\bullet)$  
is nondegenerate iff  $(S,\phi,\eta)$ is. 

\paragraph{Observation}{To check condition (\ref{gtprop}), it suffices to notice that (\ref{compat}) implies the relation: 
\be
\label{gr_serre}
tr_a(g*f)=(-1)^k tr_{b}(S^\bullet(f)*g)~~\forall f\in \Hom_\cA(a,b[k])~,~\forall g\in 
\Hom_{\cA}(b,S(a)[-k])~~,
\ee  
where $*$ is the composition of morphisms in $\cA^\bullet$. }

\subsection{Shift-equivariant cyclic structures}

When $(\cG,[1])$ is a graded category with shifts, a 
$D$-cyclic structure $\psi$ on $\cG$ is called {\em shift-equivariant} if its pairings satisfy:
\be
\label{gcy2bil}
\langle f[1], g[1]\rangle_{a[1],b[1]}=(-1)^{D+1}\langle f,g\rangle_{ab}~~\forall f\in 
\Hom_\cG(a,b)~,~\forall g\in \Hom_\cG(b,a)~.
\ee
Using the selection rule (\ref{selrule}), this becomes:
\be
\label{gcy2bil_twisted}
\langle f[[1]], g[[1]]\rangle_{a[[1]],b[[1]]}=-\langle f,g\rangle_{ab}~~\forall f\in 
\Hom_\cG(a,b)~,~\forall g\in \Hom_\cG(b,a)~.
\ee
When $\cG$ is unital, the shift equivariance condition takes the following form in terms of traces:
\be
\nn
\tr_{a[1]}(f[1])=(-1)^{D+1}\tr_a(f)\Leftrightarrow \tr_{a[[1]]}(f[[1]])=-\tr_a(f)~~.
\ee
Recall that $\psi$ can be identified with a duality structure 
$([[D]],\phi)$ on $\cG$, whose pairings and traces we denote by $(~)_{ab}$ and $tr_a$ (the
traces are defined only in the unital case). Naturality and shift-equivariance amount to the conditions:
\bea
(u\circ f, v)_{a'b}&=&( u, f[[D]]\circ v)_{a,b}~~
\forall f\in\Hom_{\cG}(a',a),~u\in\Hom_{\cG}(a,b),~v\in\Hom_{\cG}(b,a'[D])~\label{gcy0nu}\\
( u, v)_{a,b}&=&( v, u[[D]])_{b,a[D]}~~, 
~~\forall u\in\Hom_{\cG}(a,b)~~\forall v \in\Hom_{\cG}(b,a[D])\label{gcy1nu}\\
( u[1],v[1])_{a[1]b[1]}&=&(-1)^{D+1}(u,v)_{ab}~~\forall 
u\in \Hom_{\cG}(a,b)~~\forall v\in \Hom_{\cG}(b,a[D])~~,\label{gcy2nu}~~~~~~~~~~~~~~~~~~~~~~~~~~~~
\eea
or, in terms of traces:
\bea
tr_a(u\circ v)&=&(-1)^{\deg u~\deg v}tr_b(v[[D]] \circ u)~,~\forall v\in \Hom_{\cG}(a,b)~,~\forall 
u\in\Hom_{\cG}(b,a[D])~~\label{cy1}~~~~~~~~~\\
tr_{a[1]}(u[1])&=&(-1)^{D+1}tr_a(u)~,~\forall u\in 
\Hom_{\cG}(a, a[D])~~.~~~~~~~~~~~~\label{cy2}~~
\eea
When $\cG$ is unital, the shift-equivariance condition (\ref{cy2}) means that $([[D]],\phi,(\eta^{(s)})^D)$ is a graded 
duality structure on $(\cG, [1])$. Thus: 

\

{\em A shift-equivariant $D$-cyclic structure on a unital graded category with shifts $(\cG,[1])$ amounts 
to a graded duality structure of the form $([[D]],\phi, (\eta^{(s)})^D))$ on $(\cG, [1])$.}

\

Restricting to degree zero morphisms gives a $D$-cyclic structure on $\cG^0$ which satisfies the shift-equivariance
condition: 
\be
\nn
(u[1],v[1])_{a[1]b[1]}=(-1)^{D+1}(u,v)_{ab}\Leftrightarrow tr_{a[1]}(u[1])=(-1)^{D+1}tr_a(u)~~. 
\ee
This justifies the following:

\paragraph{Definition}{A shift-equivariant $D$-cyclic structure on a unital associative category with shifts 
is a graded duality structure  $(S,\phi,\eta)$ such that  $(S,\eta)= s^D=([D],(\eta^{(s)})^D)$ as graded functors. 
In this case, $(S,\phi)$ is a $D$-cyclic structure on $(\cA,[1])$.} 

\

Giving a shift-equivariant $D$-cyclic structure on $(\cA,[1])$ amounts 
to giving bilinear forms
$(~)_{ab}:\Hom_{\cA}(a,b)\otimes \Hom_{\cA}(b,a[D])\rightarrow \C$ 
which obey the conditions: 
\bea
(u\circ f, v)_{a'b}&=&( u, f[D]\circ v)_{a,b}~~
\forall f\in\Hom_{\cG}(a',a),~u\in\Hom_{\cG}(a,b),~v\in\Hom_{\cG}(b,a'[D])~\label{gcy0nu0}\\
( u, v)_{a,b}&=&( v, u[D])_{b,a[D]}~~, 
~~\forall u\in\Hom_{\cG}(a,b)~~\forall v \in\Hom_{\cG}(b,a[D])\label{gcy1nu0}\\
( u[1],v[1])_{a[1]b[1]}&=&(-1)^{D+1}(u,v)_{ab}~~\forall 
u\in \Hom_{\cG}(a,b)~~\forall v\in \Hom_{\cG}(b,a[D])~~,\label{gcy2nu0}~~~~~~~~~~~~~~~~~~~~~~~~~~~~
\eea 
or, equivalently, to giving traces $\tr_a: \Hom_{\cA}(a,b[D])\rightarrow \C$ which 
satisfy:
\bea
tr_a(u\circ v)&=&(-1)^{\deg u~\deg v}tr_b(v[D] \circ u)~,~\forall v\in \Hom_{\cG}(a,b)~,~\forall 
u\in\Hom_{\cG}(b,a[D])~~\label{cy10}~~~~~~~~~\\
tr_{a[1]}(u[1])&=&(-1)^{D+1}tr_a(u)~,~\forall u\in 
\Hom_{\cG}(a, a[D])~~.~~~~~~~~~~~~\label{cy20}~~
\eea 

\

Given a shift-equivariant $D$-cyclic structure $([D],\phi, (\eta^{(s)})^D)$ on $(\cA,[1])$, its graded completion 
gives a shift-equivariant $D$-cyclic structure on $(\cA^\bullet, [1])$.  Indeed, we have 
$s^\bullet=([D],(\eta^{(s)})^D)^\bullet=([[D]],(\eta^{(s)})^D)$. Thus: 

\

{\em Giving a shift-equivariant $D$-cyclic structure on an
associative category with shifts $(\cA,[1])$ amounts to giving a
shift-equivariant $D$-cyclic structure on the graded category with shifts
$(\cA^\bullet,[1])$. Moreover, the cyclic structure on $(\cA,[1])$ is
non-degenerate iff the cyclic structure on $(\cA^\bullet, [1])$ is
nondegenerate.}

\

Hence the notion of shift-equivariant cyclic structure is well-behaved
under the inverse equivalences $^\bullet$ and $^0$ of Subsection
\ref{sec:shift_invar}.

\paragraph{Observation} Given a shift-equivariant $D$-cyclic structure on a unital 
associative or unital graded associative category with shifts, it is easy to see that 
its idempotent completion is again a shift-equivariant $D$-cyclic structure on 
the idempotent-completed category.

\subsection{Equivalence of cyclic structures.}

Given a unital category $\cA$, we let $Z(\cA)$ denote its center,
defined as the unital associative $\C$-algebra of endomorphisms of the
identity functor $\id_{\cA}$. Its elements are given by collections
$f=(f_a)_{a\in \Ob\cA}$ with $f_a\in \Hom_{\cA}(a,a)$ such that
$f_b\circ u=u\circ f_a$ for all $u\in \Hom_{\cA}(a,b)$ and any objects
$a,b$ of $\cA$. The invertible elements under multiplication form the
group of automorphisms $\Aut(\id_\cA)$.  They are given by collections
$f$ as above with the supplementary condition that $f_a\in
\Aut_{\cA}(a)$ for all $a$. When $\cA$ has shifts, an element $f\in
\Aut(\id_\cA)$ is called {\em shift-invariant} if $f_{a[1]}=f_a[1]$
for all $a$; such elements form a subgroup $\Aut_{\rm si}(\id_\cA)$.

Let $\cT^D(\cA,[1]_{\cA})$ be the set of all $D$-cyclic
structures on $(\cA,[1]_{\cA})$ and $\cT^D_{\rm
se}(\cA,[1]_{\cA})$, $\cT^D_{\rm nd}(\cA,[1]_{\cA})$ be the
subsets of those $D$-cyclic structures which are shift-equivariant
respectively nondegenerate (the latter is defined when $\cA$ is Hom-finite). 

We say that two $D$-cyclic structures  $tr$, $tr'$ on $(\cA,[1])$ 
are {\em equivalent} if there exists $f\in
\Aut(\id_\cA)$ such that $tr'_a(u):=tr_a(u\circ
f_a)=tr_a(f_{a[D]}\circ u)$ for all $u\in \Hom_{\cA}(a,a[D])$. In this case, 
we write  $tr'\approx tr$, which defines an equivalence relation on $\cT(\cA,[1])$.

When $tr$ and $tr'$ are shift-equivariant, we say that they are {\em
graded equivalent} if there exists $f\in \Aut_{\rm si}(\id_\cA)$ with
the property above. In this case, 
we write  $tr'\approx_{\gr} tr$, giving an equivalence relation on 
$\cT_{\rm se}(\cA,[1])$. It is clear that $\approx$ and $\approx_{\rm gr}$ are 
compatible with nondegeneracy and that $tr\approx_{\gr}tr'\Rightarrow tr\approx tr'$. 
We set $T^D (\cA,[1]):=\cT^D(\cA,[1])/\approx$ 
as well as  $T^D_{\rm se} (\cA,[1]):=\cT^D_{\rm se}(\cA,[1])/\approx_{\gr}$ and 
$T^D_{\rm nd} (\cA,[1]):=\cT^D_{\rm nd }(\cA,[1])/\approx$.

The Yoneda lemma implies that any two {\em non-degenerate} $D$-cyclic structures $\tr,\tr'$ on 
$(\cA,[1])$ are equivalent through a uniquely-determined $f\in \Aut(\id_\cA)$. When 
the cyclic structures are also shift-invariant, we must have $f\in \Aut_{\rm si}(\id_\cA)$, so 
in this case equivalence implies graded equivalence.
Thus $T^D_{\rm nd} (\cA,[1])$ and $[\cT^D_{\rm nd }(\cA,[1])\cap\cT^D_{\rm se}(\cA,[1])] /\approx_{\gr}$
have a single element.

\subsection{Transport of cyclic structures}

\paragraph{Morphisms of graded functors.}{Let $(\cA,[1]_{\cA})$ and $(\cB, [1]_{\cB})$ be two unital associative
categories with shifts.  Given two graded functors
$F,G:(\cA,[1]_{\cA})\rightarrow (\cB,[1]_{\cB})$, a {\em morphism of
graded functors} from $F$ to $G$ is a natural transformation
$\phi:F\rightarrow G$ such that the following diagram commutes for all
$a\in \Ob\cA$:
\be
\nn
\begin{array}{ccc}
 F(a[1]) & \stackrel{\eta_a^F}{\longrightarrow } & F(a)[1] \\
\phi_{a[1]}~\downarrow~~~~  &~ & ~~~~~~\downarrow ~\phi_a[1]\\
G(a[1]) & \stackrel{\eta_a^G}{\longrightarrow} & G(a)[1]~
\end{array}
\ee
An isomorphism of graded functors is a morphism of graded functors
such that all $\phi_a$ are isomorphisms.  We say that two graded
functors $F,G$ are {\em graded isomorphic} if there exists an
isomorphism of graded functors from $F$ to $G$; in this case, we write
$F\approx_{\gr} G$. This defines an equivalence relation on the class
of all graded functors from $(\cA,[1]_{\cA})$ to $(\cB,[1]_{\cB})$.
We say that $F$ and $G$ are {\em ungraded isomorphic} if they are
isomorphic as usual functors; we write this weaker equivalence
relation as $F\approx G$. Of course, we have $F\approx_{\gr}
G\Rightarrow F\approx G$.}

\paragraph{Graded equivalence and weak graded equivalence.}

A graded functor $F:(\cA,[1]_{\cA})\rightarrow (\cB,[1]_{\cB})$ is
called an {\em equivalence of graded categories} (or graded equivalence) if there exists a
graded functor $H:(\cB,[1]_{\cB})\rightarrow (\cA,[1]_{\cA})$ such
that $F\circ H\approx_{\gr} \id_{\cB}$ and $H\circ
F\approx_{\gr}\id_{\cA}$.  If there exists a graded equivalence from
$(\cA,[1]_{\cA})$ to $(\cB,[1]_{\cB})$, then we write
$(\cA,[1]_{\cA})\approx_{\gr} (\cB,[1]_{\cB})$ and say that
$(\cA,[1]_{\cA})$ and $(\cB,[1]_{\cB})$ are {\em graded-equivalent}.

A {\em weak graded equivalence} is a graded functor
$F:(\cA,[1])\rightarrow (\cB,[1])$ with the property that that there
exist a graded functor $H:(\cB,[1])\rightarrow (\cA,[1])$ such that
$F\circ H\approx \id_{\cB}$ and $H\circ F\approx \id_{\cA}$ (notice
that these relations involve usual isomorphism of functors rather than
isomorphism of graded functors).  In particular, $F$ is an equivalence
of unital associative categories. Given two graded functors 
$F,G$ from $\cA$ to $\cB$, we have the obvious implication 
$F\approx_{\gr} G \Rightarrow F\approx G$.

\paragraph{Pull-back of cyclic structures}

Let $(\cA,[1]_{\cA})$ and $(\cB, [1]_{\cB})$ be two unital associative
categories with shifts and $F:(\cA,[1])\rightarrow (\cB,[1])$ a graded
functor, with grading isomorphism $\eta^F:F\circ [1]_{\cA}\rightarrow [1]_{\cB}\circ F$. 
Given a $D$-cyclic structure $tr$ on $\cB$, define maps
$tr^F_a:\Hom_{\cA}(a,a[D])\rightarrow \C$ through: \be \nn tr_a^F
(u):=tr_{F(a)}((\eta_a^F)^D\circ F(u))~~.  \ee Using naturality of
$\eta^F$ and the cyclicity property of $tr$, it is not hard to check
that $tr^F_a$ define a $D$-cyclic structure on $(\cA,[1]_{\cA})$. We
call this the {\em pull-back} of $tr$ through $F$. 
When $tr$ is shift-equivariant, then
it is not hard to check that $tr^F$ is again shift-equivariant.  When
$F$ is fully faithful and $tr$ is nondegenerate, then $tr^F$ is
nondegenerate. Define a map $F^*:{\cal
T}^D(\cB,[1]_{\cB})\rightarrow \cT^D(\cA,[1]_{\cA})$ through
$F^*(tr):=tr^F$. The remarks above show that $F^*(\cT^D_{\rm
se}(\cB,[1]_{\cB}))\subset \cT^D_{\rm se}(\cA,[1]_{\cA})$; when
$F$ is fully faithful, we also have $F^*(\cT^D_{\rm
nd}(\cB,[1]_{\cB}))\subset \cT^D_{\rm nd}(\cA,[1]_{\cA})$.
The pull-back of cyclic structures has the following properties, 
whose detailed proofs we leave to the reader:
 
I. Given graded functors $F:(\cA,[1]_{\cA})\rightarrow
(\cB,[1]_{\cB})$ and $G:(\cB, [1]_{\cB})\rightarrow (\cC, [1]_{\cC})$
and a $D$-cyclic structure $tr$ on $(\cC, [1]_{\cC})$, we have:
\be
\nn
tr^{G\circ F}=(\tr^G)^F~~.
\ee 
Thus $(G\circ F)^*=F^*\circ G^*$.

II. Given two graded functors $F,G :(\cA,[1]_{\cA})\rightarrow
(\cB,[1]_{\cB})$ and a $D$-cyclic structure on $(\cB,[1]_{\cB})$, we have the implication:
\be
\nn
F\approx_{gr} G\Rightarrow tr^F=tr^G~~.
\ee
In particular, a graded equivalence $(\cA,[1]_{\cA})\approx_{\gr} (\cB,[1]_{\cB})$ induces a 
bijection\footnote{Namely $F^*\circ H^*=\id_{\cT^D(\cB,[1]_{\cB})}$ and $H^*\circ
F^*=\id_{\cT^D(\cA,[1]_{\cA})}$ for a graded quasi-inverse 
pair of graded functors $\cA {\tiny \begin{array}{c} \stackrel{F}{\longrightarrow} \\ \stackrel{\longleftarrow}{G}\end{array}}\cB$.} 
$\cT^D(\cA,[1]_{\cA})\approx \cT^D(\cB,[1]_{\cB})$.
This bijection is compatible with restriction to the subsets of shift-equivariant respectively
nondegenerate cyclic structures, giving bijections $\cT_{\rm
se}^D(\cA,[1]_{\cA})\approx \cT^D_{\rm se}(\cB,[1]_{\cB})$ and
$\cT^D_{\rm nd}(\cA,[1]_{\cA})\approx \cT^D_{\rm
nd}(\cB,[1]_{\cB})$. 

III. Let $F:(\cA,[1])\rightarrow (\cB,[1])$ be a {\em faithful} graded functor and $tr, tr'$ be two $D$-cyclic
structures on $(\cB,[1])$. Then: 
\be
\nn
tr\approx tr'\Rightarrow tr^F \approx tr'^F. 
\ee
It follows that $F$ and $G$ determine the same map ${\hat F}={\hat G}$ from $T^D(\cB,[1]_{\cB})$ to 
$T^D (\cA,[1]_{\cA})$. 

\

{\em Sketch of proof}. Indeed, if $g\in \Aut(\id_{\cA})$ satisfies 
$tr'_A(U)=tr_A(U\circ g_A)$ for all $U\in \Hom_{\cB}(A,A[D])$, then  the element $f$ of $\Aut(\id_{\cA})$ defined through: 
\be
\nn
F(f_a)=g_{F(a)}~~\forall a\in \Ob\cA
\ee
satisfies $\tr'^F_a(u)=\tr^F(u\circ f_a)$ for all $u\in \Hom_{\cA}(a,a[D])$. It follows that $F^*$ descends to a well-defined 
map ${\hat F}:T^D (\cB,[1]_{\cB})\rightarrow T^D (\cA,[1]_{\cA})$.

IV. Let  $F,G:(\cA,[1])\rightarrow (\cB,[1])$ be two weak graded equivalences and $tr$ a $D$-cyclic structure on $(\cB, [1]_{\cB})$. 
Then:
\be
\nn
F\approx G\Rightarrow tr^F\approx tr^G~~.
\ee

\

\noindent {\em Sketch of proof.} Fixing a usual isomorphism of functors $\phi:F\rightarrow G$ (which need not be an 
isomorphism of graded functors), we have $\tr^F_a(u)=\tr^G_a(u\circ f_a)=\tr^G_a(f_{a[D]}\circ u)$, 
where $f\in \Aut(\id_{\cA})$ is determined by the formula: 
\be
\nn
G(f_{a[D]})=\phi_{a[D]}\circ [(\eta^F_a)^D]^{-1}\circ (\phi_a[D])^{-1}\circ (\eta_a^G)^D~~.
\ee

\

V. A weak graded equivalence $F:(\cA,[1])\rightarrow (\cB,[1])$ induces a bijection
${\hat F}:T^D (\cB,[1]_{\cB})\stackrel{\approx}{\rightarrow} T^D (\cA,[1]_{\cA})$. 
Indeed, we have $F\circ H\approx \id_{\cB}$ and $H\circ F\approx \id_{\cA}$ for 
some graded functor $H: (\cB,[1])\rightarrow (\cA,[1])$, which gives 
${\hat F}\circ {\hat G}=\id_{T^D(\cB,[1])}$ and ${\hat G}\circ {\hat F}=\id_{T^D(\cA,[1])}$.

\subsection{The triangulated case}
\label{sec:tria_duality}

Let $\cT$ be a triangulated category, endowed with its shift
functor $[1]$. A {\em pre-Serre duality  structure} on $\cT$ is a graded
duality structure $(S,\phi,\eta)$ such that $S$ is exact (in particular, it preserves
finite direct sums). When $\cT$ is Hom-finite, a {\em Serre
duality structure} on $\cT$ is a pre-Serre structure which is
nondegenerate as a duality structure. In this case, $S$ is called a
{\em Serre functor} \cite{BK_Serre}. Given an integer $D$, the
category $\cT$ is called {\em $D$-Calabi-Yau} if it admits a
Serre structure $(S,\phi,\eta)$ such that $(S,\eta)\approx
(s^D,(\eta^s)^D)$ as graded functors. In this case, $(S,\phi,\eta)$ is
a nondegenerate and shift-equivariant $D$-cyclic structure on ${\cal
T}$. In this subsection, a (co)homological functor on $\cT$ means
a linear $\Vect$-valued (co)homological functor which preserves direct
sums.

Let $\cT$ be a triangulated category and ${\cal U}\subset
\Ob\cT$, and set $\Z{\cal U}:=\{a[n]|a\in {\cal U},n\in \Z\}$.
We recall the following:

\paragraph{Lemma}   Assume that ${\cal U}$ triangle generates $\cT$ or that $\cT$ 
is idempotent complete and ${\cal U}$ Karoubi generates $\cT$.
Let $F,G:\cT\rightarrow \Vect$ be two homological or two
cohomological functors on $\cT$ and $\phi:F\rightarrow G$ a
natural transformation.  Then $\phi$ is an isomorphism of functors iff
the linear map $\phi_a\in \Hom_\C(F(a),G(a))$ is bijective for all
$a\in \Z{\cal U}$.

\

\noindent {\em Proof.} ~The direct implication is obvious, while the inverse implication is a 
trivial application of the five-lemma. In the Karoubi case we also make use of the additivity of $F$ and $G$. 

\

\noindent This implies the following criterion:

\paragraph{Proposition} Assume that ${\cal U}$ triangle generates $\cT$ or that $\cT$ 
is idempotent complete and ${\cal U}$ Karoubi generates $\cT$.
Then a graded duality structure $(S,\phi)$ on $\cT$ is strictly
non-degenerate iff the linear map $\phi_{ab}:\Hom_{\cal
T}(a,b)\rightarrow \Hom_\cT(b,S(a))^{\rm v}$ is bijective for all
$a,b\in \Z{\cal U}$.

\

\noindent {\em Proof} ~The direct implication is obvious. For the
inverse implication, let $F,G:\cT^{\rm op}\times \cT
\rightarrow \Vect$ be the linear bifunctors defined through
$F(a,b)=\Hom_\cT(a,b)$ and $G(a,b)=\Hom_\cT(b,S(a))^{\rm v}$
(with the obvious actions on morphisms). Fixing $a\in \Z{\cal U}$
gives homological functors $F_a:=F(a,\cdot)$ and $G_a:=\Hom_{\cal
T}(\cdot ,S(a))^{\rm v}$ on $\cT$, related by the natural
transformation $\phi_a:=\phi_{a,\cdot}$ Applying the lemma, we find
that $\phi_a$ is an isomorphism of functors for all $a\in \Z{\cal U}$
so $\phi_{ab}$ is bijective for all $a\in \Z{\cal U}$ and $b\in {\cal
T}$.  Now fix $b\in \cT$ and consider the cohomological functors
$F^b:=F(\cdot,b)$ and $G^b:=\Hom_\cT(b ,S(\cdot))^{\rm v}$ on
$\cT$. These are related by the natural transformation
$\phi_{\cdot,b}$, which gives gives bijections $\phi_{ab}$ when $a\in
\Z{\cal U}$.  A second application of the lemma shows that $\phi_{ab}$
is a bijection for all $a,b\in \Ob\cT$.

Taking $S=s^D$ (with $s$ the twisted shift functor of $\cT$), the
proposition translates as follows into the language of cyclic
structures.

\paragraph{Corollary}  Assume that ${\cal U}$ triangle generates $\cT$ or that $\cT$ 
is idempotent complete and ${\cal U}$ Karoubi generates $\cT$.
Then a shift-equivariant $D$-cyclic structure on $(\cT,[1])$
specified by the bilinear forms $(~,~)_{ab}$ is
non-degenerate iff $(~,~)_{ab}:\Hom_\cT(a,b)\times
\Hom_\cT(b,a[D])\rightarrow \C$ are non-degenerate for
all $a,b\in \Z{\cal U}$. 

\

\noindent Notice that the condition in the last proposition is
equivalent to nondegeneracy of the pairing
$\langle~, ~\rangle_{ab}:\Hom_{\cT^\bullet}(a,b)\times \Hom_{{\cal
T}^\bullet}(b,a)\rightarrow \C$ for all $a,b\in {\cal U}$.

\section{Symmetric $\infty$-inner products on an \ainf algebra}
\label{sec:cyc_geom}

In this appendix, we consider the notion of symmetric $\infty$-inner
product on an \ainf algebra $A$.  After reviewing the geometric
description of \ainf algebras in terms of formal noncommutative
Q-manifolds (FNQ-manifolds) \cite{konts_formal, Lazarev, nc}, we
define a symmetric $\infty$-inner product on $A$ as a noncommutative
pre-symplectic form on the associated FNQ-manifold.  A symmetric
$\infty$-inner product can also be viewed as a countable sequence of
multilinear maps on $A$ which are compatible with the \ainf products
and obey certain graded symmetry conditions, showing that this notion
is a particular case of that considered in \cite{Tradler}.  The
$\infty$-inner product is called flat when all higher multilinear maps
vanish, so one is left with a bilinear pairing on $A$, which is graded
symmetric and compatible with the \ainf products; this corresponds to
a cyclic pairing as considered in the body of the paper as well as in
\cite{HLL, Costello, nc}.  Hence $\infty$-inner products generalize
cyclic pairings. Their main advantage over the latter is that a
morphism of \ainf algebras pulls-back \inf-inner products to
\inf-inner products, a statement which fails in general for cyclic
pairings.  After discussing pull-backs of symmetric $\infty$-inner
products through \ainf morphisms, we recall the noncommutative Darboux
theorem of \cite{konts_formal} and introduce nondegeneracy and
homological nondegeneracy conditions, showing that homological
nondegeneracy is preserved when pulling back through \ainf
quasi-isomorphisms.  We also show that a dGA endowed with a cyclic
pairing admits a cyclic minimal model, i.e a minimal model on which
the pairing transports to a flat \inf-product. This result generalizes
the `gauge-fixing' construction of a cyclic minimal model (Section
\ref{sec:minmodel}) and is used in Section \ref{sec:dG}.

\subsection{\ainf algebras as formal noncommutative Q-manifolds}

\paragraph{The bar construction.}

Consider a $\Z$-graded $A_\infty$ algebra $A$ with (degree one) suspended infinity products 
$r_n:A[1]^{\otimes n}\rightarrow A[1]$ ($n\geq 1$). As usual, we let $|x|$ denote 
the degree of homogeneous elements $x\in A$, and ${\tilde x}:=|x|-1$ denote the suspended degree 
(the degree of $x$ as an element of $A[1]$). We also let $m_n$ be defined through 
$r_n:=s\circ m_n\circ (s^{-1})^{\otimes n}$, where $s:A\rightarrow A[1]$ is the suspension operator. 

It is well-known that the \ainf products are equivalently encoded by a degree $+1$ codifferential 
$\delta$ on the graded coassociative coalgebra ${\bar T}(A[1]):=
\oplus_{n\geq 1} A[1]^{\otimes n}$, known as the reduced tensor coalgebra of the vector 
space $A[1]$. The suspended products $r_n$ can be recovered through: 
\be
\nn
r_n:=p_1\circ \delta|_{A[1]^{\otimes n}}~~,
\ee
where $p_1:{\bar T} (A[1])\rightarrow A[1]$ is the projection on the first component of 
${\bar T} A[1]$. The differential graded coalgebra $BA:=({\bar T}(A[1]),\delta)$ 
is known as the bar dual of $A$. 
As explained in \cite{Hasegawa}, the bar construction 
provides a functor $B:{\rm Alg}_\infty\rightarrow {\rm Cogc}$ from the category ${\rm Alg}_\infty$ of \ainf algebras 
to the category ${\rm Cogc}$ of cocomplete\footnote{A coassociative coalgebra $(C,\delta)$ 
is called cocomplete if $\cup_{n} C_n=C$, where 
$C_n=\ker\Delta^{(n+1)}$ are the components of the so-called 
primitive filtration $C_1\subset C_2\subset \ldots \subset C$. Here $\Delta^{(2)}=\Delta$ is 
the comultiplication of $C$ and $\Delta^{(n+1)}:=(\id_C^{\otimes n-1}\otimes \Delta)\circ 
\Delta^{(n)}:C\rightarrow C^{\otimes (n+1)}$ for all $n\geq 2$.}
differential graded coalgebras. This functor induces an equivalence 
between ${\rm Alg}_\infty$ and the full subcategory ${\rm Cogtr}$ 
of ${\rm Cogc}$ consisting of those 
objects whose underlying graded coalgebra is isomorphic with the reduced tensor coalgebra of a 
graded vector space. The category ${\rm Cogc}$ admits a Quillen model structure with respect 
to which all objects are cofibrant. With respect to this structure, an 
object of ${\rm Cogc}$ is fibrant iff it lies in ${\rm Cogtr}$. Hence the category 
${\rm Alg}_\infty$ 
of \ainf algebras identifies via the bar construction with the full subcategory ${\rm Cogtr}$ of 
fibrant objects in ${\rm Cogc}$. Similarly, the category ${\rm Alg}_\infty$  admits a `model 
structure without limits'\footnote{I.e. all axioms of a model category are satisfied except for 
the existence of finite limits and colimits, which is replaced by a weaker axiom (see 
\cite{Hasegawa}).} 
whose weak equivalences are the \ainf quasi-isomorphisms, and whose 
cofibrations/fibrations are the \ainf morphisms $\varphi:A_1\rightarrow A_2$ such that 
$\varphi_1$ is 
a monomorphism/epimorphism. The bar construction maps these into 
weak equivalences, respectively into cofibrations/fibrations of ${\rm Cogc}$ \cite{Hasegawa}. 

Two morphisms
$F,G: C_1\rightarrow C_2$ of differential graded coalgebras are called {\em  
(classically) homotopy equivalent} if there exists a degree -1 linear map 
$H:C_1 \rightarrow C_2$ such that: 
\be
\nn 
\Delta_2 \circ H=(F\otimes H+H\otimes G)\circ \Delta_1 ~~{\rm and}~~
F-G=\delta_2\circ H+H\circ \delta_1~~, 
\ee
where $\Delta_i$ are the comultiplications on $C_i$. 
When $C_1$ and $C_2$ belong to ${\rm Cogtr}$, it is shown in \cite{Hasegawa} that $F$ and $G$ 
are homotopy equivalent iff they are left homotopy equivalent in the sense of model 
categories. Moreover, the bar functor $B$ interchanges 
homotopy equivalent morphisms of \ainf algebras with homotopy equivalent morphisms of coalgebras. 
These results imply that homotopy equivalence of \ainf algebras is an equivalence relation, and 
that a morphism of \ainf algebras is a quasi-isomorphism iff it is a homotopy equivalence. In 
particular, any quasi-isomorphism of \ainf algebras admits an inverse up to homotopy. 
Moreover, the bar dual of the minimal model of an \ainf algebra is a minimal model 
in ${\rm Cogc}$ (in the sense of model categories) of the bar dual of the original \ainf algebra.
These results give a complete description of \ainf algebras in the language of cocomplete differential 
graded coalgebras. 

\paragraph{Description through formal noncommutative $Q$-manifolds.}

The geometric interpretation of \ainf algebras arises by further dualizing this picture 
\cite{konts_formal, Lazarev}. 
We say that a graded associative $\C$-algebra is {\em formal} if it is the inverse limit of an 
inverse system of nilpotent and finite-dimensional associative graded 
$\C$-algebras. Such  algebras 
are topological algebras with respect to the inverse limit topology. They form a category whose 
morphisms are continuous morphisms of graded associative algebras. Given such an algebra $B$
we let $\Der_l(B)$ denote the space of its {\em continuous} left derivations. 
As explained in the appendix 
of \cite{Lazarev}, the category of cocomplete graded coalgebras is antiequivalent with the 
category of formal graded algebras, an 
antiequivalence being given by taking the vector space dual $C\rightarrow C':=\Hom_\C(C,\C)$, and  
inverse antiequivalence given by taking the dual as a topological vector space, 
an operation which we denote by $^*$.  
A {\em formal noncommutative Q-manifold}
(FNQ-manifold) is a pair $(B,Q)$ where $B$ is a formal graded $\C$-algebra whose dual 
coalgebra $B^*$ belongs to ${\rm Cogtr}$ (i.e. is a reduced tensor coalgebra) and 
$Q\in \Der_l(B)$ 
is a {\em homological derivation}, i.e. 
a continuous left derivation of degree $-1$ 
which squares to zero. A morphism of FNQ-manifolds is a 
morphism of formal graded algebras which commutes with the homological derivations. 
Applying the dualization functor, we find that the category ${\rm Cogtr}$ of fibrant cocomplete 
differential graded coalgebras is antiequivalent with the category of FNQ-manifolds, 
with inverse anti-equivalence given by taking the topological dual. The results of 
\cite{Hasegawa} recalled above now 
translate trivially into the dual language of formal NQ-manifolds. 
In particular, we find the following dual description of quasi-isomorphisms of \ainf algebras. 
Given two formal noncommutative Q-manifolds $B_2$ and $B_1$, two morphisms
$f,g:B_2 \rightarrow B_1$  of noncommutative Q-manifolds 
are called {\em homotopy equivalent} if there exists a degree $+1$ linear 
map $h:B_2 \rightarrow B_1$ such that: 
\be
\nn 
h(xy)=(-1)^{\tilde x} f(x)h(y)+h(x)g(y) ~~{\rm and}~~f-g=Q_1\circ h+h\circ Q_2~~. 
\ee
The morphism $f$ is called 
 a homotopy equivalence if it becomes invertible in the category obtained by 
taking homotopy equivalence classes of morphisms. The discussion above shows that 
quasi-isomorphisms (a.k.a homotopy equivalences) of \ainf algebras from $A_1$ to $A_2$ correspond to homotopy equivalences of 
formal noncommutative Q-manifolds from $(B A_2[1])'$ to $(B A_1[1])'$.

\subsection{Symmetric \inf-inner products}

\paragraph{Noncommutative Cartan calculus.}

Given an \ainf algebra $A$, let $B:=B'A[1]:= (BA[1])'=\prod_{n\geq
1}\Hom_\C(A[1]^{\otimes n},\C)$ denote the corresponding formal
NQ-manifold, endowed with the homological vector field $Q:=\delta^{\rm
v}$. Notice that $B$ is bigraded\footnote{All gradings are in the
sense of direct {\em product} decompositions.}; we denote the grading
induced from $A[1]$ by a tilde and place it in first position; the
'tensor grading' is placed in the second position.  Let $\Omega B$ be
the formal dGA of noncommutative forms over $B$, whose differential we
denote by $D$.  Consider the Karoubi complex ${\cal C}(B)=\Omega
B/[\Omega B, \Omega B]$ of $B$, whose differential (induced from $B$)
we again denote by $d$. Notice that $\Omega B$ is trigraded, with the
rank grading placed in first position, grading induced from $A[1]$
(tilde grading) in second position and grading induced from the tensor
grading of $B$ in third position. We let ${\cal C}^n(B)$ be the
homogenous components of ${\cal C}(B)$ with respect to the form rank
grading. Notice that each ${\cal C}^n(B)$ carries the bigrading
induced from $B$.  We let $\pi_c:\Omega B\rightarrow {\cal C}(B)$ be
the canonical surjection, and use the notation
$(\omega)_c:=\pi(\omega)$ for any $\omega\in \Omega B$. We also let
$\Der_l^g(B)$ (resp. $\Der_l^{(h,g)}(\Omega B)$) be the spaces of
continuous left derivations of $B$ (resp $\Omega B$) which are
homogeneous of tilde degree $g$ respectively of bidegree $(h,g)$ with
respect to the (rank, tilde) bigrading.  Given a homogeneous
derivation $\theta\in \Der_l^{\tilde \theta}(B)$, we have well-defined
derivations $i_\theta\in \Der_l^{-1,{\tilde \theta}}(\Omega B)$ and
$L_\theta\in \Der_l^{0,{\tilde \theta}}(\Omega B)$ which play the role
of contraction and Lie derivative. These are uniquely determined by
the conditions (in the conventions of \cite{nc}):
\be
\nn
i_\theta(x)=0~~,~~i_\theta(dx)=\theta(x)~~\forall x\in B
\ee
and: 
\be
\nn
L_\theta(x)=\theta(x)~~,~~L_\theta(dx)=d\theta(x)~~\forall x\in B~~.
\ee
They descend to well-defined
operators on the Karoubi complex, which we denote by the same letters. The 
operators $d, L_\theta$ and $i_\theta$ satisfy the classical identities:
\be
L_\theta=\left[i_\theta, d\right] ~,~ \left[L_\theta, i_\gamma\right]=i_{\left[\theta, \gamma\right]}~,~
\left[L_\theta, L_\gamma\right]=L_{\left[\theta, \gamma\right]}~,~\left[i_\theta, i_\gamma\right]=\left[L_\theta, d\right]=0 ~~.\nn
\ee
In particular, one finds $[L_Q,d]=0$ and 
$L_Q^2=\frac{1}{2}
[L_Q,L_Q]=\frac{1}{2}L_{[Q,Q]}=0$, where $[~,~]$ denotes the graded commutator of continuous 
left derivations (which is again a continuous left derivation). Since $L_Q$ preserves 
form rank, this allows us to consider 
the homology of $L_Q$ on ${\cal C}^n(B)$ and on the subspaces 
${\cal C}^n(B)_{cl}$ of closed $n$-forms; we will denote these by $H_{L_Q}(\ldots)$, where 
$\ldots$ stands for the corresponding complex. 
We let $Z_{L_Q}({\cal C}^n(B))$ etc. denote the 
corresponding spaces of $L_Q$-cycles. A morphism $\phi:B_1\rightarrow B_2$ of formal 
noncommutative Q-manifolds (i.e. a continuous morphism of the underlying topological algebras) 
induces a morphism of differential graded algebras $\phi_*:\Omega B_1\rightarrow \Omega B_2$. 
It is easy to check that this obeys all expected properties, in particular 
$d_2\circ \phi_*=\phi_*\circ d_1$ and $L_{Q_2}\circ \phi_*=
\phi_*\circ L_{Q_1}$. The last identities imply that $\phi_*$ descends to a well-defined linear map
from $H_{L_{Q_1}}({\cal C}(B_1))$ to $H_{L_{Q_2}}({\cal C}(B_2))$, denoted by 
${\hat \phi}$. Moreover, ${\hat \phi}$ maps $H_{L_{Q_1}}({\cal C}(B_1)_{cl})$ into $H_{L_{Q_2}}({\cal C}(B_2)_{cl})$.
We also have:
\bea
L_{\phi \circ \theta \circ \phi^{-1}}&=&\phi_* \circ L_\theta \circ \phi_*^{-1}\nn\\
i_{\phi \circ \theta \circ \phi^{-1}}&=&\phi_* \circ i_\theta \circ \phi_*^{-1}\nn~~,
\eea
for any associative algebra automorphism $\phi$ of $B$.

\paragraph{Geometric approach to symmetric \inf-inner products.}

A {\em noncommutative pre-symplectic~} form on $B$ is a closed
two-form $\omega\in {\cal C}^2(B)_{cl}$ which is homogeneous (of degree
$-{\tilde \omega}$) with respect to the grading induced from $A[1]$
(tilde grading).  Given such a form, the pair $(B,\omega)$ is called a {\em presymplectic 
FNQ-manifold}. A symplectomorphism $\phi:(B_1,\omega^{(1)})\rightarrow (B_1,\omega^{(2)})$ between 
presymplectic FNQ-manifolds is a morphism of $NQ$-manifolds such that $\phi_*(\omega^{(1)})=\omega^{(2)}$.

Given a pre-symplectic form $\omega$, we can expand it as
$\omega=\oplus_{n\geq 0}{\omega_n}$, where $\omega_n$ are its
homogeneous components with respect to the tensor product grading.  Explicitly, one
has:
\be
\label{omega_expansion}
\omega_{n-2}=\frac{1}{n}\omega_{a_1\ldots a_{i-1} \underline {a_i} a_{i+1}\ldots a_{n-1}
\underline{a_n}} (s^{a_1}\ldots s^{a_{i-1}} ds^{a_i} s^{a_{i+1}} \ldots s^{a_{n-1}} ds^{a_n})_c
\ee
where $(s^a)$ is a topological basis of $A_1[1]'$ 
suspended dual to a basis $(e_a)$ of $A$ (we use implicit 
summation over repeated indices). Here 
$\omega_{a_1\ldots a_{i-1} \underline {a_i} a_{i+1}\ldots a_{n-1}
\underline{a_n}}$ are graded-cyclic 
complex coefficients, and the underline on the indices $a_i$ and $a_n$ indicates the position of $ds^{a_i}$ 
and $ds^{a_n}$ in the noncommutative 
differential monomial appearing in (\ref{omega_expansion}). These coefficients can be used to 
define homogeneous linear maps $\omega_{i,n}:A[1]^{\otimes n }\rightarrow \C[{\tilde \omega}]$ 
via the relations: 
\be
\nn
\omega_{i,n}(e_{a_1}\otimes \ldots \otimes e_{a_n}):=
-\omega_{a_1\ldots a_{i-1} \underline {a_i} a_{i+1}\ldots a_{n-1}\underline{a_n}}~~,
\ee
where the sign is chosen for agreement with \cite{nc}. Notice that $\omega_{i,n}$ satisfy 
the graded antisymmetry properties: 
\be
\label{gco}
\omega_{i,n}(x_1\otimes \ldots \otimes x_n)=(-1)^{1+({\tilde x_1}+\ldots +{\tilde x}_i)
({\tilde x}_{i+1}+\ldots +{\tilde x}_n)}
\omega_{n-i,n}(x_{i+1}\otimes \ldots \otimes x_n\otimes x_1\otimes \ldots \otimes x_n)
\ee
Defining $\langle ~\rangle_{i,n}:A^{\otimes n }\rightarrow \C[{\tilde
\omega}+n]$ via $\omega_{i,n}=\langle ~\rangle_{i,n}\circ
(s^{-1})^{\otimes n}$ for all $n\geq 2$ and $1\leq i\leq n-1$ gives a
countable sequence of homogeneous multilinear maps on $A$ which
satisfy a graded symmetry property derived from (\ref{gco}).  The
integer $D:={\tilde \omega}+2$ will be called the {\em dimension} of
this collection of multilinear maps.

For $n=2$, this gives a single pairing $\langle ~\rangle_{1,2}$ which
is a graded-symmetric bilinear form of degree $-D$ on $A$ and can be
described invariantly as follows \cite{nc}.  Expand
$\omega_0=-\frac{1}{2}\sum_{i}{(df_idg_i)_c} $, with $f_i,g_i\in
A[1]'$. Then $\langle ~\rangle_{1,2}$ is given by:
\be
\nn
\langle x,y\rangle_{1,2}=(-1)^{\tilde x}\sum_i{f_i(xg_i(y))}~~\forall x,y\in A~~. 
\ee

A pre-symplectic form is called $Q$-{\em compatible} if $L_Q\omega=0$. In this case, 
the collection of maps $\langle~\rangle_{i,n}$ satisfies a complicated series of compatibility conditions 
with the products $r_n$, which can be obtained by computing $L_Q\omega$. The collection 
$(\langle~\rangle_{i,n})$ associated to a $Q$-compatible presymplectic form will be called 
a {\em symmetric \inf-inner product} on $A$. The pair $(A, \langle~,~\rangle_{\cdot, \cdot})$ will be called 
a {\em symmetric \ainf algebra}. 

A lengthy direct computation shows that a 
symmetric \inf-inner product is an \inf-inner product in the sense of \cite{Tradler}, subject to 
the graded symmetry conditions derived from (\ref{gco}) (which are not required in loc. cit.) As explained there, 
giving such data amounts to giving a morphism of \ainf bimodules from $A$ to 
$A^{\rm v}$, where $A$ and $A^{\rm v}$ are viewed as $A$-bimodules in the obvious manner. 

\paragraph{Cyclic structures as flat symmetric \inf-inner products.} 

We say that $\omega$ and its \inf-inner product are {\em flat} if $\omega=\omega_0$, i.e. 
$\omega_n=0$ for all $n\geq 1$. In this case, all multilinear 
forms $\langle~\rangle_{i,n}$  vanish except for the pairing $\langle ~\rangle_{1,2}$, 
which for simplicity we shall denote by $\langle ~\rangle$. 
Moreover,  equation $L_Q\omega=0$ reduces \cite{nc} 
to the cyclicity conditions (\ref{alg_cyc}). Hence:

\

\noindent {\em Giving a $D$-cyclic pairing on $A$ amounts to giving
the following equivalent data:

\noindent (a) a flat symmetric \inf-inner product on $A$ of dimension
$D$.

\noindent (b) a flat pre-symplectic form $\omega\in {\cal
C}^2(A)_{cl}$, homogeneous of degree $-{\tilde \omega}=2-D$, which
satisfies $L_Q\omega=0$.}

\paragraph{Nondegeneracy conditions.}

A symmetric \inf-inner product and its associated presymplectic form
$\omega$ are called {\em strictly nondegenerate} if $A$ is
finite-dimensional and the map $\theta\in \Der_l(B)\rightarrow
i_\theta\omega\in {\cal C}^1(B)$ is a bijection. One can check by
direct computation that the second condition amounts to the
requirement that $\langle ~\rangle_{1,2}$ is nondegenerate.  In this
case, we say that the \inf-inner product is {\em strictly
nondegenerate} and that the triplet $(B'A[1],Q,\omega)$ is a {\em
formal noncommutative symplectic manifold}.  A basic result for this
case (originally due to \cite{konts_formal}) is the noncommutative
Darboux theorem, which states that there exists a (continuous) algebra
automorphism $\phi:B'A[1]\rightarrow B' A[1]$ such that
$\phi_*(\omega)$ is a flat symplectic form (notice that $\phi$ need
not commute with $Q$).  Defining $Q_1:=\phi\circ Q\circ \phi^{-1}$,
this shows that $(B'A[1],Q,\omega)$ is isomorphic with
$(B'A[1],Q_1,\phi_*(\omega))$ as noncommutative formal symplectic
manifolds, i.e. as symmetric \ainf algebras.  Because of this result,
the theory of strictly nondegenerate symmetric \ainf algebras reduces
to the theory of strictly nondegenerate cyclic \ainf algebras.

In terms of the products $m_n$ (defined through $r_n=s\circ m_n\circ
(s^{-1})^{\otimes n}$), the first cyclicity condition in (\ref{gco})
takes the form $\langle m_1(x),y\rangle_{1,2}+(-1)^{|x|}\langle
x, m_1(y)\rangle_{1,2}=0$, which implies that $\langle
~,~\rangle_{1,2}$ descends to a graded symmetric pairing on $H(A)$. We
say that $\omega$ and the corresponding symmetric \inf-inner product
are {\em homologically nondegenerate} if $A$ is compact and this
pairing induced on $H(A)$ is nondegenerate. In the flat case, this
notion reduces to homological nondegeneracy of cyclic pairings.

\subsection{Pull-back of symmetric \inf-inner products}

Any morphism of \ainf algebras 
$\varphi:A_2\rightarrow A_1$ allows one 
to pull back a symmetric \inf-inner product $\langle~,~\rangle_{\cdot, \cdot}$ from $A_1$ to $A_2$. 
Indeed, $\varphi$ corresponds to a morphism $\phi$ of formal noncommutative 
Q-manifolds from $\B A_1[1]$ to $\B A_2[1]$, and $\phi_*$ takes closed forms to closed forms and 
$L_{Q}$-closed forms to $L_Q$-closed 
forms. We define the pullback of $\langle~,~\rangle_{\cdot, \cdot}$ through $\varphi$ to be the 
symmetric \inf-inner product  $\langle~,~\rangle^{(\phi)}_{\cdot, \cdot}$ associated with $\phi_*(\omega)$, where $\omega$ is the presymplectic 
form on $B' A_1[1]$ associated with $\langle~,~\rangle_{\cdot, \cdot}$ 
A {\em symmetric \ainf morphism} (or morphism of symmetric \ainf algebras) 
$\varphi:(A_2,\langle~,~\rangle^{(2)}_{\cdot, \cdot})\rightarrow (A_1,\langle~,~\rangle^{(1)}_{\cdot, \cdot})$ is an \ainf morphism 
such that $\langle~,~\rangle^{(2)}_{\cdot, \cdot}=[\langle~,~\rangle^{(1)}]^{(\phi)}_{\cdot, \cdot}$ This corresponds to a 
symplectomorphism (in the opposite direction) between the associated presymplectic FNQ-manifolds. A {\em symmetric \ainf 
quasi-isomorphism} is a morphism of symmetric \ainf algebras which is an \ainf quasi-isomorphism. 
The proposition below shows that pull-back through an \ainf quasi-isomorphism preserves homological nondegeneracy. 

\paragraph{Proposition} {Let $A_1$ and $A_2$ be two $A_\infty$ algebras and  $\varphi:A_2\rightarrow A_1$ an \ainf quasi-isomorphism. 
If $\langle~,~\rangle^{(1)}_{\cdot, \cdot}$ is a homologically nondegenerate symmetric \inf-inner product on $A_1$, then its pullback
through $\varphi$ is homologically nondegenerate. }

\

\noindent {\em Proof}. Let 
$\omega^{(1)}\in Z_{L_{Q_1}}({\cal C}^2(\B A_1[1])_{cl})$ and  $\omega^{(2)}:=\phi_*(\omega^{(1)})\in Z_{L_{Q_2}}({\cal C}^2(\B A_2[1])_{cl})$
be the $Q_j$-compatible presymplectic forms associated with  $\langle~,~\rangle^{(1)}_{\cdot, \cdot}$ and 
its pull-back. 
We have to show that $\omega^{(2)}$ is homologically nondegenerate if $\omega^{(1)}$ is. 
Consider the expansion in components homogeneous with respect to the tensor grading $\omega^{(1)}=\sum_{n\geq 0}{\omega^{(1)}_n}$,
where: 
\be
\nn
\omega^{(1)}_{n-2}=\omega^{(2)}_{a_1\ldots a_{i-1} \underline {a_i} a_{i+1}\ldots a_{n-1}
\underline{a_n}} (s^{a_1}\ldots s^{a_{i-1}} ds^{a_i} s^{a_{i+1}} \ldots s^{a_{n-1}} ds^{a_n})_c
\ee
is further expanded in a topological basis $(s^a)$ of $A_1[1]'$. Then $\omega^{(2)}=\sum_{n\geq 0}{\omega^{(2)}_n}$
with: 
\be
\nn
\omega^{(2)}_{n-2}=\omega^{(2)}_{a_1\ldots a_{i-1} \underline {a_i} a_{i+1}\ldots a_{n-1}
\underline{a_n}} (\phi(s^{a_1})\ldots \phi(s^{a_{i-1}}) d\phi(s^{a_i}) \phi(s^{a_{i+1}}) 
\ldots \phi(s^{a_{n-1}}) d\phi(s^{a_n}))_c
\ee
Expanding: $\phi(s^a)=
\sum_{m\geq 1}{\phi^a_{\alpha_1\ldots \alpha_m} \sigma^{\alpha_1}\ldots \sigma^{\alpha_m}}$ 
with respect to a topological basis $(\sigma^\alpha)$ of $A_2[1]'$, we find 
$\omega^{(2)}_0=\frac{1}{2}\omega^{(2)}_{\underline{\alpha}\underline{\beta}}(d\sigma^\alpha d\sigma^\beta)_c$ , 
where $\omega^{(2)}_{\underline{\alpha}\underline{\beta}}=\omega^{(1)}_{\underline{a}\underline{b}}
\phi^a_\alpha \phi^b_\beta$. Hence the bilinear form on $A_2$ induced by the `constant term' of $\omega^{(2)}$ takes the form: 
\be
\label{form_rel}
\langle x,y\rangle^{(2)}_0=\langle \varphi_1(x),\varphi_1(y)\rangle^{(1)}_0~~
\forall x, y \in A_2
\ee
where $\langle x,y\rangle^{(1)}_0$ is the corresponding form on $A[1]$ induced by $\omega^{(1)}$ 
and $\varphi_1:A_2\rightarrow A_1$ is the first component of the $A_\infty$ morphism 
$\varphi$. The latter has the expansion $\varphi_1(\epsilon_\alpha)=\phi^a_\alpha e_a$
in the bases $(e_a), (\epsilon_\alpha)$ of $A_1, A_2$ suspended dual to $(s^a), (\sigma^\alpha)$.
Since $\varphi$ is an $A_\infty$ quasi-isomorphism, we know that $\varphi_1$ is a 
quasi-isomorphism of cochain complexes from $(A_2,m^{(2)}_1)$ to $(A_1,m^{(1)}_1)$. 
Moreover, the pairing $\langle ~,~\rangle^{(1)}$ descends to a non-degenerate pairing on 
$H_{m_1^{(1)}}(A_1)$ since $\omega^{(1)}$ is homologically non-degenerate. 
Together with relation (\ref{form_rel}), these observations imply that 
$\langle ~,~\rangle^{(2)}$ descends to a non-degenerate pairing on 
$H_{m_1^{(2)}}(A_2)$. Thus $\omega^{(2)}$ is homologically non-degenerate. 

\paragraph{Cyclic minimal models as flat symmetric minimal models.}

A {\em symmetric minimal model} of the symmetric \ainf algebra
$(A,\langle~,~\rangle_{\cdot,\cdot})$ is a minimal\footnote{Minimal as
an \ainf algebra.} symmetric \ainf algebra $(A_{\rm
min},(~,~)_{\cdot,\cdot})$, together with a symmetric \ainf
quasi-isomorphism $\varphi: (A_{\rm
min},(~,~)_{\cdot,\cdot})\rightarrow
(A,\langle~,~\rangle_{\cdot,\cdot})$. The minimal model theorem for
\ainf algebras implies that any symmetric \ainf algebra has a
symmetric minimal model. The Darboux theorem implies that any {\em
flat} compact and homologically nondegenerate symmetric \ainf algebra
admits a {\em flat} symmetric minimal model: \

\paragraph{Proposition}{Let $A$ be a compact \ainf algebra endowed with a homologically nondegenerate cyclic pairing $\langle~\rangle$. 
Then there exists a finite-dimensional minimal model $A_{\rm min}$ of
$A$ and an \ainf quasi-isomorphism $\varphi:A_{\rm min}\rightarrow A$
such that the pulled back symmetric \inf-inner product is a
nondegenerate cyclic pairing on $A_{\rm min}$.}

\

\noindent{\em Proof}. Let $\omega$ be the presymplectic form on
$B'A[1]$ determined by $\langle~\rangle$.  Pick any minimal model
$A_m$ of $A$ and any quasi-isomorphism $\varphi_m: A_m\rightarrow A$
(for example, take any inverse up to homotopy of a quasi-isomorphism
from $A$ to $A_m$).  By the previous proposition, pulling back through
$\varphi_m$ gives a $Q_m$-compatible presymplectic form
$\omega_m:=\phi_m^*(\omega)$ which is homologically
nondegenerate. Since $A_m$ is minimal, $\omega_m$ is in fact
symplectic. Applying the noncommutative Darboux theorem, we pick any
$\varphi_0 \in \Aut(A_m)$ and define $\varphi:=\varphi_0\circ
\varphi_m$, $Q_{\rm min}:=\phi_0\circ Q_m\circ \phi_0^{-1}$,
$\omega_{\rm min}:=\phi_0^*(\omega_m)=\phi_*(\omega)$. Then the \ainf
algebra $A_{\rm min}$ defined by the formal noncommutative Q-manifold
$(B' A_{\rm min}[1], Q_{\rm min})$ is symplectic with flat symplectic
form $\omega_{\rm min}$.

\

A minimal model endowed with a constant pre-symplectic form as in the
proposition will be called a {\em cyclic minimal model} of
$(A,\langle~,~\rangle)$.  Since minimal models are unique up to \ainf
isomorphism, it is clear that two cyclic minimal models are related by
an isomorphism of cyclic \ainf algebras, i.e. an \ainf isomorphism
which interchanges cyclic structures under pull-back.

\paragraph{Observation}{It follows from the results of \cite{KS} that a 
cyclic minimal model is determined by the $L_Q$-homology class of $\omega$ 
up to isomorphism of cyclic \ainf algebras.}


\begin{thebibliography}{100}

\bibitem{CIL1}{C.~I.~Lazaroiu,
 ``On the structure of open-closed topological field theory in two dimensions,''
Nucl.\ Phys.\ B {\bf 603}  (2001) 497 [arXiv:hep-th/0010269].}
%%CITATION = HEP-TH 0010269;%%

\bibitem{Moore}{G.~W.~Moore,
``Some comments on branes, G-flux, and K-theory,''
Int.\ J.\ Mod.\ Phys.\ A {\bf 16} (2001) 936 [arXiv:hep-th/0012007].}
%%CITATION = HEP-TH 0012007;%%

\bibitem{HLL}{M.~Herbst, C.~I.~Lazaroiu and W.~Lerche,
``Superpotentials, A(infinity) relations and WDVV equations for open
topological strings,'' JHEP {\bf 0502} (2005) 071 [arXiv:hep-th/0402110].}
%%CITATION = HEP-TH 0402110;%%

\bibitem{Costello}{K. Costello, ``Topological conformal field theories and 
Calabi-Yau categories'', math.QA/0412149.} 

\bibitem{nc}{C. I. Lazaroiu, ``On the non-commutative geometry of topological D-branes'', 
JHEP 0511 (2005) 032 [hep-th/0507222]}

\bibitem{CIL4}{C.~I.~Lazaroiu, ``String field theory and brane superpotentials,'' 
JHEP {\bf 0110} (2001) 018 [arXiv:hep-th/0107162].}
%%CITATION = HEP-TH 0107162;%%

\bibitem{CIL5}{C.~I.~Lazaroiu,
``D-brane categories,'' Int.\ J.\ Mod.\ Phys.\ A {\bf 18} (2003) 5299
[arXiv:hep-th/0305095].}

\bibitem{HMS}{M.~Kontsevich, ``Homological algebra of mirror symmetry'', Proc.
  Internat. Congr. Math., Z\"urich, Switzerland 1994 (Basel), vol.~1,
  Birkh\"auser Verlag, 1995, 120--139.}

\bibitem{Hasegawa}{ K. Lefevre-Hasegawa, 
``Sur les A-infini cat\'egories'', Ph.D. Thesis, Universite Paris 7 (2002)
[math.CT/0310337]}

\bibitem{Fukaya_mirror}{
K.~Fukaya, ``Floer homology and mirror symmetry. II'', in {\em Minimal
  surfaces, geometric analysis and symplectic geometry} (Baltimore, MD,
  1999), Adv. Stud. Pure Math. {\bf 34}, Math. Soc. Japan, Tokyo,
  2002, pp.~31--127.}

\bibitem{BK}{A. I. Bondal, M. M. Kapranov, ``Enhanced triangulated categories'', 
Math USSR Sbornik, 1991, 70 (1), 93-107.}

\bibitem{Gaberdiel}{M.~R.~Gaberdiel, B.~Zwiebach, 
``Tensor constructions of open string theories I: Foundations,''
  Nucl.\ Phys.\ B {\bf 505} (1997) 569  [arXiv:hep-th/9705038].}

\bibitem{CIL2}{C.~I.~Lazaroiu,
``Generalized complexes and string field theory,''
JHEP {\bf 0106} (2001) 052 [arXiv:hep-th/0102122].}
%%CITATION = HEP-TH 0102122;%%

\bibitem{CIL3}{C.~I.~Lazaroiu,
``Unitarity, D-brane dynamics and D-brane categories,''
JHEP {\bf 0112} (2001) 031 [arXiv:hep-th/0102183].}

\bibitem{KS_old}{M.~Kontsevich, Y.~Soibelman, 
``Homological mirror symmetry and torus fibrations'', math.SG/0011041.}  

\bibitem{Keller_dG}{B. Keller, ``Deriving dG categories'', Ann. Scient. Ec. Norm Sup, 
$4^e$ serie, 27 (1994) 63-102}

\bibitem{konts_formal}{M. Kontsevich, ``Formal (non)commutative
symplectic geometry.'' , {\em The Gelfand Mathematical Seminars, 1990--1992},
173--187, Birkh\"auser Boston, Boston, MA, 1993.}

\bibitem{Ginzburg}{ V. Ginzburg, ``Noncommutative symplectic
geometry, quiver varieties and operads'', Math. Res. Lett {\bf 8} (2001) 3, pp
377-400.}

\bibitem{Lazarev}{ A. Hamilton, A. Lazarev, ``Homotopy algebras and
noncommutative geometry'', math.QA/0410621.}

\bibitem{Keller_intro}{ B. Keller, 
``Introduction to A-infinity algebras and modules'', 
Homology, Homotopy and Applications {\bf 3} (2001)1, pp 1--35
[math.RA/9910179]}

\bibitem{Keller_dGcat}{B. Keller, 
``On differential graded categories ``, 
[math.KT/0601185]}

\bibitem{Keller_functor}{B. Keller, 
``A-infinity algebras, modules and functor categories'', 
[math.RT/0510508]}

\bibitem{Seidel}{P. Seidel, `` Fukaya categories and Picard-Lefschetz theory'', 
preprint }

\bibitem{KS}{M. Kontsevich, Y. Soibelman, ``Notes on A-infinity algebras, 
A-infinity categories and non-commutative geometry. I'', [math.RA/0606241]}

\bibitem{Polishchuk}{A.~Polishchuk, ``Homological mirror symmetry with higher products'', [math.AG/9901025].}

\bibitem{Merkulov}{S.~A.~Merkulov, ``Strongly homotopy algebras of a K\"ahler manifold'', 
Internat. Math. Res. Notices {\bf 3} (1999) 153-164 [math.AG/9809172].} 

\bibitem{CIL6}{C.~I.~Lazaroiu, R.~Roiban,
``Holomorphic potentials for graded D-branes,''
JHEP {\bf 0202} (2002) 038 [arXiv:hep-th/0110288].}
%%CITATION = HEP-TH 0110288;%%

\bibitem{CIL7}{C.~I.~Lazaroiu, R.~Roiban and D.~Vaman,
``Graded Chern-Simons field theory and graded topological D-branes,''
JHEP {\bf 0204}  (2002)023 [arXiv:hep-th/0107063].}
%%CITATION = HEP-TH 0107063;%% 

\bibitem{CIL8}{C.~I.~Lazaroiu, R.~Roiban,
``Gauge-fixing, semiclassical approximation and potentials for graded
Chern-Simons theories,'' JHEP {\bf 0203} (2002) 022 [arXiv:hep-th/0112029].}
%%CITATION = HEP-TH 0112029;%%

\bibitem{Zwiebach_oc}{B. Zwiebach, ``Oriented Open-Closed String Theory Revisited'', 
Annals Phys. 267 (1998) 193-248 [hep-th/9705241].}

\bibitem{Diac}{D.-E. Diaconescu, ``Enhanced D-Brane Categories from String Field 
Theory'', JHEP 0106 (2001) 016 [hep-th/0104200] }

\bibitem{Bocklandt}{R. Bocklandt, ``Graded Calabi Yau Algebras of dimension 3'', 
[math.RA/0603558]}

\bibitem{BK_Serre}{
A. I.  Bondal,  M. M. Kapranov, ``Representable functors, Serre functors and mutations'', 
Math. USSR Izv. 1990, {\bf 35} (3), 519-541. }

\bibitem{Neeman}{A. Neeman, {\em Triangulated categories}, Annals of Mathematics Studies 
{\bf 148}, Princeton Univ. Press, 2001.}

\bibitem{BS}{P.~Balmer, M.~Schlichting, {\em Idempotent completion of triangulated categories}, 
J. Algebra 236(2001) 819-834.}

\bibitem{Tradler}{T. Tradler, ``Infinity-Inner-Products on A-Infinity-Algebras'', 
[math.AT/0108027]}

\end{thebibliography}
\end{document}